\documentclass[12pt]{article}
\usepackage{graphicx} 
\usepackage[left=1in,top=1in,right=1in,nohead,bottom=1in]{geometry}
\usepackage{graphicx}
\usepackage{amsmath, amssymb, enumerate}
\usepackage{natbib} 
\usepackage{algorithm}
\usepackage{algpseudocode}
\usepackage{url}
\usepackage[applemac]{inputenc}
\usepackage[T1]{fontenc}
\usepackage{amsfonts}
\usepackage{float}
\usepackage{tikz}
\usepackage{mathtools}

\usepackage{parskip}
\usepackage{multirow}
\usepackage{amsthm}

\newcommand\lo[1]{_{\nano{#1}}}

\def\nano{\scriptscriptstyle}

\def\L2T{L \lo 2 (T)}
\def\L2TX{L \lo 2 (T\lo X)}
\def\L2TX{L \lo 2 (T\lo Y)}

\def\eod{\end{document}}

\title{Predictions of damages from Atlantic tropical cyclones: a hierarchical Bayesian study on extremes}
\author{
Lindsey Dietz\textsuperscript{a}, Sakshi Arya\textsuperscript{b} \thanks{Corresponding author email: sakshi.arya@case.edu},  Vishal Subedi\textsuperscript{c}, \\Auroop R. Ganguly\textsuperscript{d} and Snigdhansu Chatterjee\textsuperscript{c}}
\date{%
 \textsuperscript{a} University of Minnesota\\ \textsuperscript{b} Case Western Reserve University\\ \textsuperscript{c} University of Maryland, Baltimore County\\
\textsuperscript{d} Northeastern University}

\begin{document}
\maketitle
\begin{abstract}
Bayesian hierarchical models are proposed for modeling tropical cyclone characteristics and their damage potential in the Atlantic basin. We model the joint probability distribution of tropical cyclone characteristics and their damage potential at two different temporal scales while considering several climate indices. First, a predictive model for an entire season is developed that forecasts the number of cyclone events that will take place, the probability of each cyclone causing some amount of damage, and the monetized value of damages. Then, specific characteristics of individual cyclones are considered to predict the monetized value of the damage they will cause. The proposed method successfully predicts damages for Atlantic cyclones during 2016-22, aligning closely with actual costs. Seasonal analysis estimates average annual damages for the United States and demonstrates high accuracy in predicting storm damages. Robustness studies are conducted and excellent prediction power is demonstrated across different data science models and evaluation techniques.
\end{abstract}

\section{Introduction}
\label{sec: intro}
Tropical cyclones or hurricanes are among the foremost natural phenomena that regularly cause great harm to human communities and infrastructure \citep{dietz2016advanced, elsner1999hurricanes,rappaport2000loss,field2018ipcc}. Many studies have been conducted on the physics of these storms \citep{emanuel2003tropical, liang2015sudden, vidale2021impact, tamizi2021physics}, their frequencies, intensities and potential for causing damage and the dependence of these on climatic features like teleconnections and sea-surface temperatures \citep{Elsner3, Elsner10,Elsner11, Elsner18, Villarini, Dailey, Elsner28, Elsner50}.  However, the relationship between economic loss and a tropical cyclone's size, intensity, storm surge,  rainfall and other important climatic factors, is complex and difficult to model explicitly \citep{Pielke2}. There is a need for employing robust statistical methodologies that can leverage the observable and quantifiable properties of tropical cyclones and related climate conditions \textit{to predict}  the risks and damages that tropical cyclones can cause. 


In this paper, we provide the Bayesian hierarchical models for predicting monetary damages for tropical cyclones at two different temporal scales. First, to aid preparation for each tropical cyclone season, we develop a predictive model that forecasts the number of cyclone events that will take place, the probability that a given cyclone will inflict damages, and the monetized value of damages for that season. Then, to aid immediate damage mitigation interventions, we predict the probability that a given tropical cyclone  may be damage-inflicting, and the amount of damage that it can cause. While our study is restricted to the Atlantic basin, conceptually it can be extended to any tropical cyclone basin.

\begin{figure}
    \centering
\includegraphics[width=\textwidth]{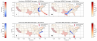}
    \caption{\textbf{The trajectory of four tropical cyclones in the North Atlantic basin, along with wind-speed (circles), central pressure (crosses), and associated monetary damages (title).}\\
    Left: Hurricanes Florence (I) and Matthew (II) made landfall in the Carolinas. Right: Hurricanes Harvey (III) and Irma (IV) made landfall in Texas Louisiana, and Florida, respectively. The darker red circles and crosses reflect more severe storm characteristics, especially close to landfall.}
    \label{fig: intro_individual}
\end{figure}

\textbf{Individual Predictions: }Past studies have established the sensitivity of annual damage to individual extreme events, thus emphasizing the need for studying individual cyclone events for better estimating the risk of extreme losses and for better financial planning \citep{halverson2018costliest,mudd2014assessing,Blake,ye2020dependence}. The relationship between maximum wind speed (\emph{maxWS}) and minimum central pressure (\emph{minCP}) in tropical cyclones has been studied for several decades \citep{Atkinson1977,WangWu2004,Kieu2010,chavas2017physical}.  It has been claimed that while the most accurate and reliable estimate of tropical cyclone intensity is the {minCP}, destructive potential is better related to {maxWS} \citep{Knaff2007}.  The frequency, intensity, and size of hurricanes are naturally also influenced by climatic factors like changes in sea surface temperature (SST), El Ni{\~n}o events and so on \citep{goldenberg1996physical,zhao2010analysis,mudd2014assessing,villarini2010modeling,patricola2016degree, wang2016hurricane, pant2019wind, Emmanuel2,lin2010modeling, rezapour2014classification, wang2021sensitivity}.
 It is of interest to understand the stress due to climate change on relationships between various characteristics of a cyclone, climatic factors, and the risks associated with tropical cyclones to human life and property, ecology, biodiversity, and various other vulnerabilities \citep{WangWu2004,Maclay2008,Mei2012,pruitt2019call,rappaport2000loss}. For example, in Figure \ref{fig: intro_individual}, we plot the storm trajectories of four different Atlantic tropical hurricanes, all of which started as high-intensity storms but varied in the amount of damage they caused eventually. It can be noted from the figure that some key factors that play a role are the minimum central pressure, maximum wind speed, and, location of landfall. Note that Harvey (2017) had attained its maximum wind speed and minimum central pressure close to making landfall, and also hit the highly populated regions of Texas and Louisiana, thus leading to enormous losses to life and property. Whereas, on the other hand, Florence made landfall in the relatively less populated regions of the Carolinas and had already dissipated in terms of the wind speed and central pressure as they made landfall-- thus causing relatively less monetary damages. Therefore, we propose to jointly model monetary damages for individual storms, along with {minCP}, {maxWS}, and other factors such as location and climate indices.  For this purpose, we present a hierarchical generalized extreme value probability distribution (GEV) framework that has not been studied before, coupling the maximum wind speed (maxWS), minimum central pressure (minCP), and financial damages data of tropical cyclones, while controlling for climate indices and other factors such as location of regions where the cyclones were recorded (details are in the Methods section). Our use of extreme value probability distributions also extends several recent studies on extreme climate phenomena and related economic analyses \citep{economou2014spatio, waylen1991modeling,chikobvu2015modelling,miniussi2020analyses, Elsner28,chavas2017physical}.

\begin{figure}[h!]
\centering
\begin{tabular}{c c}
\multicolumn{2}{c}{\includegraphics[width=0.9\textwidth]{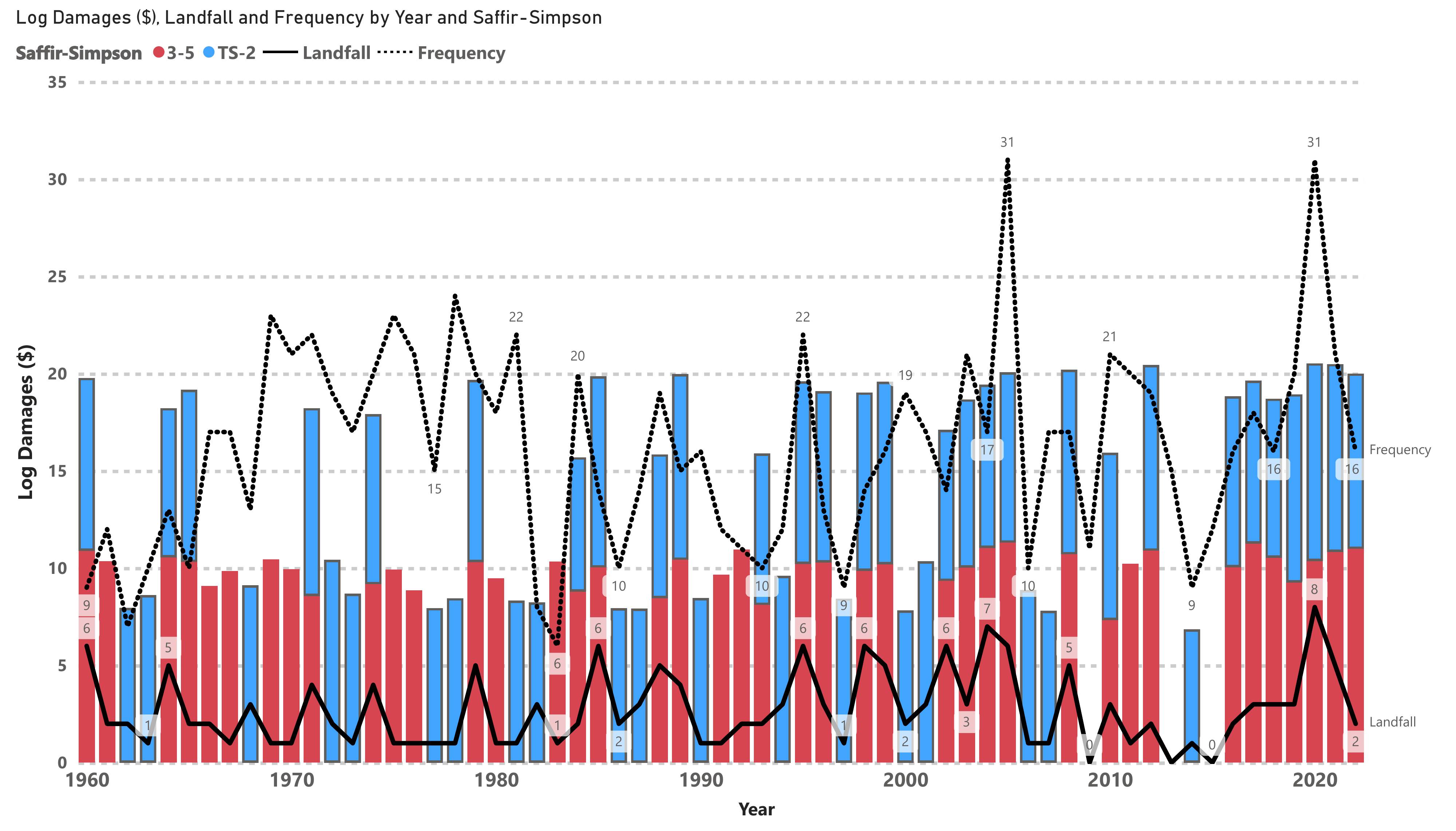} }      \\
{\includegraphics[width=0.3\textwidth]{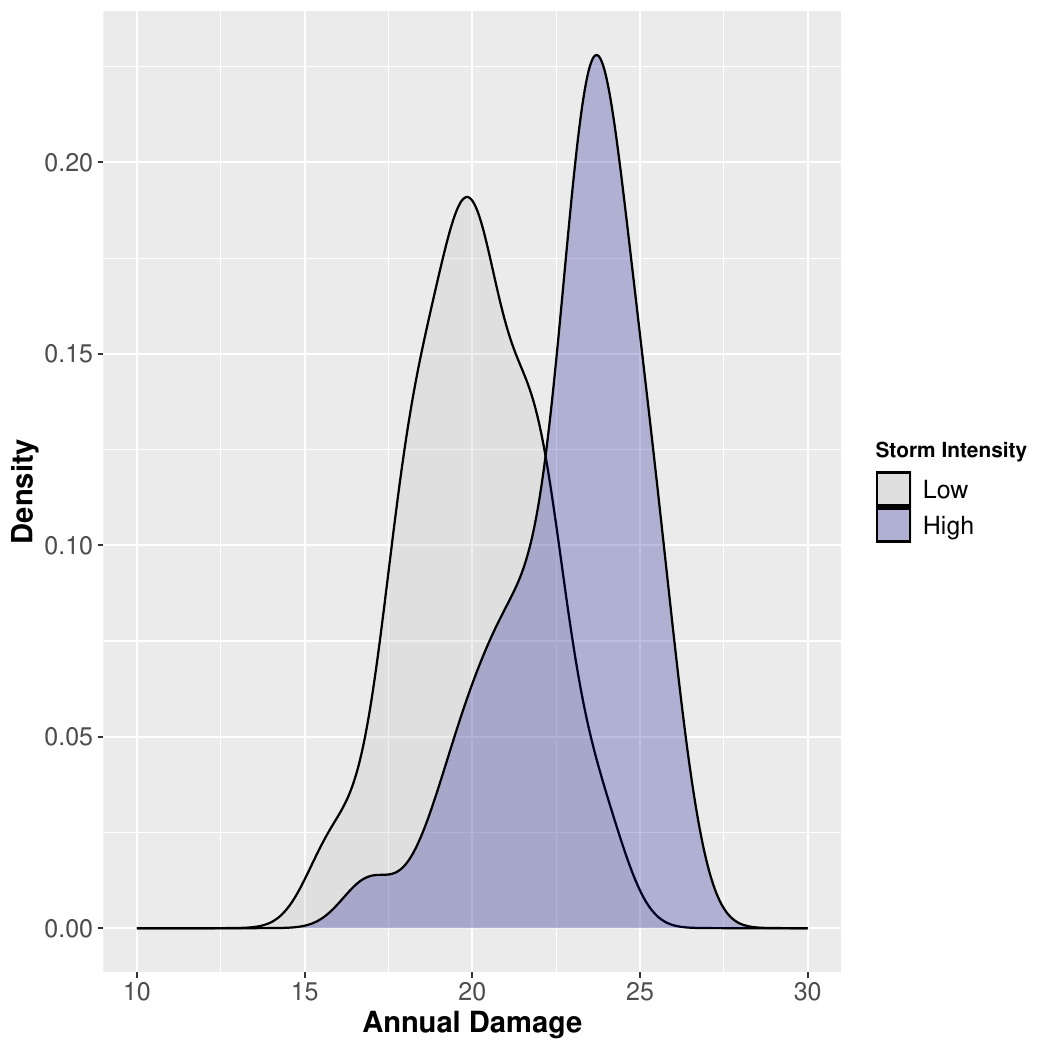}}     & \includegraphics[width=0.5\textwidth]{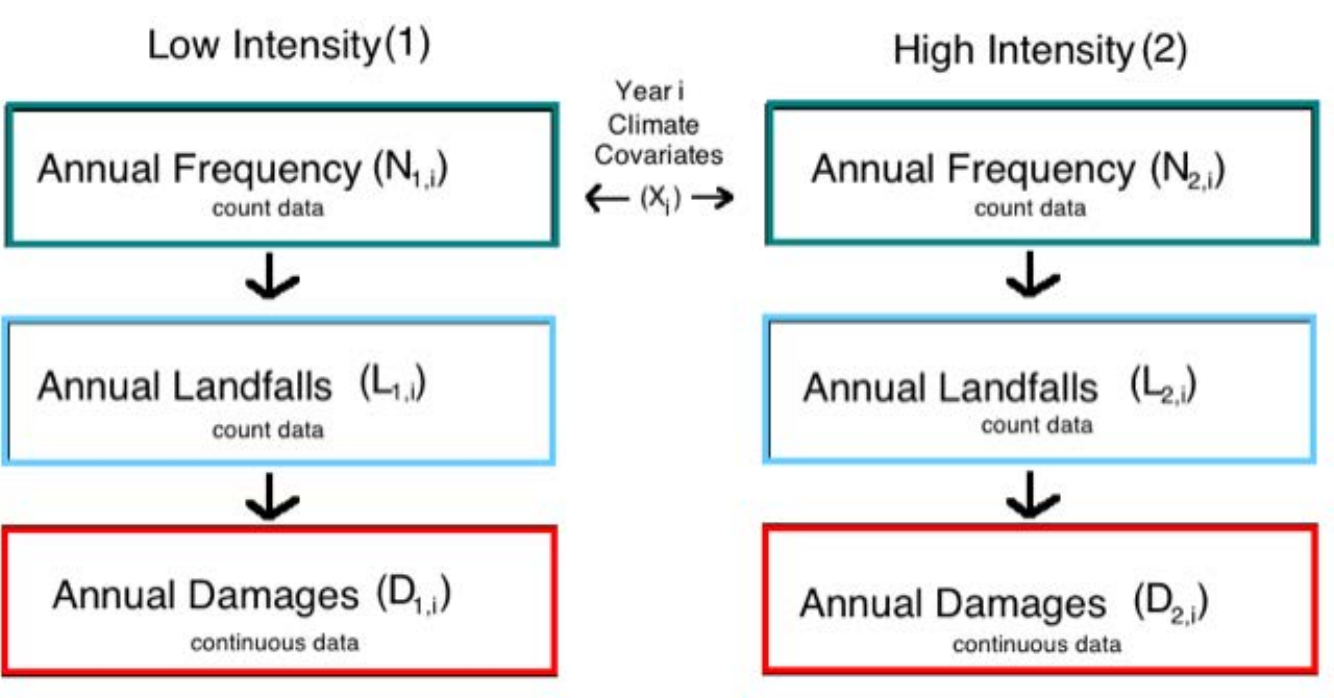}
\end{tabular}
 \caption[Seasonal Predictions]{\textbf{Seasonal predictions of damages as functions of annual landfalls and frequency of occurrence by storm intensity.} (A) In this plot, it can be observed that the monetary damage in a given year tends to increase as the number of storms that occurred that year and those that made landfall are large, for both low intensity (blue) and high-intensity storms (red). (B) Estimated Densities of Annual Log Damage in US dollars for Cyclones in Low and High-Intensity Categories illustrate a bimodality, with high-intensity storms expected to cause greater damages. (C) A flow chart for the proposed Bayesian hierarchical model, motivated by observations in (A) and (B).} 
        \label{fig:intro_seasonal}
\end{figure}

\textbf{Seasonal predictions: } Historical changes in damages are a result of meteorological factors (climate change or as a result of human activity) and socio-economic factors (increased population in hurricane-prone areas and increased prosperity). Therefore, to assess the effect of climate change on hurricane trends, most studies correct for socio-economic influences by normalizing the damage data \citep{pielke1998normalized,grinsted2019normalized}.  A substantial number of these studies did not find a significant increase in hurricane damage since 1900 suggesting that changes in the climate have not led to noticeable increases in hurricane damage in the past \citep{pielke1998normalized,schmidt2009tropical,Pielke2,Pielke,estrada2015economic,weinkle2018normalized,vecchi2021changes};. However, some did find increases since the 1970s \citep{miller2008exploration,nordhaus2010economics,knutson2021climate}. However, the common consensus is that the Atlantic basin has substantial year-to-year and decade-to-decade variability in tropical cyclone activity levels and corresponding losses.
In this light, it has been claimed \citep{Pielke2} that model-based prediction may not be able to improve upon what is expected from long-term historical record of U.S. tropical cyclone landfalls and damages.  However, our findings indicate otherwise.  We address the issue of `hurricane droughts'\citep{Hall,hall2015frequency} in our Bayesian framework for the seasonal scale, i.e. several years without landfall.  This type of `drought' may make a standard time series analysis of historical records and other classical statistical approaches relatively complex and inefficient, however, the proposed Bayesian framework of this paper is unaffected by such volatilities. Keeping in mind the bimodal nature of the damage distribution as shown in Figure \ref{fig:intro_seasonal} (bottom left), we use hierarchical Bayesian models for jointly modeling the annual frequency of occurrence, number of tropical storms making landfalls and damages separately for low intensity and high-intensity storms, as described in the flow chart schematic in Figure \ref{fig:intro_seasonal} (bottom right) and as described in Section \ref{Sec:Seasonal_Model}. It can be noted in Figure \ref{fig:intro_seasonal}(top panel) that the higher the frequency of storms that make landfall in a given season, the higher the expected monetary damages (note, the y-axis is $\log$-damages so even a small increase denotes a millions in monetary damage). In addition, we include other natural and anthropogenic features in our model, building on and extending several data science-driven approaches for modeling tropical cyclone frequency and intensity  \citep{Elsner3,loehle2020hurricane, Elsner10,Elsner11, Villarini, Elsner28,Elsner50, hodges2014sun,vecchi2014seasonal}.

In addition, we conduct thorough robustness studies with the probabilistic inferential and prediction frameworks. This is done by using, ($i$) several choices of prior distributions, ($ii$)  empirical Bayesian and frequentist frameworks as alternatives to the proposed hierarchical Bayesian model, and, ($iii$) different mathematical optimization approaches. Details are reported in the Appendix. These additional studies ensure that the inferences and predictions are not sensitive to the choice of the data science model or technology used, but instead reflect what the data tells us. 

 We present our findings in two parts. First, the hierarchical Bayesian techniques used for the seasonal model are presented in Section \ref{Sec:Seasonal_Model}, followed by the hierarchical generalized extreme value Bayesian framework for predictions of damages from individual tropical cyclones in Section \ref{Sec:Storm_Model}.  In Section \ref{sec: seasonalpredictions} we present the results for the season-level prediction for the number of cyclones, the probability that any given cyclone will inflict some damage, and the monetized value of such damages from the hierarchical Bayesian model. Section \ref{Sec:Seasonal_PosteriorPredictiveDistribution} contains technical specifications about the predictive distributional models. The predictions from 2017, 2020 and 2022, which were highly active seasons are reported. Additional predictions are reported in the Appendix. Then the results of individual damage-causing cyclones are presented in Section \ref{Sec:Storm_Prediction}, using a hierarchical Bayesian extreme value distributional framework. We evaluate the prediction framework for cyclones of the 2017, 2020, and 2022 seasons. Discussions and comments on the obtained results are in the section following the results. Finally, the description of the data and data sources that we use are described in Section \ref{Sec:Data}.
 Appendices contain alternative data science and computational models, and additional predictions.

\section{Methods}


\subsection{Methods: Bayesian Modeling of Seasonal Cyclone Activity}
\label{Sec:Seasonal_Model}

We classify the cyclones for every season into two groups based on the Saffir-Simpson scale, those with low intensity (denoted by $C = 1$) and those with high intensity (C = 2), thus ensuring that there is a reasonable number of cyclones in each group in most years. Also, such a grouping is compatible with the bimodal distribution of damages that is evidenced from the data, see  Figure \ref{fig:intro_seasonal} (bottom left) on the logarithm of nonzero damages. Modes on the logarithmic scales correspond approximately to damages around 197 million and {4.74 billion in 2022 dollars} for the low and high-intensity categories respectively. 

In each season, we consider three aspects of  Atlantic tropical cyclone activity. For each season/year $i$, these are the number of cyclones ($N_{C, i}$), the number of tropical cyclones that make landfall  $(L_{C, i})$, and the valuation of the damages $(D_{C, i})$. We also consider several climatic features ($X_i$) that may be associated with Atlantic tropical cyclones, these include sea surface temperatures, natural phenomena like sunspots and solar magnetic disturbances, and different climate indices like the Atlantic multi-decadal oscillation index (AMO), the north Atlantic oscillation index (NAO), the southern oscillation index (SOI), the Nino3.4 anomaly index \citep{Elsner3,loehle2020hurricane,Elsner18}.

 We find that the distribution of annual low-intensity cyclone frequencies is over-dispersed, and hence use a negative binomial parameterization for it, as in \eqref{lowintensityfreq_new_FB}.  In \eqref{lowintensityfreq_new_FB}, we parameterize the negative binomial density for observation $i$ with $p_i$ and $r$. 
The latter is the (over)dispersion parameter, which in the Poisson distribution equals 1 (no overdispersion). The parameter for observation $i$, $p_i=r/(r+\lambda_i)$, is referred to as the success parameter, where $\log{\lambda_i} = \sum_j \beta_j X_{ij}$.   In each group, the frequency of cyclones that inflict economic damages (or make landfall) is captured using a binomial distribution, conditional on the total number of cyclones in each group for a given year, see \eqref{lowintensityland_new_FB}. The actual valuation of cyclone damages is modeled using a mixture distribution with a mass at zero and a lognormal distribution, as in \eqref{lowintensitydam_new_FB}. The precise modeling details are given below.  We use the notations $\boldsymbol{0}_{q}$ for a vector of 0s of length $q$ and $I_{q}$ for an identity matrix of size $q \times q$ where $q$ depends on the number of covariates in the model being fit.  The notation $[X]$ for a random variate $X$ denotes its distribution.  
\begin{align}
[N_{1,i}|X_i,r,p_i,\boldsymbol{\beta}_1] &\sim \text{NegBinom}(r, p_i) \label{lowintensityfreq_new_FB}\\
p_i &= \frac{r}{r + \lambda(x_i)} \nonumber \\
\log(\lambda(x_i)) &=x_i\boldsymbol{\beta}_1; \boldsymbol{\beta}_1 \in \mathbb{R}^{q_1} \nonumber\\
[L_{1,i}|N_{1,i} = n_{1,i},X_i,\phi,\boldsymbol{\beta}_1,\theta_1] &\sim \text{Binomial}(n_{1,i},\theta_1)\label{lowintensityland_new_FB} \\
[D_{1,i}|L_{1,i},N_{1,i} = n_{1,i},X_i,\phi,\boldsymbol{\beta}_1,\theta_1,\mu_{1},\sigma_{1}]&\sim \nonumber \\
(1-(1-\theta_1)^{n_{1,i}}) * \text{Lognormal}(\mu_{1},&\sigma_{1}) + (1-\theta_1)^{n_{1,i}} * 0
\label{lowintensitydam_new_FB}\\
[{\boldsymbol\beta_{1}}] & \sim \mathcal{N}(\boldsymbol{0}_{q_1}, 10^{5} I_{q_1})\label{lowintensitynorm1_new_FB},
\end{align}
with priors , $[\theta_{1}] \sim \text{Beta}(1,1), [\mu_1] \sim \mathcal{N}(0,10^{5}), \frac{1}{\sigma_1^2}  \sim \text{Gamma}(1,1)$ and $[r]  \sim \text{Unif}(0,70)$, respectively. Note, the use of a uniform prior with an upper bound of 70 for $r$ is not restrictive as the negative binomial tends to the Poisson as $r \rightarrow \infty$.

The hierarchical specification for the high-intensity cyclones is similar to that of low intensity except that the cyclone frequencies are modeled as Poisson distribution with mean parameter, $\gamma(\mathbf X)=\mathbf X\boldsymbol\beta_2$, as in \eqref{highintensityfreq_new_FB}. 
\begin{align}
[N_{2,i}|X_i,\boldsymbol{\beta}_2] &\sim \text{Poisson}(\lambda(x_i)) \label{highintensityfreq_new_FB}\\
&\log(\lambda(x_i))=x_i\boldsymbol{\beta}_2; \boldsymbol{\beta}_2 \in \mathbb{R}^{q_2} \nonumber\\
[L_{2,i}|N_{2,i},X_i,\boldsymbol{\beta}_2,\theta_2]& \sim \text{Binomial}(n_{2,i},\theta_2) \label{highintensityland_new_FB}\\
[D_{2,i}|L_{2,i},N_{i2},X_i,\boldsymbol{\beta}_2,\theta_2,\mu_{2},\sigma_{2}]&\sim \nonumber \\
(1-(1-\theta_2)^{n_{i2}}) * \text{Lognormal}&(\mu_{2},\sigma_{2}) + (1-\theta_2)^{n_{i2}} * 0 \label{highintensitydam_new_FB}\\
[{\boldsymbol\beta}_{2}] & \sim \mathcal{N}_2(\boldsymbol{0}_{q_2} , 10^5 I_{q_2}) \label{highintensitynorm1_new_FB}
\end{align}
with priors, $[\theta_{2}]\sim \text{Beta}(1,1), [\mu_2] \sim \mathcal{N}(0,10^{5})$ and $[1/{\sigma_2^2}] \sim$ \text{Gamma}(1,1), respectively.

To ensure the results and inference we obtain from the data are not sensitive to modeling assumptions, we repeated the analysis using several alternative statistical models and data science formalism. {In the supplementary materials 
we report an empirical Bayesian modeling approach (Section \ref{Sec:Seasonal_EB})
and a different computational approach that uses data cloning. The results of all these alternative data modeling approaches are also in the supplementary materials (Appendix \ref{Sec:Seasonal_DClone}), and are all substantially identical, confirming the robustness of the results to the modeling framework.}

\subsection{Methods: Bayesian Modeling of Individual Cyclones}
\label{Sec:Storm_Model}
We consider a hierarchical Bayesian model to jointly model
a tropical cyclone's minimum central pressure (minCP), maximum windspeed (maxWS)  and the monetary value of the damages that the cyclone caused. Let $Z_1$ represents log(minCP), $Y_1$ represents log(maxWS) and $Y_2$ is for log(damages). Let $X$ denote the design matrix with 11 covariates, namely, average latitude, average longitude, start month, year of occurence, NAO, SOI, AMO, ANON 3.4, Atlantic SST, and sunspots. In other words, $X \in \mathbb{R}^{N \times p}$ for $p = 11$, with the first column of ones for the intercept.  We use the notation GEV for the \textit{generalized extreme value distribution}
 \citep{de2006extreme} and consider the following model,
\begin{align}
Z_1| X &\sim \text{GEV}(\mu_{z_1}(X), \sigma_{z_1}, \xi_{z_1}) \label{mod1_main}\\
Y_1|(Z_1, X) &\sim \text{GEV}(\mu_{y_1}(Z_1,X), \sigma_{y_1}, \xi_{y_1}) \label{mod2_main}\\
Y_2|(Z_1, Y_1, X) &\sim \text{GEV}(\mu_{y_2}(Z_1, Y_1, X), \sigma_{y_2}, \xi_{y_2}) \label{mod3_main},
\end{align}
where, for $\alpha \in \mathbb{R}^p, \beta \in \mathbb{R}^{p+1}$ and $\gamma \in \mathbb{R}^{p+2}$.
\begin{align*}
\mu_{z_1}(X) &=  X \alpha\\
\mu_{y_1}(Z_1, X) &= [Z_1, X]\beta\\
\mu_{y_2}(Z_1, Y_1, X) &= [Z_1, Y_1, X] \gamma.
\end{align*}
The joint density can then be written as,
\begin{align*}
f(Z_1, Y_1, Y_2| X, \theta)&=  f(Y_2|Y_1, Z_1, X) f(Y_1|Z_1,X) f(Z_1|X)\\
&= \dfrac{1}{\sigma_{Y_2}} (t(Y_2))^{\xi_{Y_2} + 1} \exp(-t(Y_2)) \dfrac{1}{\sigma_{Y_1}} (t(Y_1))^{\xi_{Y_1} + 1}\\
&  \quad \quad \times\exp(-t(Y_1)) \dfrac{1}{\sigma_{z_1}} (t(z_1))^{\xi_{z_1} + 1} \exp(-t(Z_1)),
\end{align*}
where,
\begin{align*}
t(x) &= 
\begin{cases}
(1 + \xi (\frac{x - \mu}{\sigma}))^{-1/\xi} & \xi \neq 0\\
\exp\left(- \frac{x - \mu}{\sigma} \right) & \xi = 0.
\end{cases}
\end{align*}
Then the log-likelihood can be given by,
\begin{align*}
l(\theta|Y_1, Y_2, Z_1, X) &= \sum_{i=1}^n -\log(\sigma_{y_{2i}}) + (\xi_{y_{2i}} + 1) \log(t(y_{2i})) - t(y_{2i}) \\
& \quad \quad - \log(\sigma_{y_{1i}}) + (\xi_{y_{1i}} + 1) \log(t(y_{1i})) - t(y_{1i}) \\
&\quad \quad \quad - \log(\sigma_{z_{1i}}) + (\xi_{z_{1i}} + 1) \log(t(z_{1i})) - t(z_{1i}).
\end{align*}
Our choice of priors are essentially 
non-informative with variances chosen to ensure proper coverage of the sample space and reasonably good acceptance rates in the Metropolis-Hastings algorithm. 
These are:
\begin{align*}
\alpha_i &\overset{iid}{\sim} \text{N}(0,10^2), i = 1,\hdots, p\\
\beta_i & \overset{iid}{\sim} \text{N}(0,10^3), i = 1,\hdots, p + 1\\
\gamma_i & \overset{iid}{\sim} \text{N}(0,10^2), i = 1,\hdots, p + 2\\
\sigma_{z_1}, \sigma_{Y_1}, \sigma_{Y_2} &\overset{iid}{\sim} \text{IG}(\alpha = 1, \beta = 3)\\
 \xi_{z_1} &\sim \text{Unif}(-1,1)\\
 \xi_{Y_1}, \xi_{Y_2} &\overset{iid}{\sim} \text{Unif}(-0.55, 0.5).
\end{align*}
The notations $\text{N}, \text{IG}$ and $\text{Unif}$ respectively stand for the Normal/Gaussian distribution, the inverse Gamma distribution and the Uniform distribution.

For the Bayesian computations, we use the Metropolis Hastings sampling scheme to obtain the Markov Chain Monte Carlo (MCMC) chain for $N= 10^6$  steps and the step-sizes are chosen to achieve about 20\% acceptance rate.  For starting values in the MCMC algorithm,
we use the frequentist estimates of each of individual GEV models, \eqref{mod1_main}, \eqref{mod2_main} and \eqref{mod3_main}.

\textbf{Variable selection: } Since all 11 covariates may not be relevant for each layer of the hierarchy, we select relevant variables based on the posterior results obtained from the full hierarchical model. In order to establish relevance of the covariates in the hierarchical model layers, we use the concept of 1-D depth. 
We look at the marginal posterior distribution for each parameter, and calculate the one-dimensional depth.  Let $F_\beta$ be the (one-dimensional) marginal posterior distribution of a parameter, say, $\beta$, then we compute an empirical estimate of $4 F_\beta (0) (1 - F_\beta (0))$, where we plug-in the following empirical estimator in place of $F_\beta$:
\begin{align*}
 \hat{F}_\beta(0) = \dfrac{\sum_{j = 1}^M I\{\beta^{(j)} \leq 0\}}{M},
 \end{align*} 
 where $M$ is the number of MCMC samples obtained for the parameter $\beta$, i.e., $\{\beta^{(j)}, j = 1,\hdots, M\}$. 
 The closer this value is to 0, the farther away zero (thinking about it as a hypothesis testing problem to test for $\beta \neq 0$) is in the tails of the distribution and the more relevant that variable is. The  closer this value gets to 1, that is indicative of zero being the median of the distribution. We fit this hierarchical model for the entire dataset prior to the test year to select the important variables. In Table \ref{tab: depth1D_GEVfull}, the blue and purple colored variables are relevant using this metric for the entire dataset (1960-2022). We shade in blue the variables that seem to have a depth value 0 or close to 0 (suggesting them being far off in the tails) while in purple are the variables with values between 0.2 and 0.6, which suggest that these estimates are still far in the tail ($\sim$ between the 80 and 95 percentile). Next we refit the heirarchical Bayesian model on these selected variables (blue and purple colored ones). Then, we calculate the posterior means and standard deviations for this selective model, which are then tabulated in Table \ref{tab: selectedcovs_estimatesGEV}.

\begin{table}[ht]
\centering
\begin{tabular}{r||r|r|r}
  \hline
  Variables & Min CP & Max WS & Damages\\ 
  \hline
{Intercept}   & \textcolor{blue}{0.0000} & \textcolor{blue}{0.0000} & \textcolor{blue}{0.0000}\\
Min CP (scaled)           &  NA & \textcolor{blue}{0.0000} & \textcolor{blue}{0.0177}\\
Max WS (scaled)            & NA  &  NA & \textcolor{purple}{0.2479}  \\
{Avg. Latitude} &  \textcolor{blue}{0.0000} & 0.8318 &  \textcolor{purple}{0.2290} \\ 
{Avg. Longitude}  &  \textcolor{blue}{0.0009} & 0.9698 & 0.9965 \\ 
   {StartMonth}  &  \textcolor{purple}{0.3456} & 0.9949  &  \textcolor{purple}{0.3351}\\ 
{Year} &  \textcolor{blue}{0.1122} & 0.9620 &  0.5877\\ 
  NAO  & 0.9956 &  0.9973 & 0.9991\\ 
  SOI  & 0.9912 & 1.0000 &  \textcolor{purple}{0.4244}\\ 
  {AMO}  & \textcolor{purple}{0.4748} & 0.9967 & 0.5665\\ 
  ANOM.3.4  & 0.9513 & 0.9974 & \textcolor{blue}{0.0486} \\ 
{Atl\_SST}  &  \textcolor{blue}{0.0021} & 0.9996 &  \textcolor{blue}{0.0091} \\ 
  Sunspots  & 0.9461 & 0.9830  & 0.9998\\ 
    {$\xi$ } & \textcolor{blue}{0.0000} & \textcolor{blue}{0.0000} & \textcolor{blue}{0.0018} \\ 
 {$\sigma$ } &  \textcolor{blue}{0.0000} &  \textcolor{blue}{0.0000} &  \textcolor{blue}{0.0000} \\ 
   \hline
\end{tabular}
\caption{1D data depth to measure how relevant a variable is in each of the layers of the model. }
\label{tab: depth1D_GEVfull}
\end{table}


We highlight that the shape ($\xi$) and scale ($\sigma$) parameters for all the three models are highly significant, thus validating the use of the GEV models for modeling the extreme behavior of these natural events. The negative estimate for the shape parameters signifying reverse Weibull distributions for the marginals of each of the three variables, log(minCP), log(maxWS) and log(damages), respectively. In addition, average latitude, average longitude, AMO, and Atlantic SST are statistically significant  in modeling the location parameter for minCP. Similarly, the effect of minCP ($\beta_1$) in modeling location parameter for maximum wind speed seems to be significant. The effect of maximum wind speed, minimum central pressure, average latitude, ANOM 3.4, and Atlantic SST are  significant in modeling the location parameter in the GEV model for log(damages). We also note that the significant variables are mostly the same across the three different modeling schemes that we have employed, namely, the hierarchical generalized extreme value Bayesian model, the trivariate generalized extreme value model (Section \ref{Sec:Storms_TrivariateGEV}), and the hierarchical generalized extreme value model with half-Normal damages (Section \ref{Sec:Storms_HierarchicalLogNormal}).  The diagnostics of the MCMC algorithm show that mixing and other properties of the Markov chain are all fully satisfactory. 


\begin{table}[ht]
\centering
\begin{tabular}{rrr}
  \hline
 & Posterior means & Posterior standard deviation \\ 
  \hline
  \textcolor{blue}{Min CP} & \\
  \hline
Intercept & 3.1797 & 0.0767 \\ 
  Avg. Latitude & -0.2251 & 0.0483 \\ 
  Avg. Longitude & -0.1257 & 0.0449 \\ 
  Start month & 0.0890 & 0.0643 \\ 
  Year & -0.0981 & 0.0465 \\ 
  AMO & -0.1145 & 0.0662 \\ 
  Atlantic SST & 0.2508 & 0.0484 \\
  $\xi$  & -0.9331 & 0.0470 \\ 
  $\sigma$ & 1.2931 & 0.0659 \\  
  \hline
  \textcolor{blue}{Max WS} & \\
  \hline
  Intercept & 3.7737 & 0.0900 \\ 
  Min CP (scaled) & 0.3696 & 0.0792 \\ 
  $\xi$ & -0.5403 & 0.0109 \\ 
  $\sigma$  & 1.0166 & 0.0516 \\ 
  \hline
  \textcolor{blue}{Damages} & \\
  \hline
  Intercept & 19.4191 & 0.1867 \\ 
  Max WS (scaled) & 0.8941 & 0.2968 \\ 
  Min CP (scaled) & 0.6707 & 0.2853 \\ 
  Avg. Latitude & -0.3315 & 0.1650 \\ 
  Start month & -0.2466 & 0.1687 \\ 
  Year & 0.1405 & 0.1954 \\ 
  SOI & 0.1912 & 0.2044 \\ 
  AMO & 0.2318 & 0.2503 \\ 
  ANOM.3.4 & 0.4064 & 0.1906 \\ 
  Atlantic SST & -0.7354 & 0.2391 \\ 
  $\xi$ & -0.2618 & 0.0380 \\ 
  $\sigma$ & 1.8824 & 0.0756 \\ 
   \hline
\end{tabular}
\caption{Posterior mean and standard deviation estimates for models \eqref{mod1_main},\eqref{mod2_main}, \eqref{mod3_main} for the fully hierarchical Bayesian Generalized Extreme Value (GEV) model with selected covariates.}
\label{tab: selectedcovs_estimatesGEV}
\end{table}

\section{Results}\label{sec: results}
\subsection{Predicting seasonal hurricane damages}
\label{sec: seasonalpredictions}
We group the tropical cyclones into two categories based on the Saffir-Simpson hurricane wind scale \citep{taylor2010saffir}. The first group is considered low intensity and corresponds to tropical cyclones up to category 2 in the Saffir-Simpson scale, while the second group is considered high intensity and comprises category 3-5 tropical cyclones (peak sustained winds exceeding 50 $m s^{-1}$). It is common in the literature to consider Saffir-Simpson Categories 3-5 Atlantic Hurricanes separately from the overall frequency and label them major hurricanes \citep{goldenberg2001recent,vecchi2021changes,weinkle2018normalized,Pielke2}. Historically, major hurricanes have accounted for about $80\%$ of hurricane-related damages in the United States of America (USA) despite only representing 34\% of USA tropical cyclone occurrences \citep{vecchi2021changes}.
The grouping of tropical cyclones based on whether they have low or high intensities also reflects the reality of the bimodal nature of the damage distribution depicted in Figure~\ref{fig:intro_seasonal} (lower-left). 

In each group, Gibbs sampling technique of Markov Chain Monte Carlo (MCMC) is used to estimate the posterior distribution for inference.
The expected proportions of damage-inflicting cyclones were respectively estimated at around 13.5\% and 43.4\% for the low and the high-intensity groups.  The monetary values of low and high-intensity damages indicated expected value in the log-normal scale of 20.045 (with 95\% credible interval = (19.49, 20.59)) and 23.22 (22.37, 23.64) respectively, corresponding to about 507.7 million (corresponding 95\% credible interval being (292.8, 877.2) million) and 9.76 billion (95\% credible interval = (5.17, 18.43) billion) dollars worth of damage.  Thus, on average, the U.S. should be prepared for about 12.597 billion dollars in damage each year, primarily from high-intensity tropical cyclones. Although an exact comparison is hard to make because estimates from different studies employ different methodologies and quote results based on the specific goals of their respective studies, these estimates seem to fall in the ballpark range observed in the literature \citep{Pielke,Pielke2,weinkle2018normalized}.

We present the details for the 2017, 2019, and 2022 tropical cyclone seasons as illustrative examples of Bayesian seasonal predictions. The degree of tropical cyclone activity and damages in these three years considerably vary,  thus providing an excellent spectrum of cases to evaluate the Bayesian predictive model. We use all available data (1960 onward) up to the prediction year and include the preseason covariates for both low and high-intensity tropical cyclones. Then our predictive model forecasts the number of cyclone events, the probability of each cyclone causing some amount of damage, and the monetized value of damages it will cause, for that season. Additional details about other years are in the Appendix \ref{sec: other_years_seasonal}.

\subsection{Posterior predictive distribution}
\label{Sec:Seasonal_PosteriorPredictiveDistribution}
We utilize the Bayesian specification to get an estimate of the posterior predictive distribution.  Let the distributional specification of the data $\mathbf Y=[\mathbf N, \mathbf L,\mathbf D]$ given parameters $\gamma=[\beta,\theta,\mu,\sigma,\phi]$ be $p(y|\gamma)$.  Also, let the posterior distribution of $\gamma$ given the data $\mathbf Y$ and the hyperparameter $\alpha$ be $p(\gamma|y,\alpha)$.  Then, the posterior predictive distribution for a new observation, $\tilde{\mathbf Y}$ is
\begin{align}
p(\tilde{y}|y,\alpha)&=\int p(\tilde{y}|\gamma)p(\gamma|y,\alpha)d\gamma
\end{align}
This is obtained computationally as follows:
\begin{enumerate}
	\item Let $\{\hat{\gamma_i}; i =1,\hdots,T\}$ be the MCMC set of posterior samples.
	\item Using each of these posterior estimates, we sample a new $y$ ($N,L,D$) given $\hat{\gamma}_i$ using the covariates for test data. We call this new $y$ as $y_{\text{pred}}^{(i)}$.
	\item Plot a histogram (density) using all the predicted $y$'s, i.e., $\{{y}^{(i)}_\text{pred}; i = 1,\hdots,T\}$.
\end{enumerate}

Since the above steps involve massive computations, we also explored two simpler 
approaches, outlined below:
\begin{itemize}
	\item \textbf{Empirical Bayesian prediction:}
	\begin{enumerate}
		\item Obtain the posterior sample means from the hierarchical Bayesian model fit in Section \ref{Sec:Seasonal_Model}, call them $\hat{\gamma}_{HB}$.
		\item Sample $10^5$ times from model specification $p(\tilde{y}|\hat{\gamma}_{HB})$ for each of the response variables, i.e., frequency of storms, frequency of damage-inflicting storms and value of damages. This provides a single realization of the posterior predictive distribution.
	\end{enumerate}
	\item \textbf{Fast computation Bayesian prediction:}
	\begin{enumerate}
	\item Let $\{\hat{\gamma_i}; i =1,\hdots,T\}$ be the MCMC set of posterior samples. Note, that $T$ is the number of posterior samples obtained in the MCMC output.
	\item Using each of these posterior estimates, we sample $y$ ($N,L,D$), $S = 1000$  times using the covariates for test data.  We call this new $y$ as $y_{\text{pred}_j}^{(i)}$.
	\item[3a.] Now estimate the density $p(\tilde{y}|\hat{\gamma}_i)$ using a kernel density estimate:
	\begin{align*}
	\hat{p}(\tilde{y}|\hat{\gamma}_i) = \frac{1}{Sh} \sum_{j=1}^S K \left(\frac{\tilde{y} -y_{\text{pred}_j}^{(i)}}{h}\right)
	\end{align*}
	\item[3b.] The predictive posterior distribution can then be estimated as:
	\begin{align*}
	\hat{p}(\tilde{y}| y, \alpha) = \dfrac{\sum_{i=1}^T \hat{p}(\tilde{y}|\hat{\gamma}_i)}{T} 
	\end{align*}
	\item[4.] Sample from the predictive posterior distribution $\hat{p}(\tilde{y}| y, \alpha)$.
\end{enumerate}
\end{itemize}

\begin{figure}[ht]
\centering
\begin{tabular}{ccc}
{\includegraphics[width=0.30\textwidth]{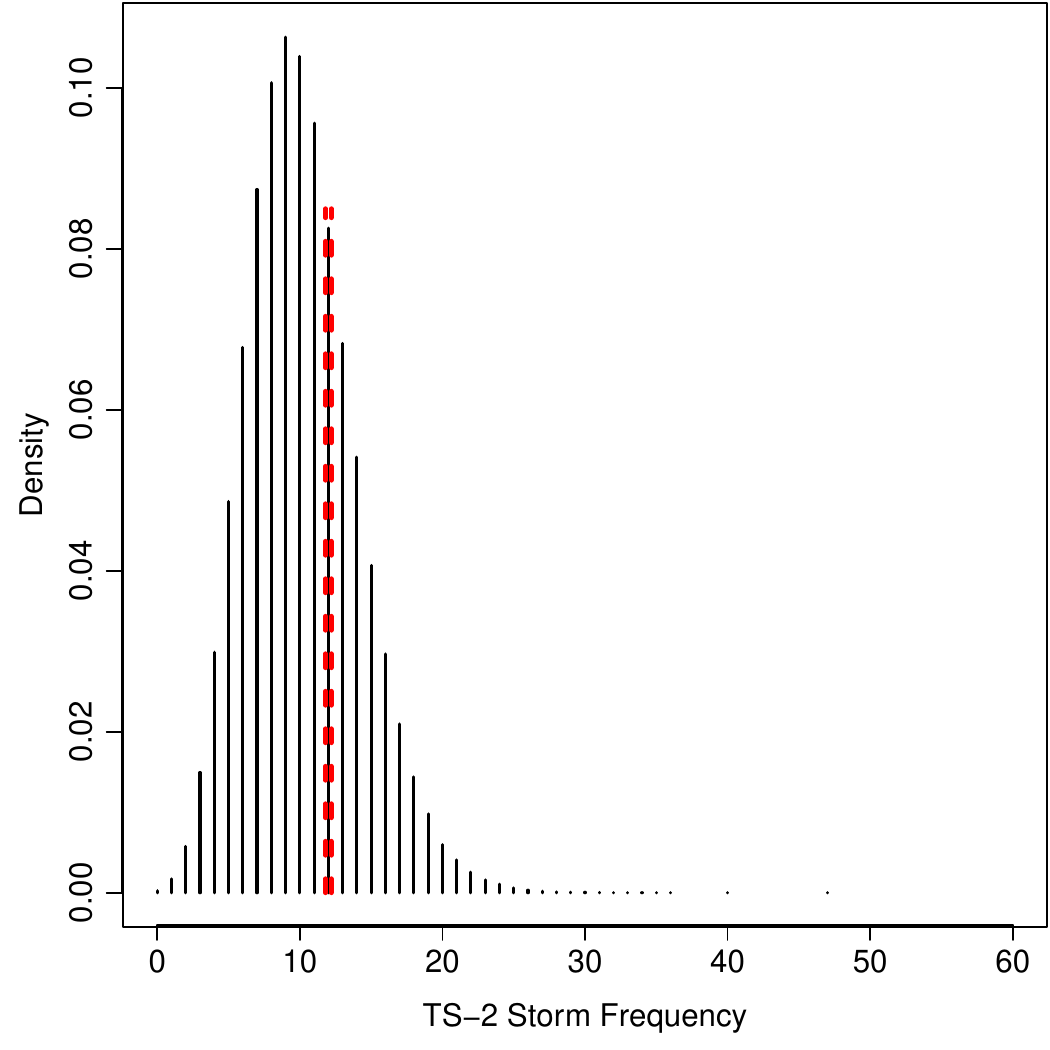}}
&
{\includegraphics[width=0.30\textwidth]{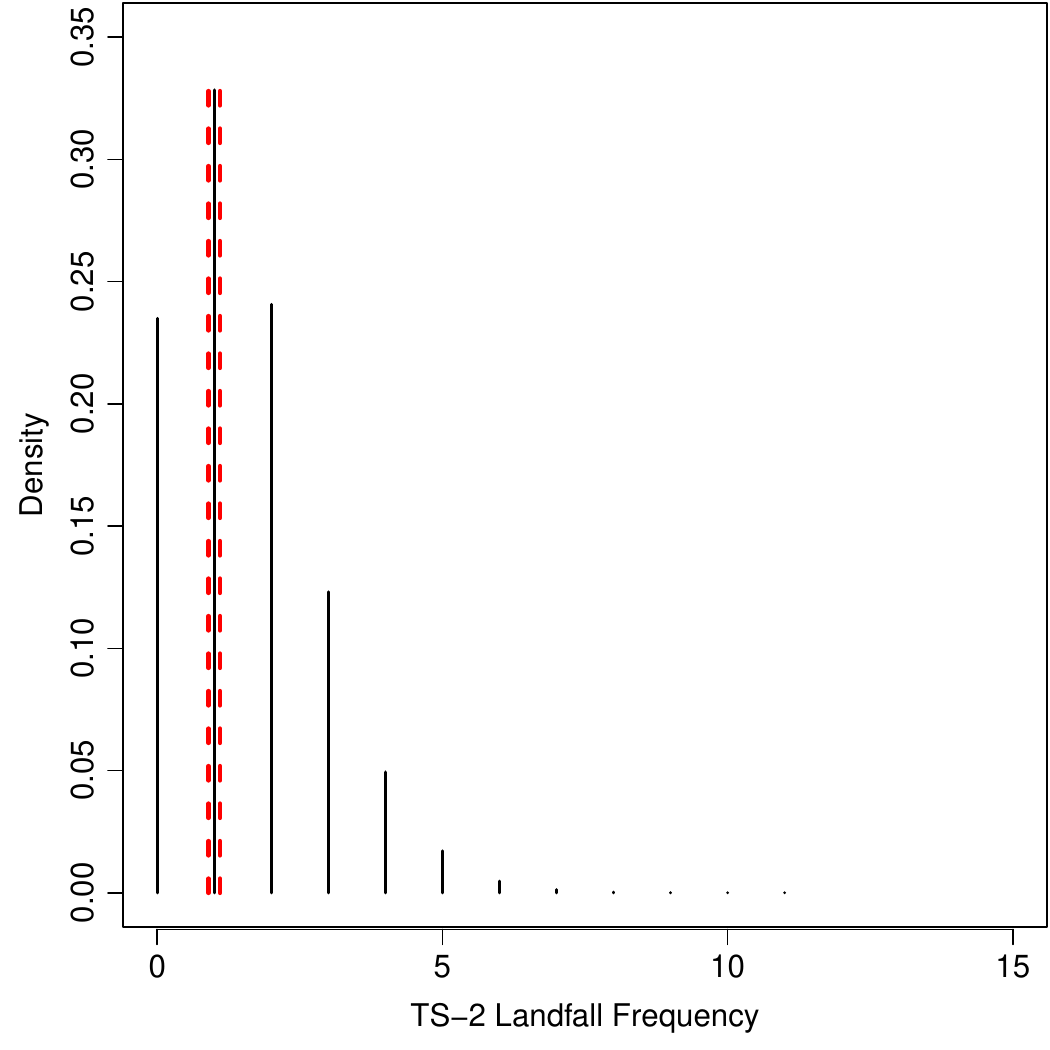}}
&
{\includegraphics[width=0.32\textwidth]{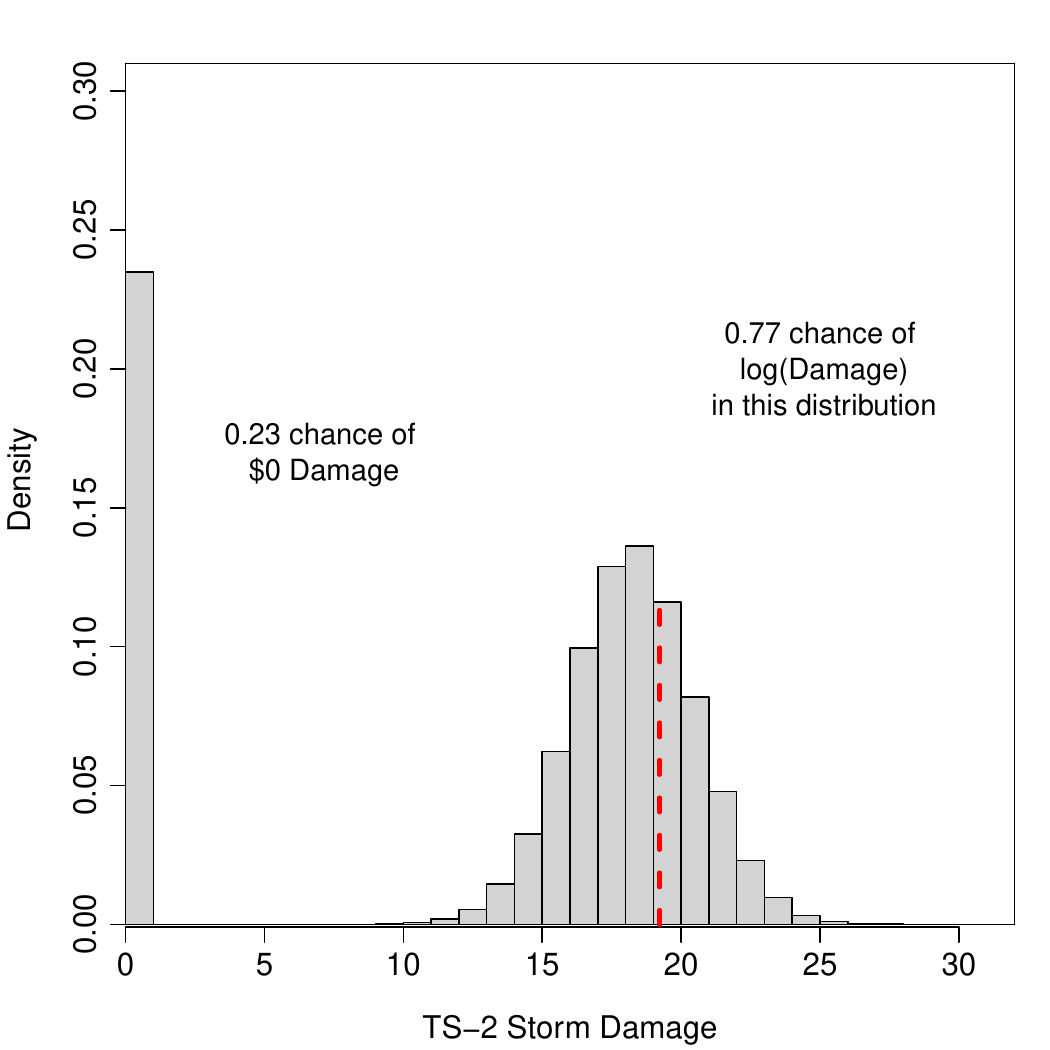}}
\\ 
{\includegraphics[width=0.30\textwidth]{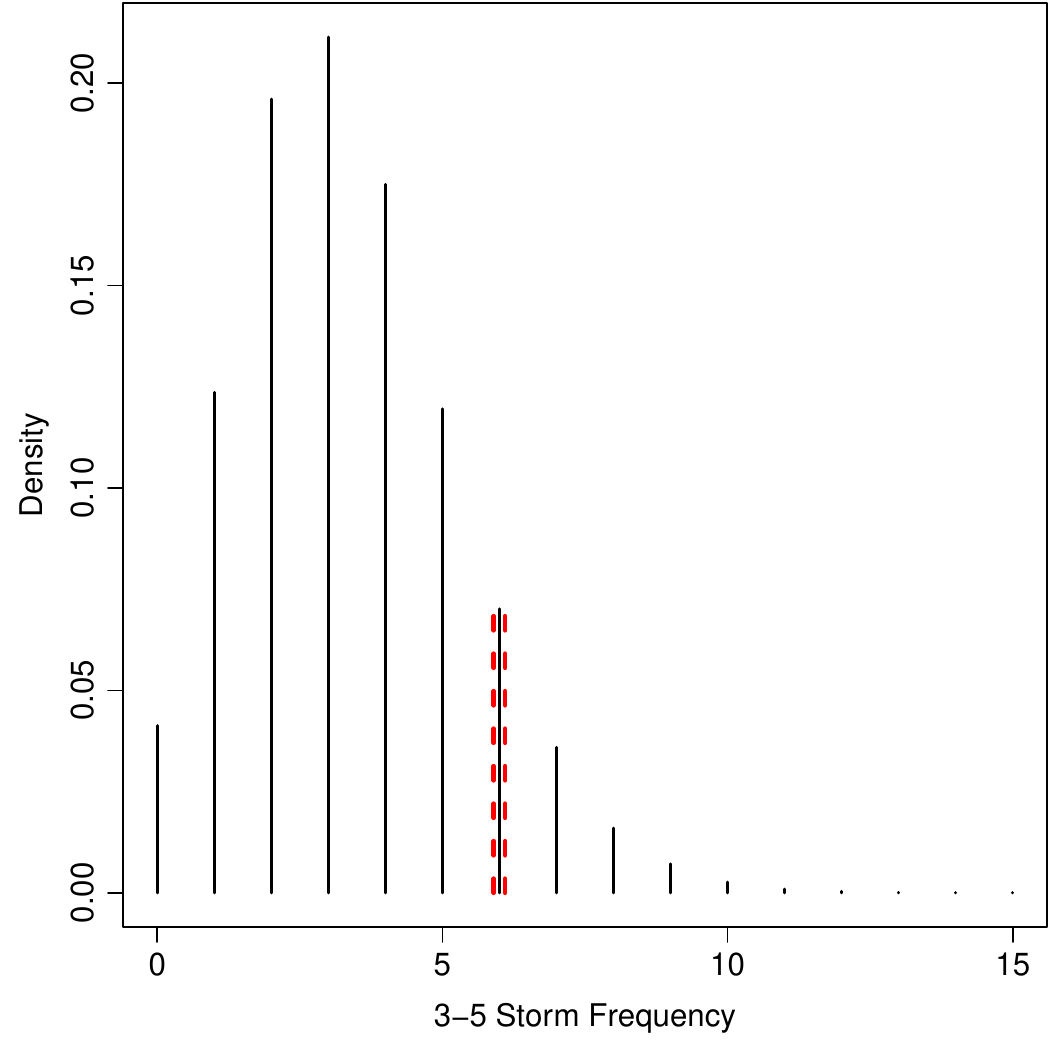}}
&
{\includegraphics[width=0.30\textwidth]{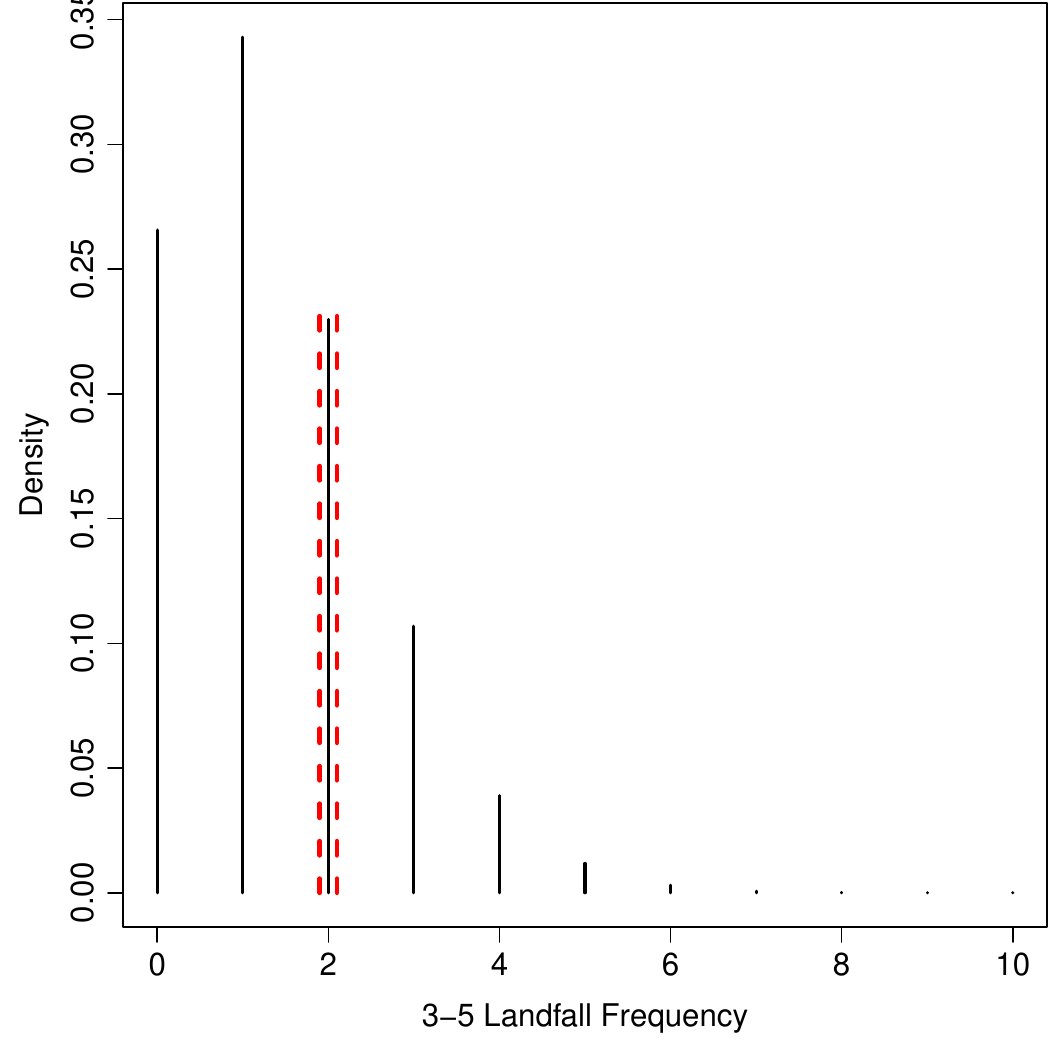}}
&
{\includegraphics[width=0.32\textwidth]{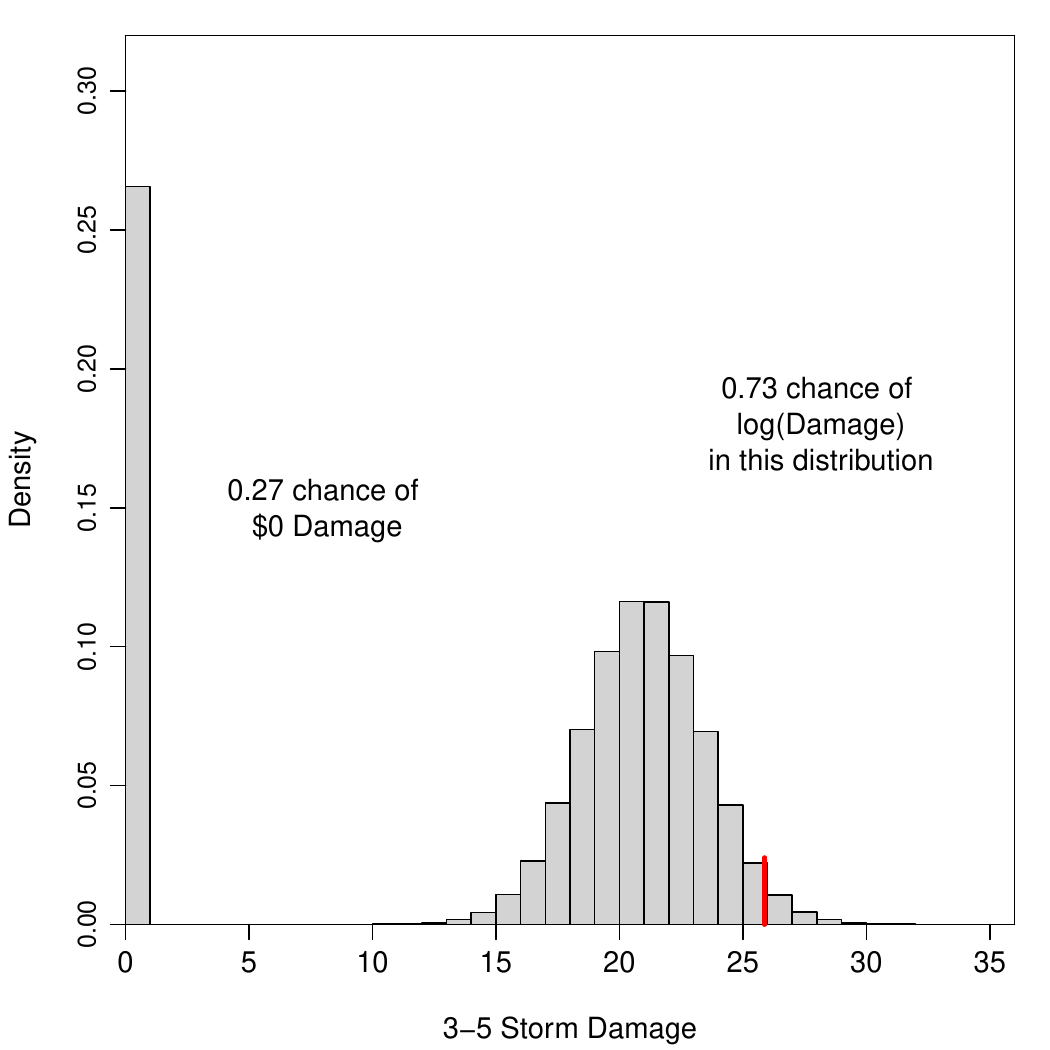}}
\end{tabular}
    \caption[Posterior Predictive Distributions for 2017 Tropical Cyclones]{Posterior Predictive Distributions for 2017 tropical cyclones. 
    The upper row is for the low-intensity case, the bottom row is for the 
    high-intensity case. The left column displays the probability mass function of the Bayesian predictive distribution for the frequency of cyclones, the middle column is the predicted probability mass function of whether a tropical cyclone may cause damage, and the right column is the predictive density for damages.
    The actual observed values are displayed with red dashed lines.}
        \label{fig:2017_pred}
 \end{figure}
\begin{figure}[ht]
\centering
\begin{tabular}{ccc}
{\includegraphics[width=0.30\textwidth]{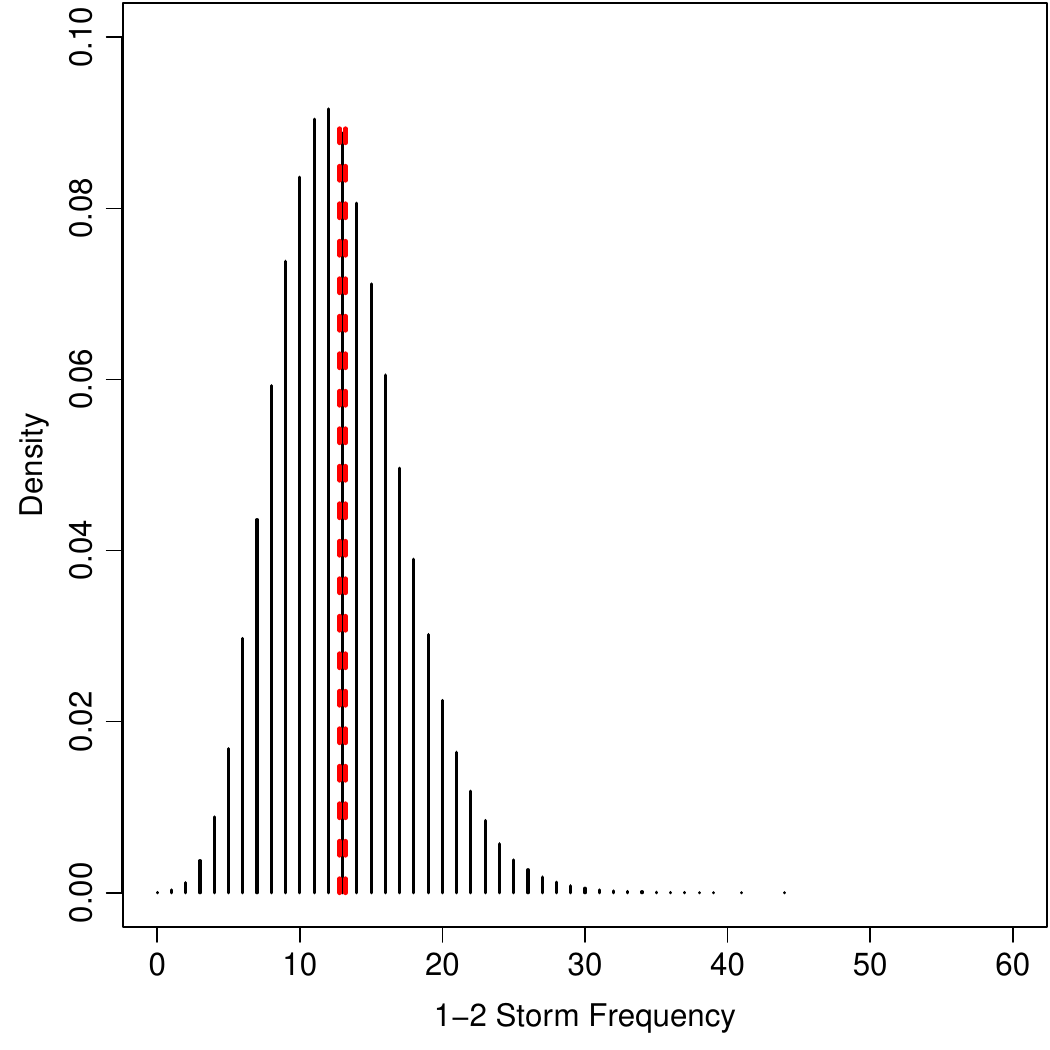}}
&
{\includegraphics[width=0.30\textwidth]{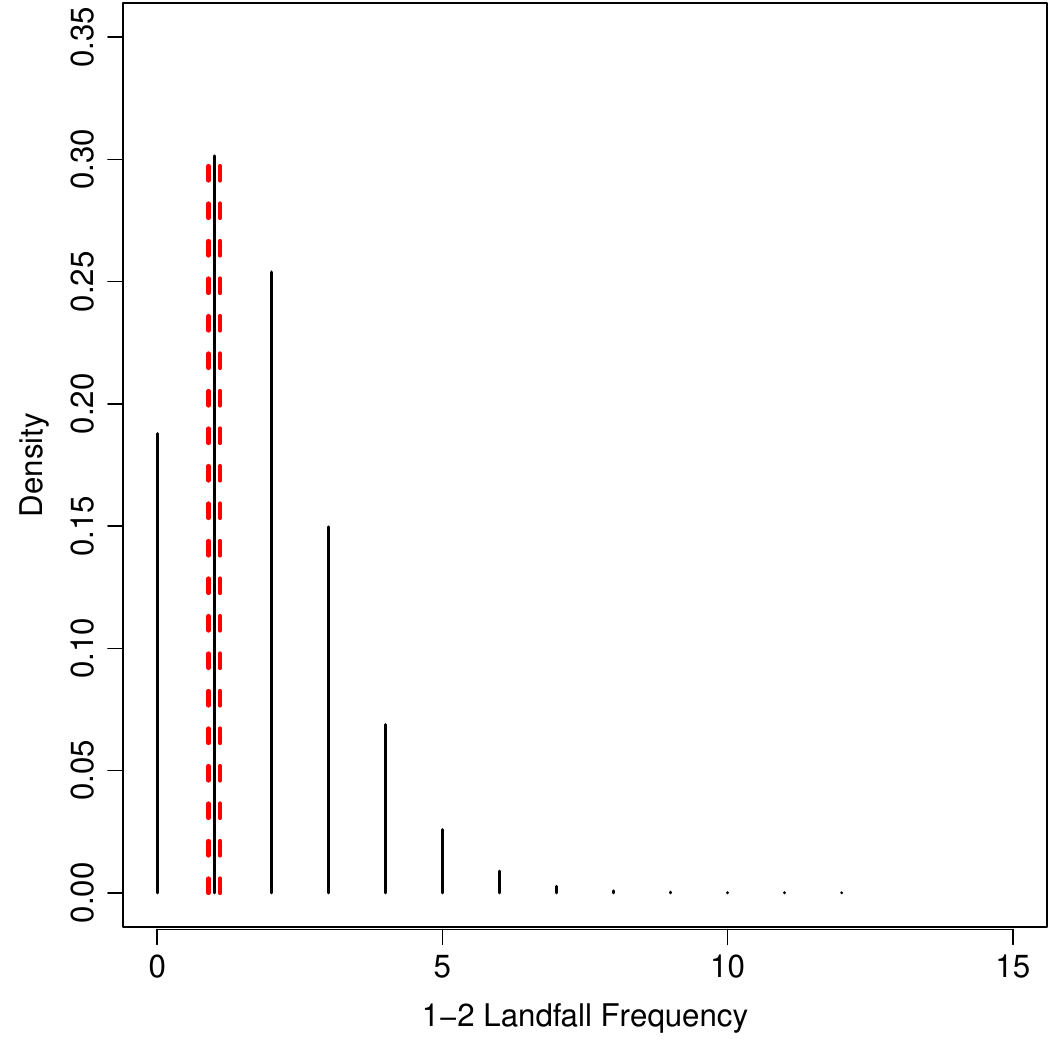}}
&
{\includegraphics[width=0.32\textwidth]{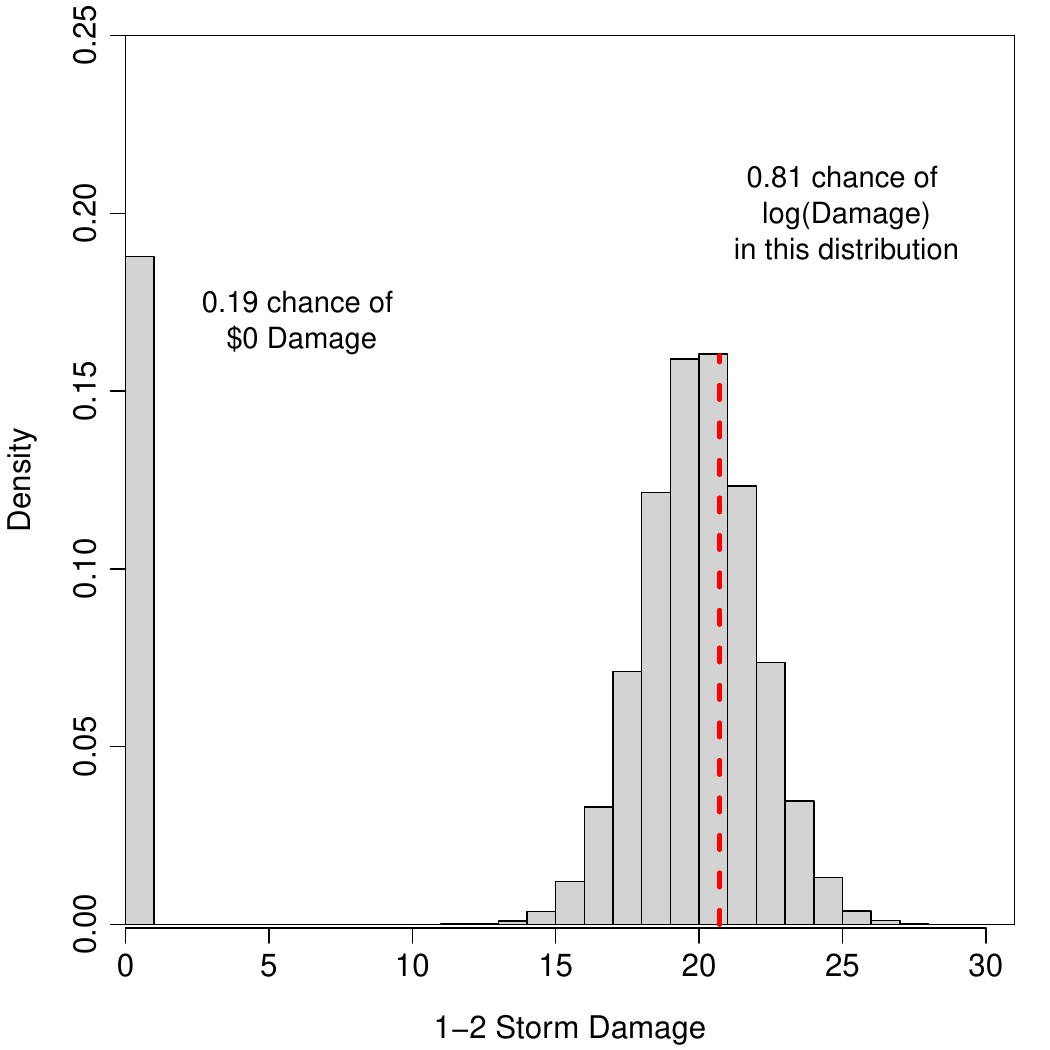}}
\\ 
{\includegraphics[width=0.30\textwidth]{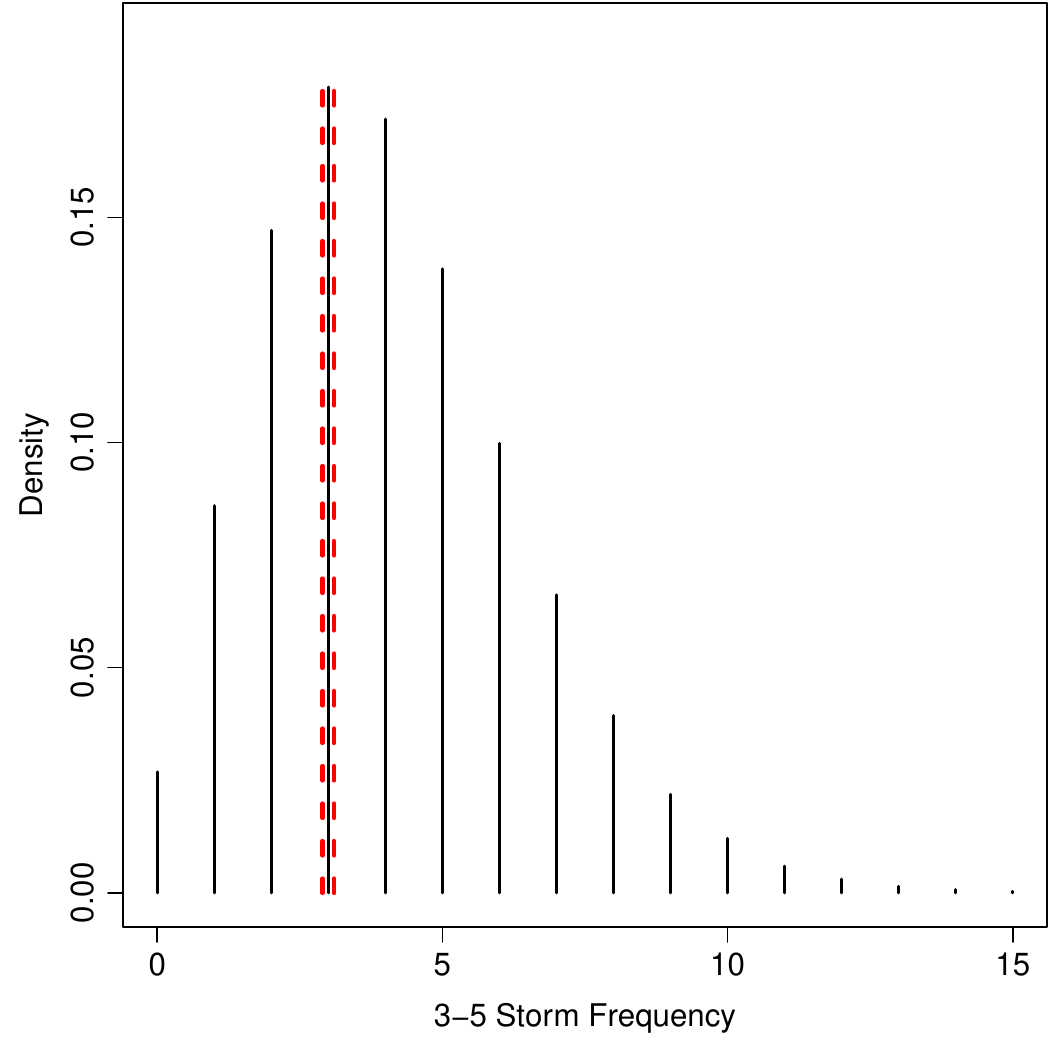}}
&
{\includegraphics[width=0.30\textwidth]{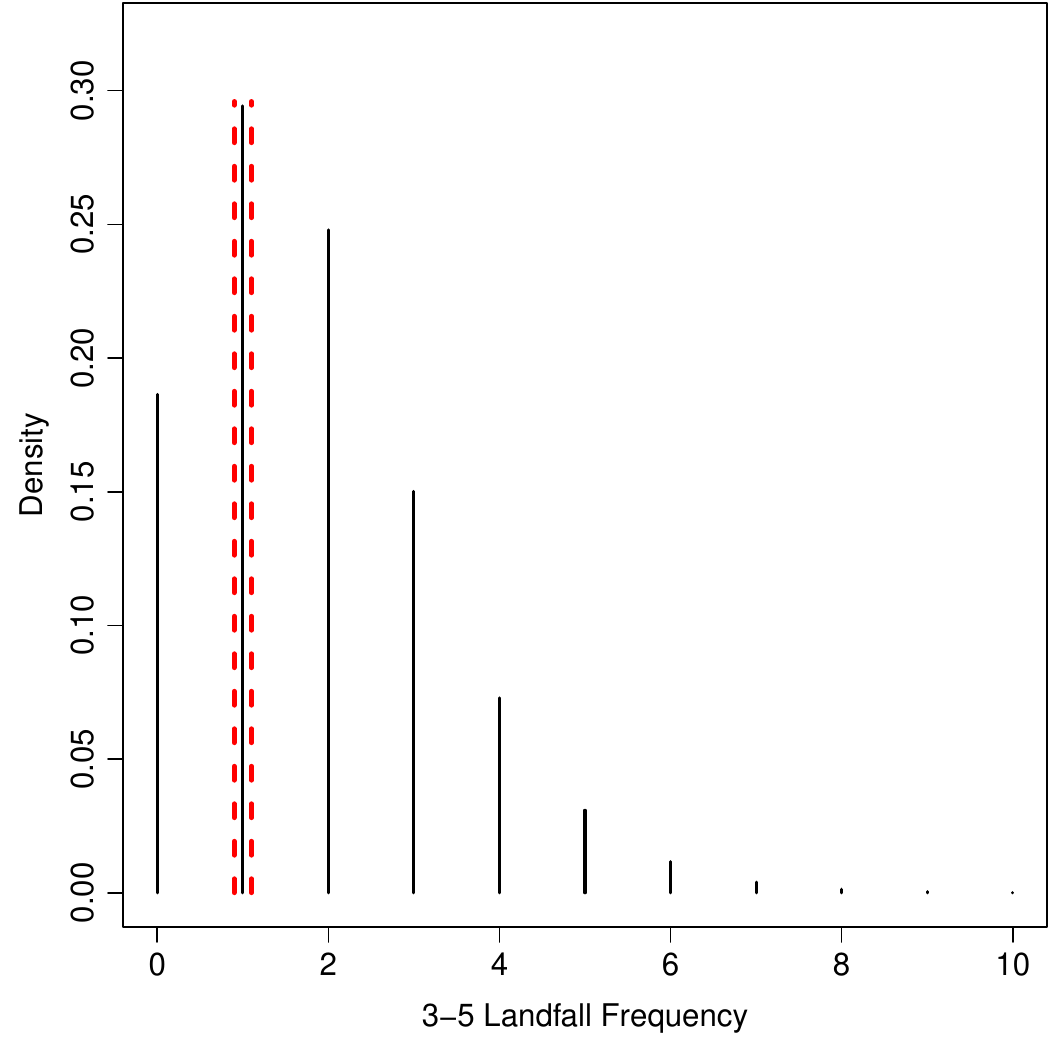}}
&
{\includegraphics[width=0.32\textwidth]{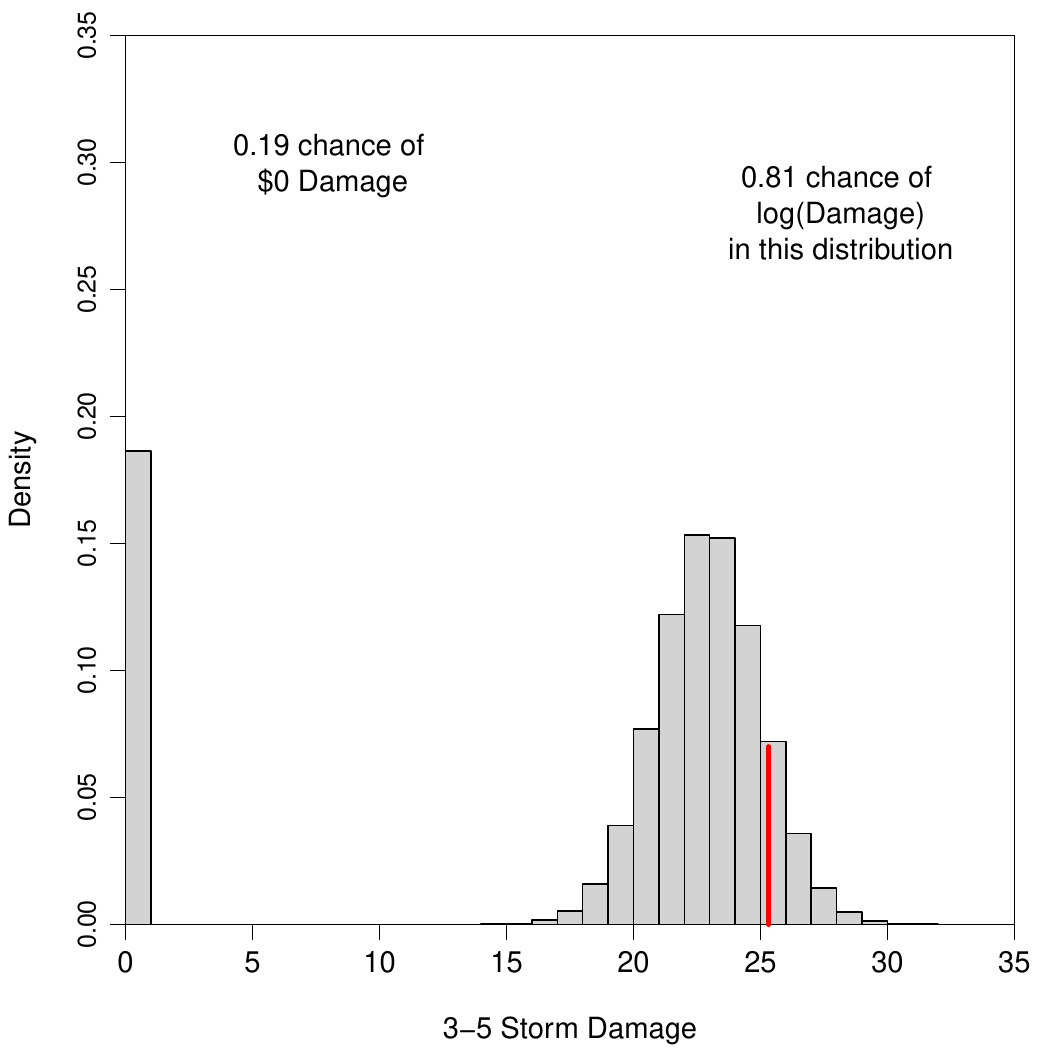}}
\end{tabular}
    \caption[Posterior Predictive Distributions for 2022 Tropical Cyclones]{Posterior Predictive Distributions for 2022 tropical cyclones. 
    The upper row is for the low-intensity case, the bottom row is for the 
    high-intensity case. The left column displays the probability mass function of the Bayesian predictive distribution for the frequency of cyclones, the middle column is the predicted probability mass function of a cyclone to cause damage, and the right column is the predictive density for damages.
    The actual observed values are displayed with red dashed lines.}
        \label{fig:2022_pred}
 \end{figure}

We display the predictive posterior mass/density functions and actual observations for 2017 in Figure~\ref{fig:2017_pred}, 2019 in 
Figure~\ref{fig:2019_pred}, and for 2022 in Figure~\ref{fig:2022_pred}. The years 2017 and 2022 are known for intense tropical cyclone activity and damages \citep{halverson2018costliest}, while 2019 was a much milder year with no damages. As seen in Figure ~\ref{fig:2017_pred}, ~\ref{fig:2019_pred}, and ~\ref{fig:2022_pred} the actual number of cyclones, landfall frequency, and damages are well within the predicted distribution for the low-intensity category as well as the high-intensity category. 
For 2019, there were no recorded damages in low-intensity as well as high-intensity cyclones. It can be seen from Figure~\ref{fig:2019_pred} (lower middle plot), that observing zero landfalling high-intensity cyclones had the highest chance at around 36-37\%. Also, the predictive distribution plots in 
Figure~\ref{fig:2019_pred} reflects the considerable chance of no damages, with about a 22\% chance for the low-intensity category and 38\% change for the high-intensity category, respectively. It can be seen from Figures~\ref{fig:2017_pred},~\ref{fig:2019_pred}, and ~\ref{fig:2022_pred} that the predictive model provides an excellent fit. The figures for some other years, along with additional figures where an empirical Bayesian or a fast Bayesian predictive modeling is used, are presented in Appendix \ref{sec: other_years_seasonal} and \ref{Sec:Seasonal_EB} respectively. 

\begin{figure}[ht]
\centering
\begin{tabular}{ccc}
{\includegraphics[width=0.30\textwidth]{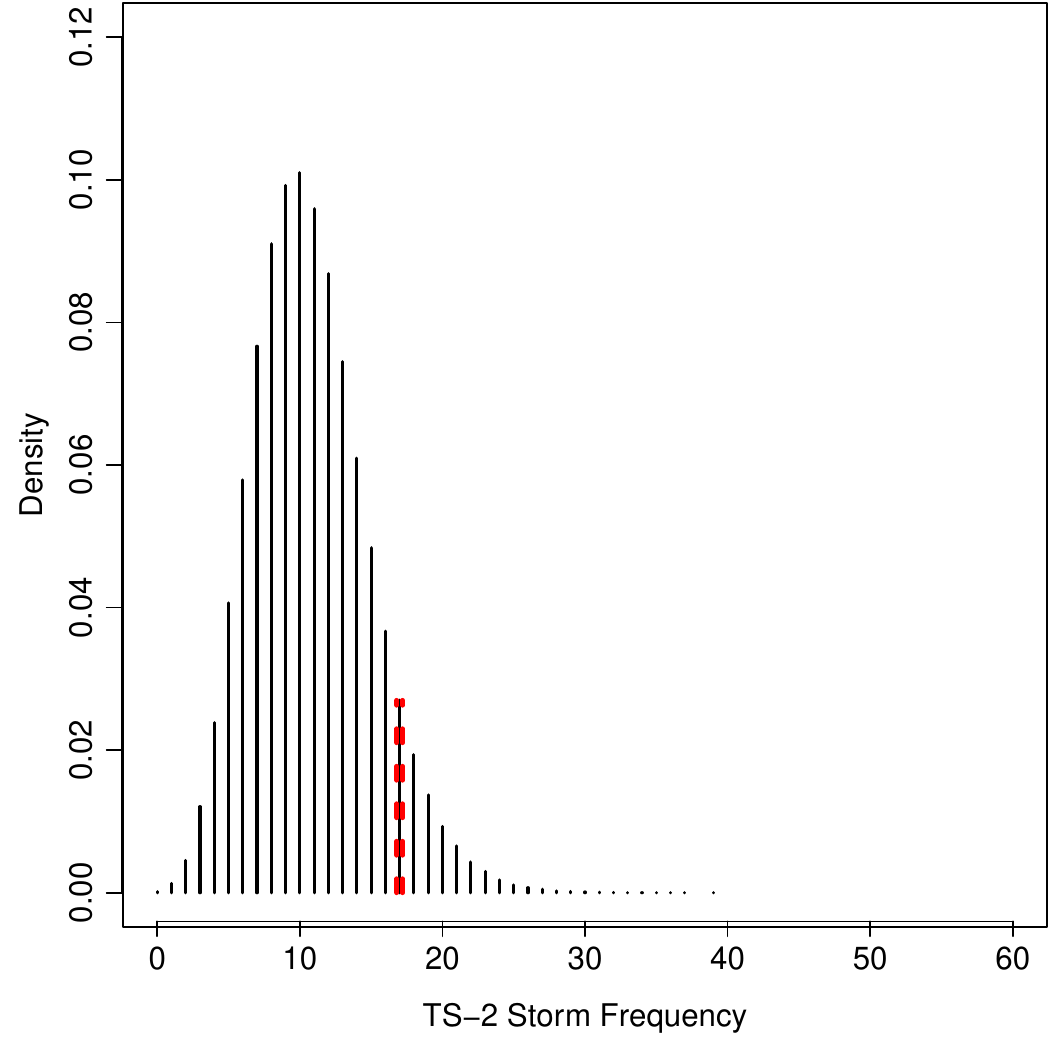}}
&
{\includegraphics[width=0.30\textwidth]{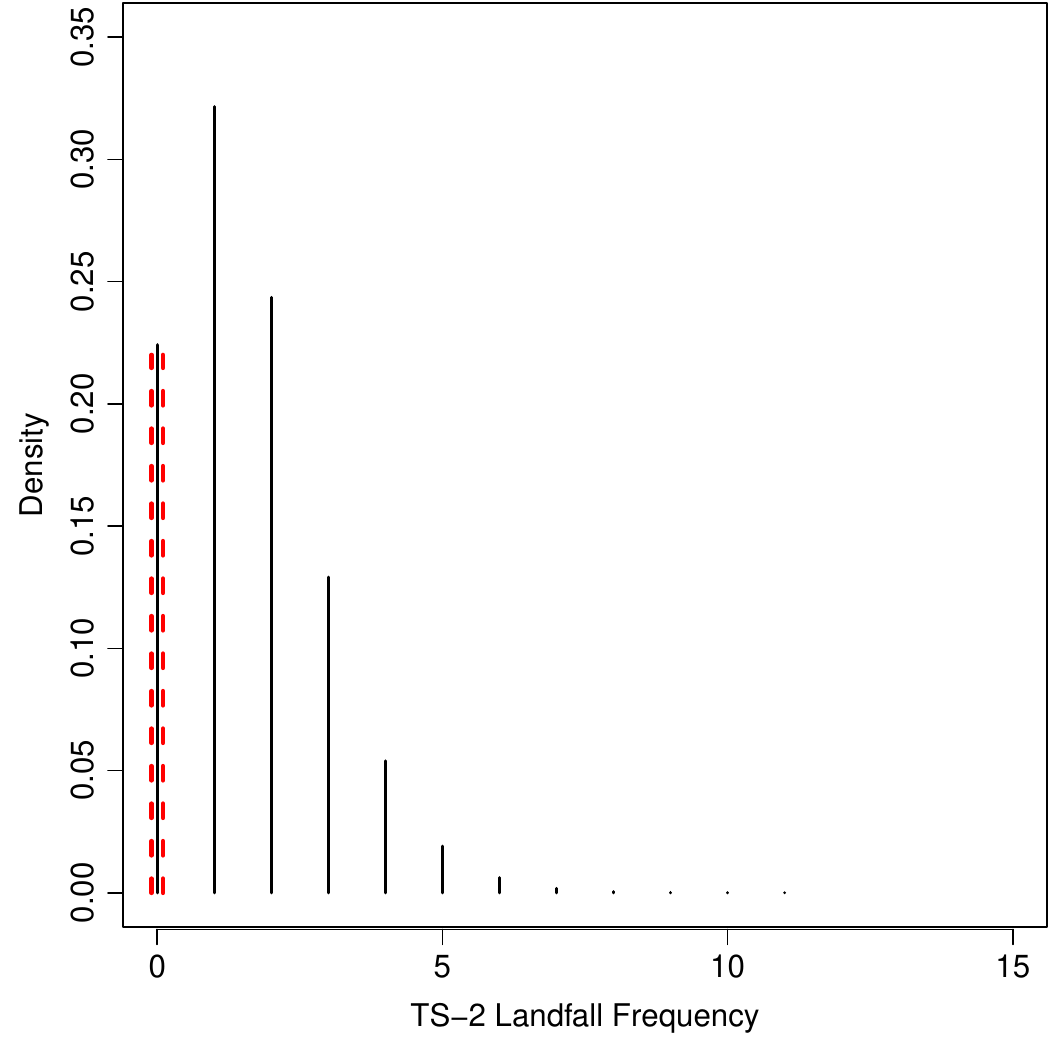}}
&
{\includegraphics[width=0.32\textwidth]{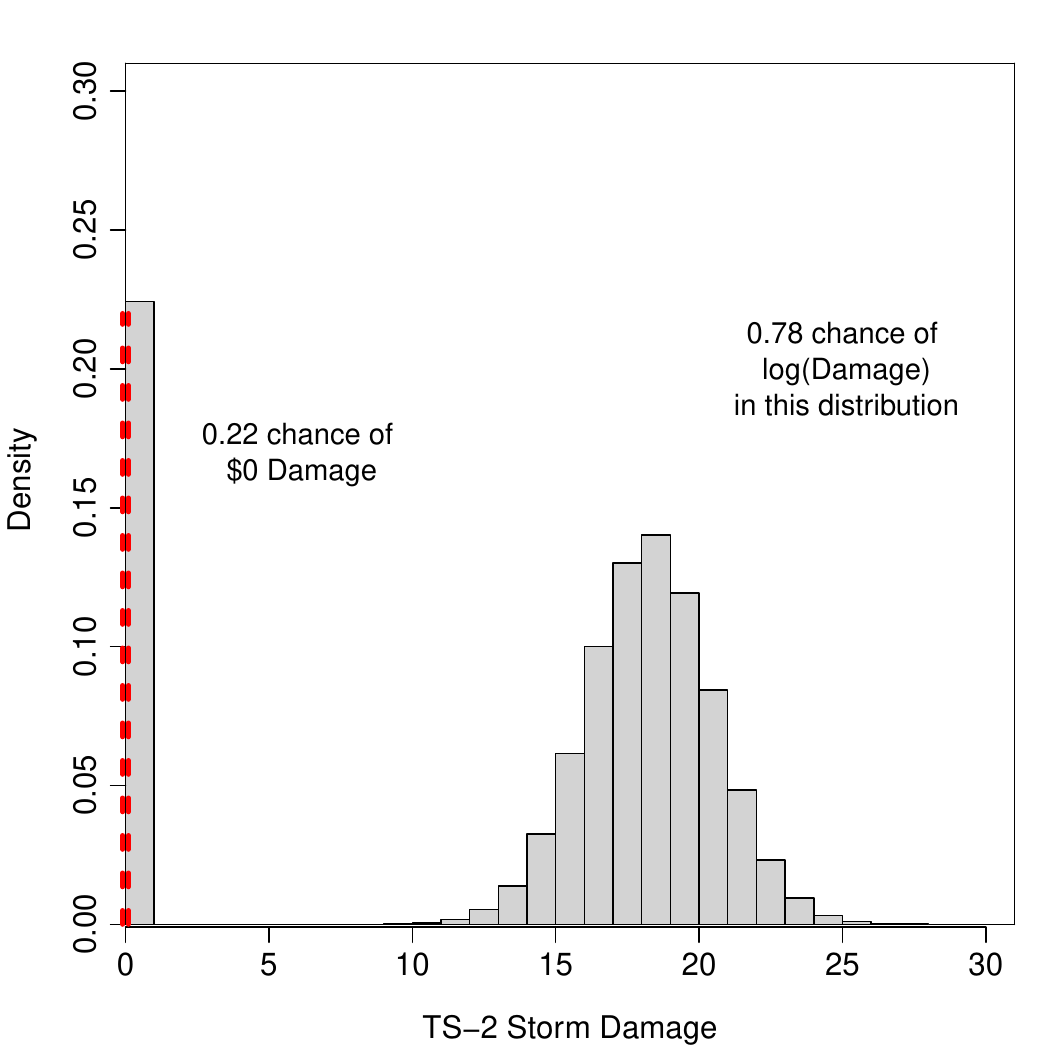}}
\\ 
{\includegraphics[width=0.30\textwidth]{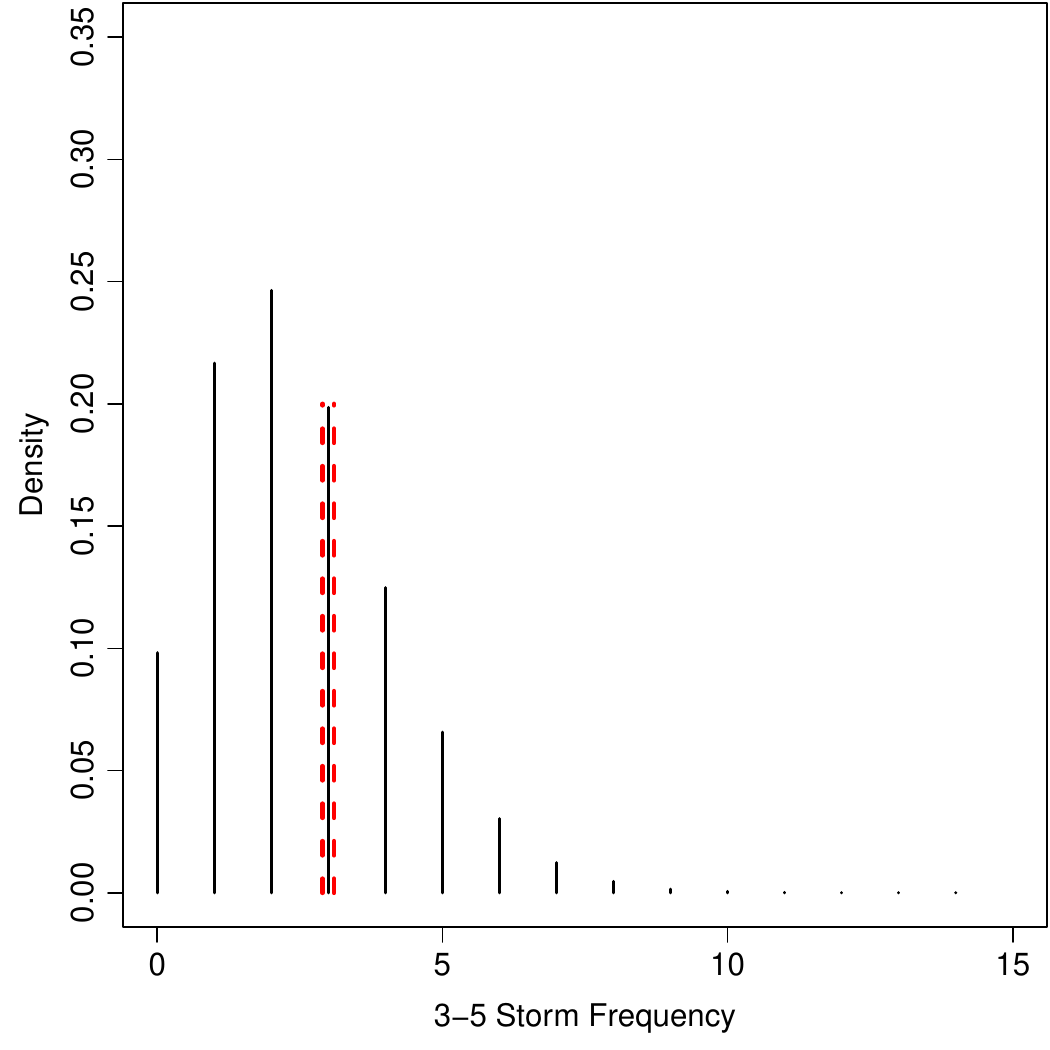}}
&
{\includegraphics[width=0.30\textwidth]{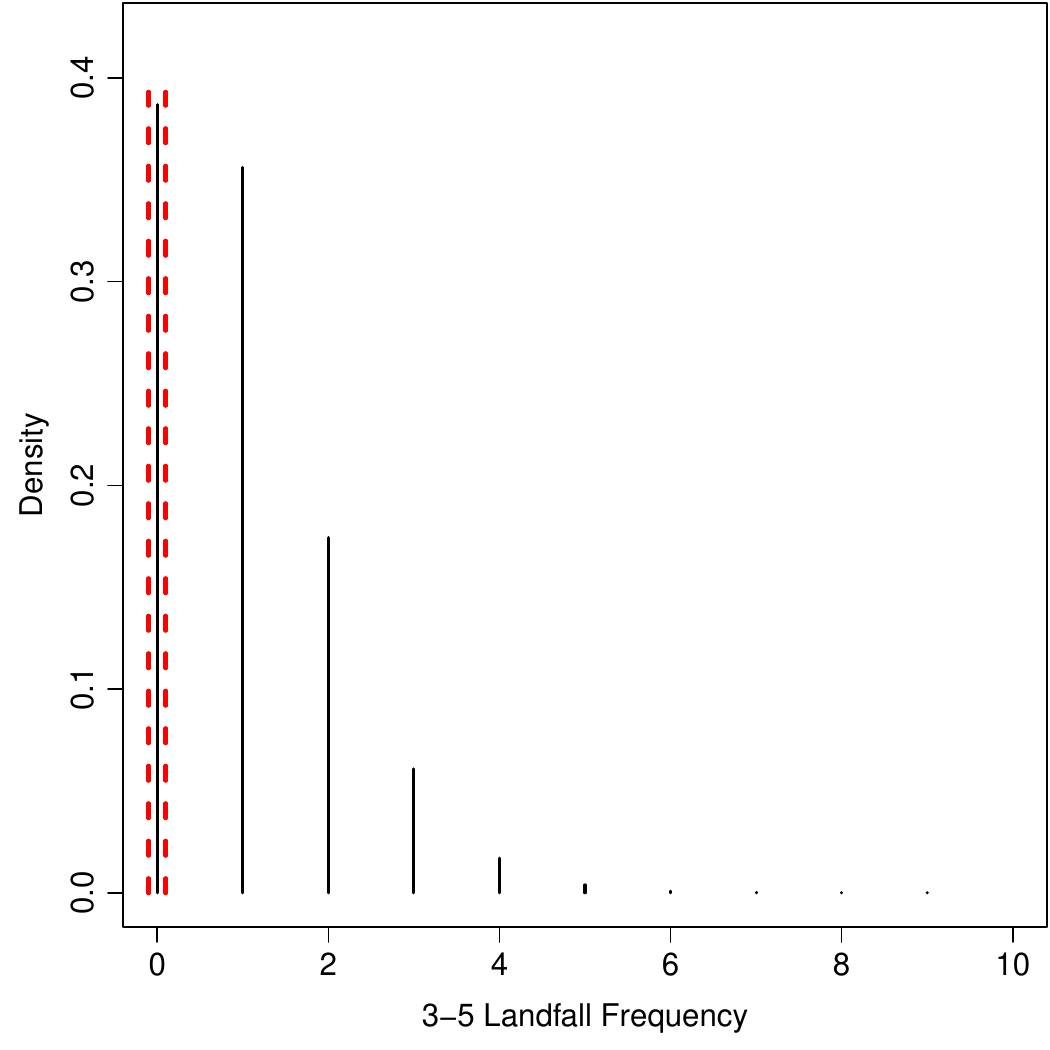}}
&
{\includegraphics[width=0.32\textwidth]{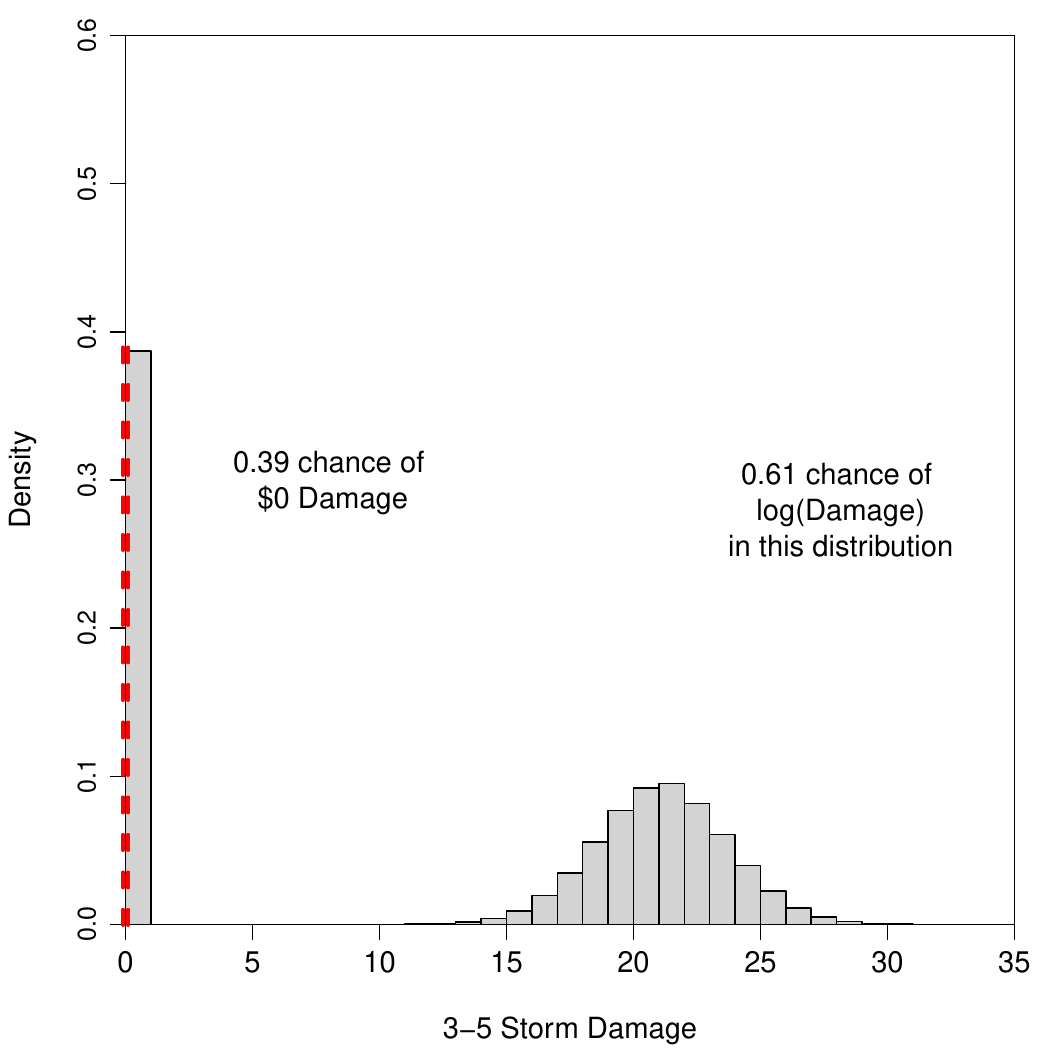}}
\end{tabular}
    \caption[Posterior Predictive Distributions for 2019 Tropical Cyclones]{Posterior Predictive Distributions for 2019 tropical cyclones. 
    The upper row is for the low-intensity case, the bottom row is for the 
    high-intensity case. The left column displays the probability mass function of the Bayesian predictive distribution for the frequency of cyclones, the middle column is the predicted probability mass function of a cyclone to cause damage, and the right column is the predictive density for damages.
    The actual observed values are displayed with red dashed lines.}
        \label{fig:2019_pred}
 \end{figure}

\subsection{Predicting individual hurricane damages}
\label{Sec:Storm_Prediction}
We use the hierarchical generalized extreme value distribution (GEV) model  to  predict the properties of the cyclones in 2017, 2020 and 2022. All these years were  active cyclone seasons with some of the costliest Atlantic hurricanes, especialy in 2017 and 2022 \citep{murakami2018dominant,emanuel2017assessing,pilkington2017real, klotzbach2021hyperactive,reinhart20232022,heidarzadeh2023normal}, thus, are good tests for the model prediction capabilities.  Since a Bayesian specification corresponds to updating the model as new data become available, we use all available data up to the prediction year and include the preseason covariates for all cyclones starting from 1960. The set of covariates considered are average latitude, average longitude, start month, year of occurence, National Atlantic Oscillation (NAO), Southern Oscillation Index (SOI), Atlantic Multidecadal Oscillation (AMO), El Ni\~no 3.4 anomaly, Atlantic Sea Surface Temperatures, and Sunspots. Amongst these, the statistically important variables are initially selected based on \textit{data depth} and consequently used in each layer of hierarchy in the predictive Bayesian model.

There are a total of 20 landfalling storms that hit continental U.S. in 2017 (three), 2020 (twelve) and 2022 (three) with non-zero damages. Some of the costliest tropical cyclones to impact the United States were in fact in 2017  and 2022:  Harvey, Irma and Nate in 2017 (with damages amounting to approximately 133 billion, 53 billion and 230 million, respectively, adjusted to 2022 consumer price index),  and Ian and Nicole in 2022 (with damages amounting to approximately 100 billion and 1 billion, respectively), whereas 2020 was a very active year as it saw 30 named storms, of which 14 became hurricanes, including 7 major hurricanes. We looked at 8 cyclones from 2020 that made landfall and the 5 others from 2017 and 2022 to  assess the performance of our models. In order to do so,  we compared where the true values for the three variables (maxWS, minCP and damages) fall on the posterior predictive distribution. For each cyclone, we find the 95\% credible intervals and check whether the actual cyclone maxWS, minCP and damages  are included in  these intervals or not. It is observed that the true minimum central pressure values and the true maximum wind speeds were within the 95\% credible interval for all the 13 cyclones. The credible interval could not capture the observed damage value for Harvey and Ian but other cyclones were within the 95\% interval. Missing two cases out of 39 with 95\% credible intervals is not surprising: a false signalling of one in every twenty instances is expected. Another reason for the miss could be that hurricane Harvey and Ian (being the costliest and the third costliest) was exceptional in the amount of damage it caused, and in such a situation, statistical risk assessment could suffer from short, incomplete and/or inaccurate past records \citep{emanuel2017assessing}.

In order to quantify exactly where the true value for each of the three variables falls on the posterior predictive distribution for the cyclones in 2017, 2020 and 2022, we calculate the percentile of the true value on the posterior predictive distribution. Let us call this $\alpha$. Then, we calculate $\delta = 2 \min\{ \alpha, 1-\alpha\}$ for each of the alpha values corresponding to each of the three variables and 8 cyclones. Note, the closer the value of $\delta$ is to 1, the closer the truth is to the posterior predictive median and if $\delta$ is closer to zero, then the truth is near the tail of the distribution. Table~\ref{tab:delta_value_hierGEV} presents the $\delta$ values for the 5 hurricanes in 2017 and 2022, and 3 randomly chosen ones from 2020.  These essentially show the proposed model is an excellent predictive tool. 

\begin{figure}[h!] 
\centering
    \caption[Posterior Predictive Distributions for Storms in 2017]{\textbf{Posterior Predictive Distributions for Storms in 2017}: Posterior Predictive Distributions for minCP, maxWS and damages of tropical storms Harvey (row a), Irma (row b) and Nate (row c), in 2017, based on the hierarchical Bayesian GEV model with selected variables.  The actual values for the storms are displayed with red dashed lines.}
        \label{fig:harvey_irma_hierGEV}
 
\begin{tabular}{ccc}
{(a)\includegraphics[width=0.30\textwidth]{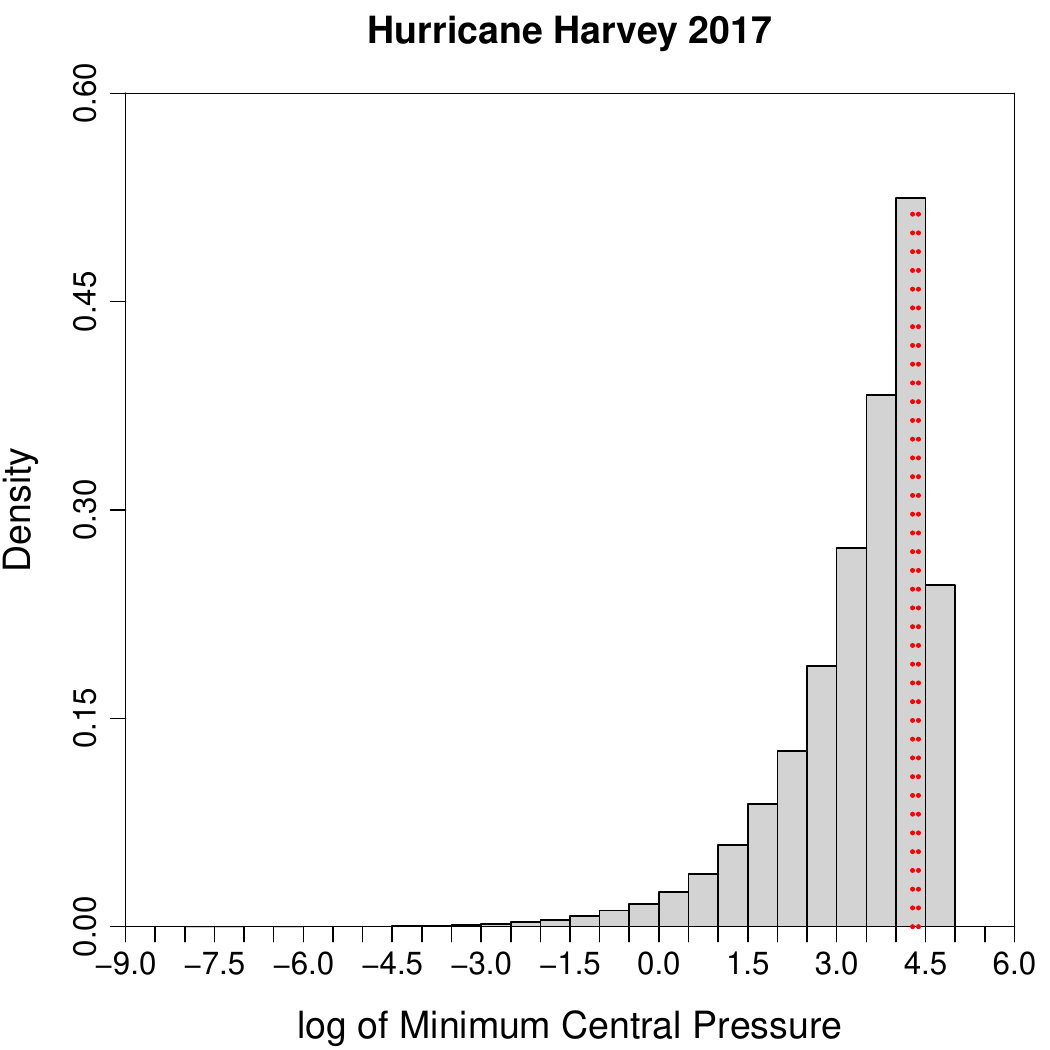}}
&
{\includegraphics[width=0.30\textwidth]{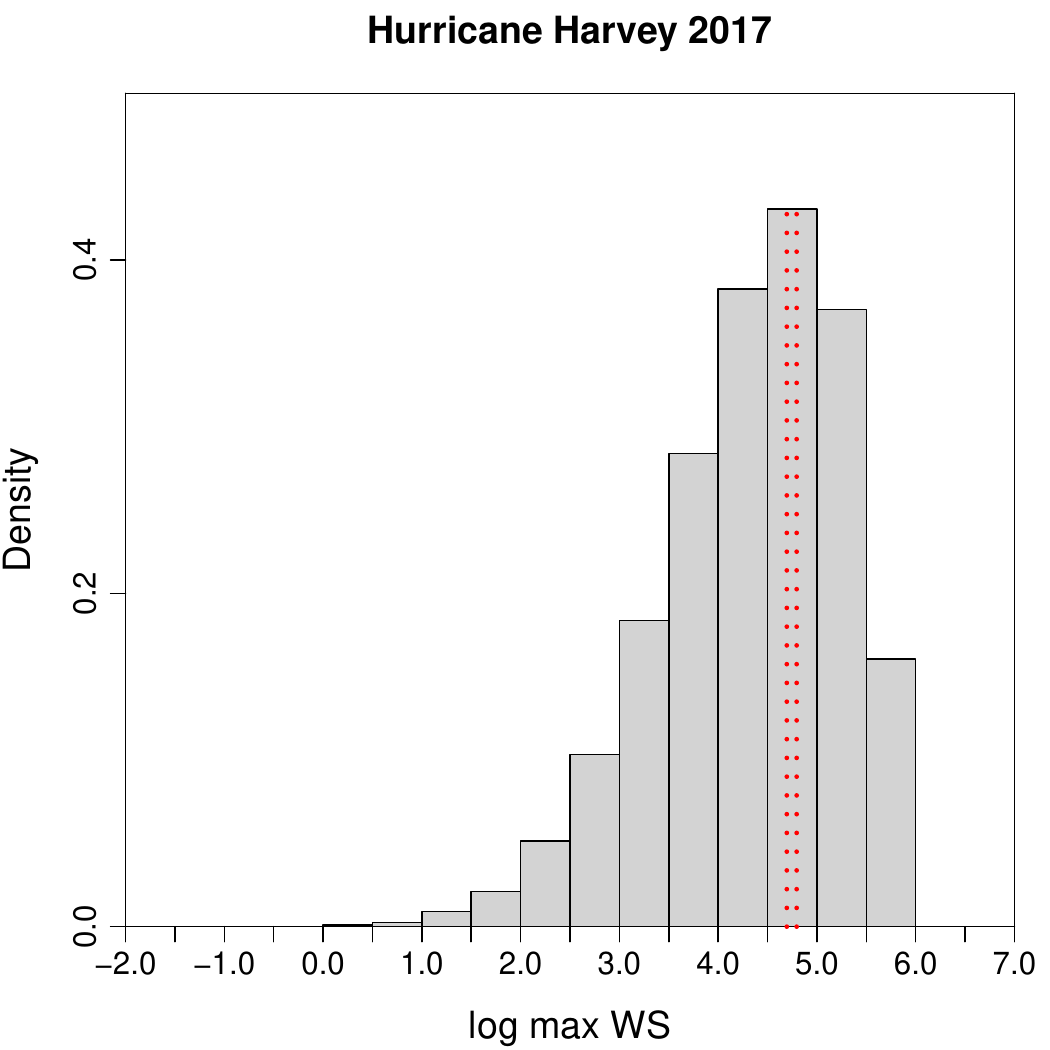}}
&
{\includegraphics[width=0.30\textwidth]{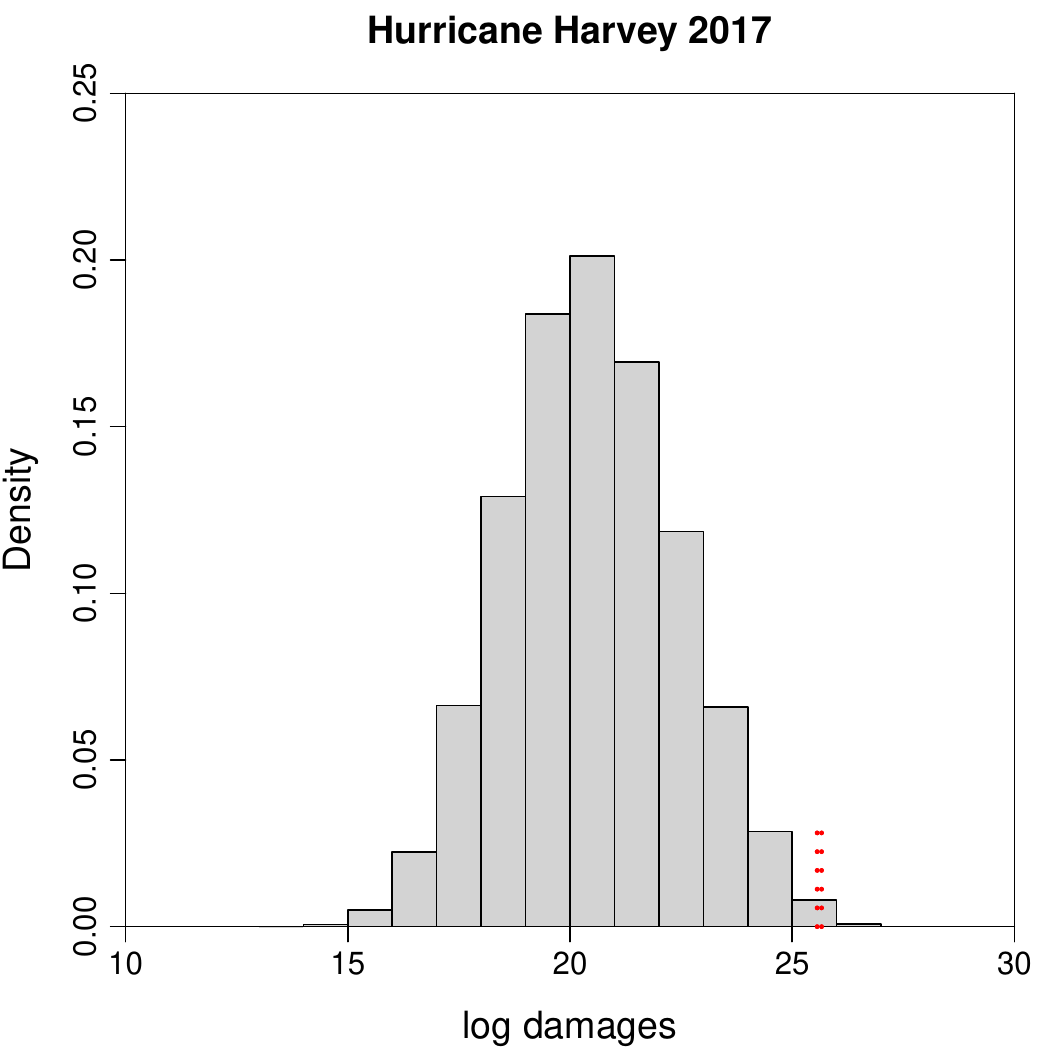}}
\\
{(b)\includegraphics[width=0.30\textwidth]{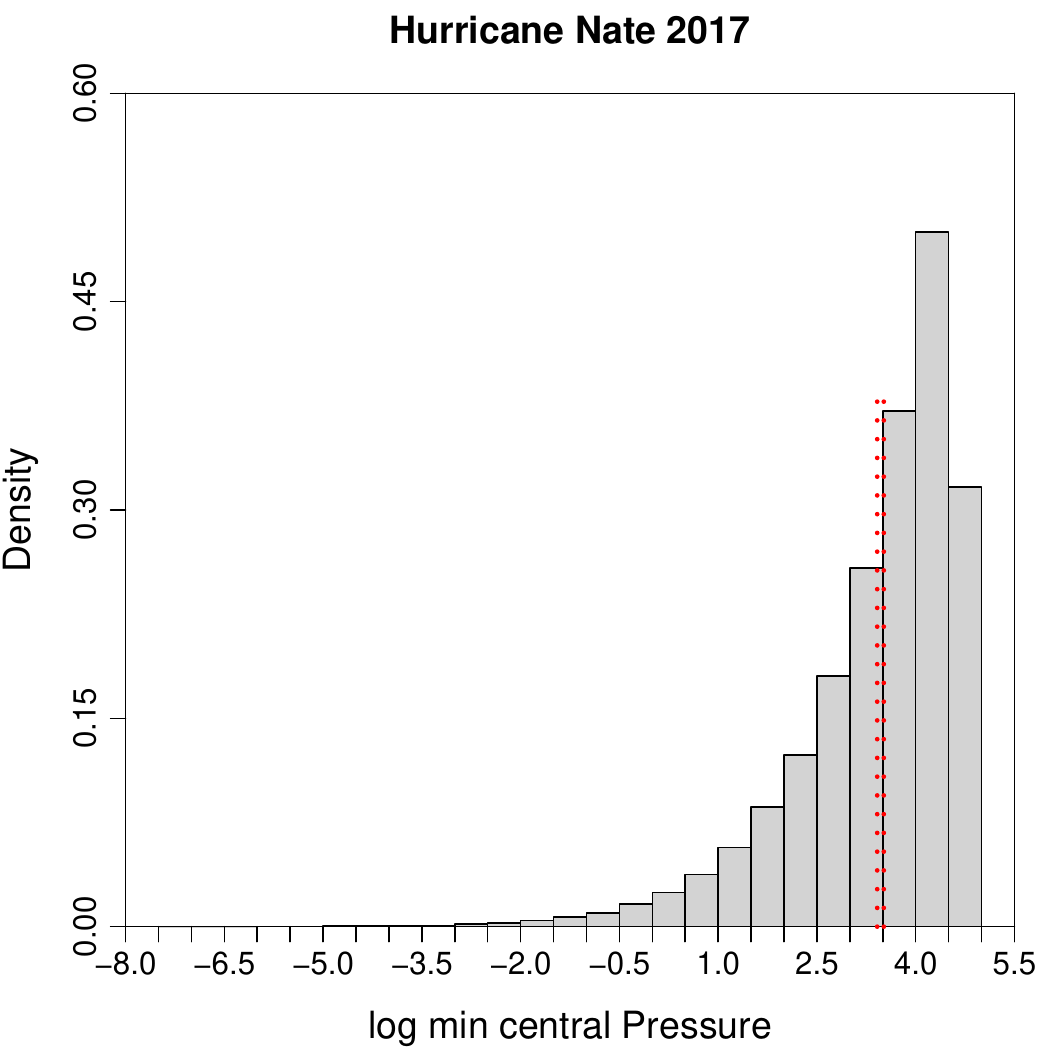}}
&
{\includegraphics[width=0.30\textwidth]{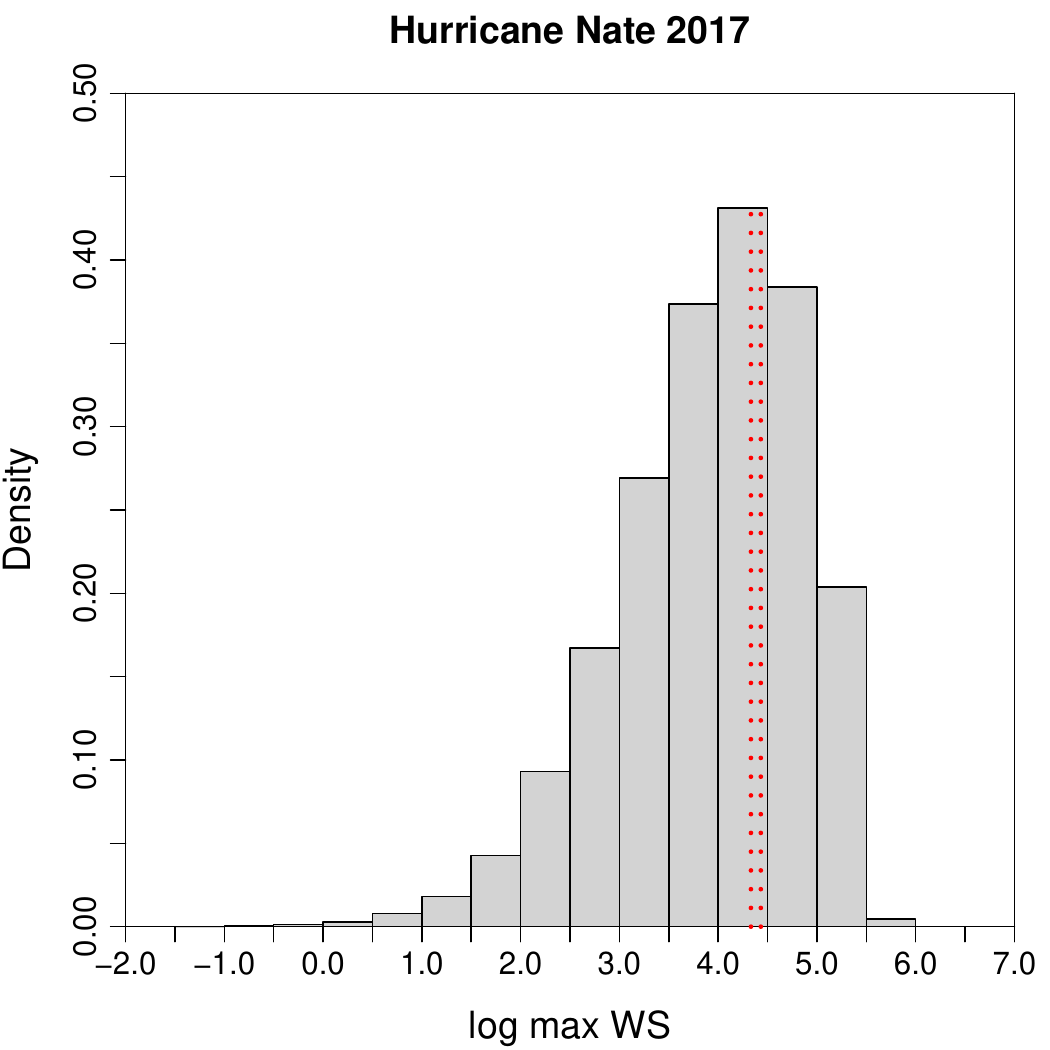}}
&
{\includegraphics[width=0.30\textwidth]{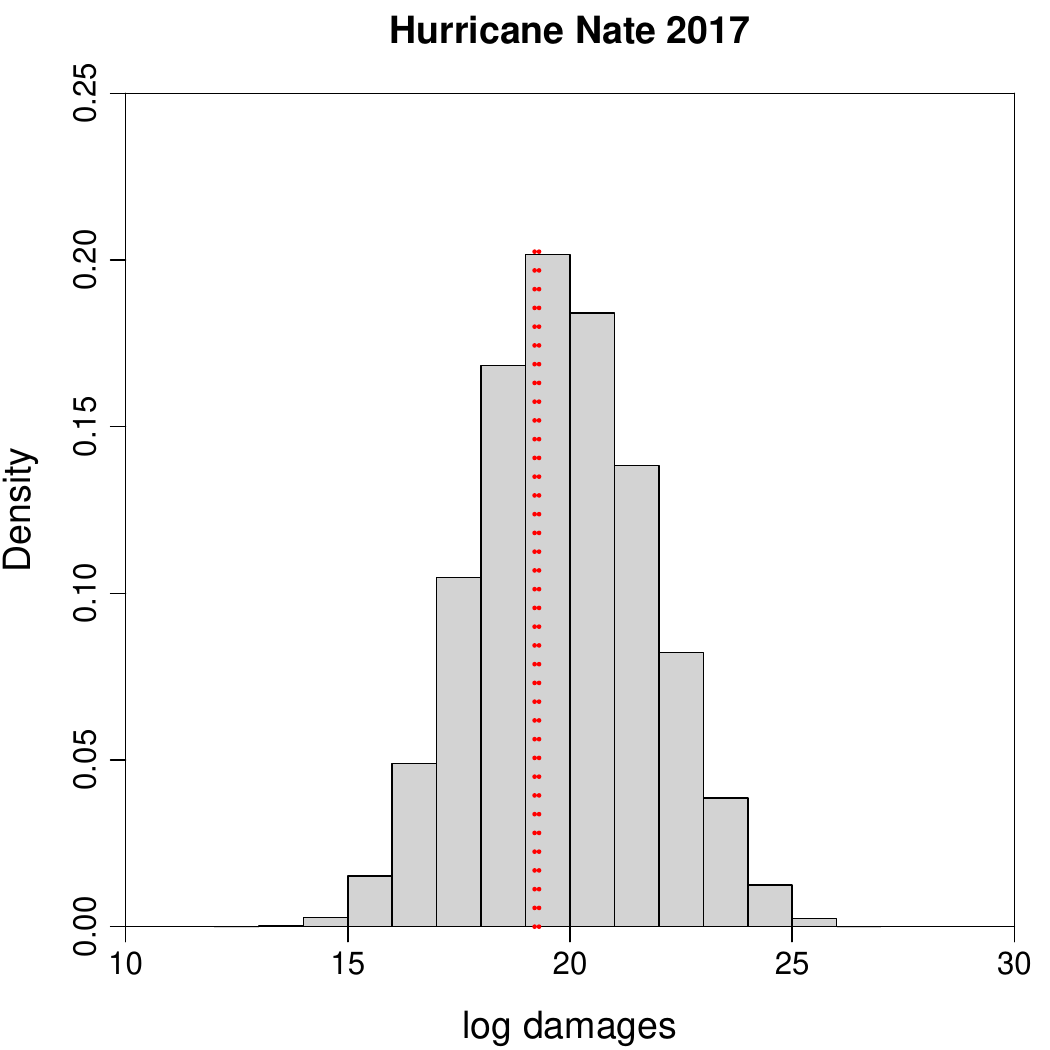}}\\
{(c)\includegraphics[width=0.30\textwidth]{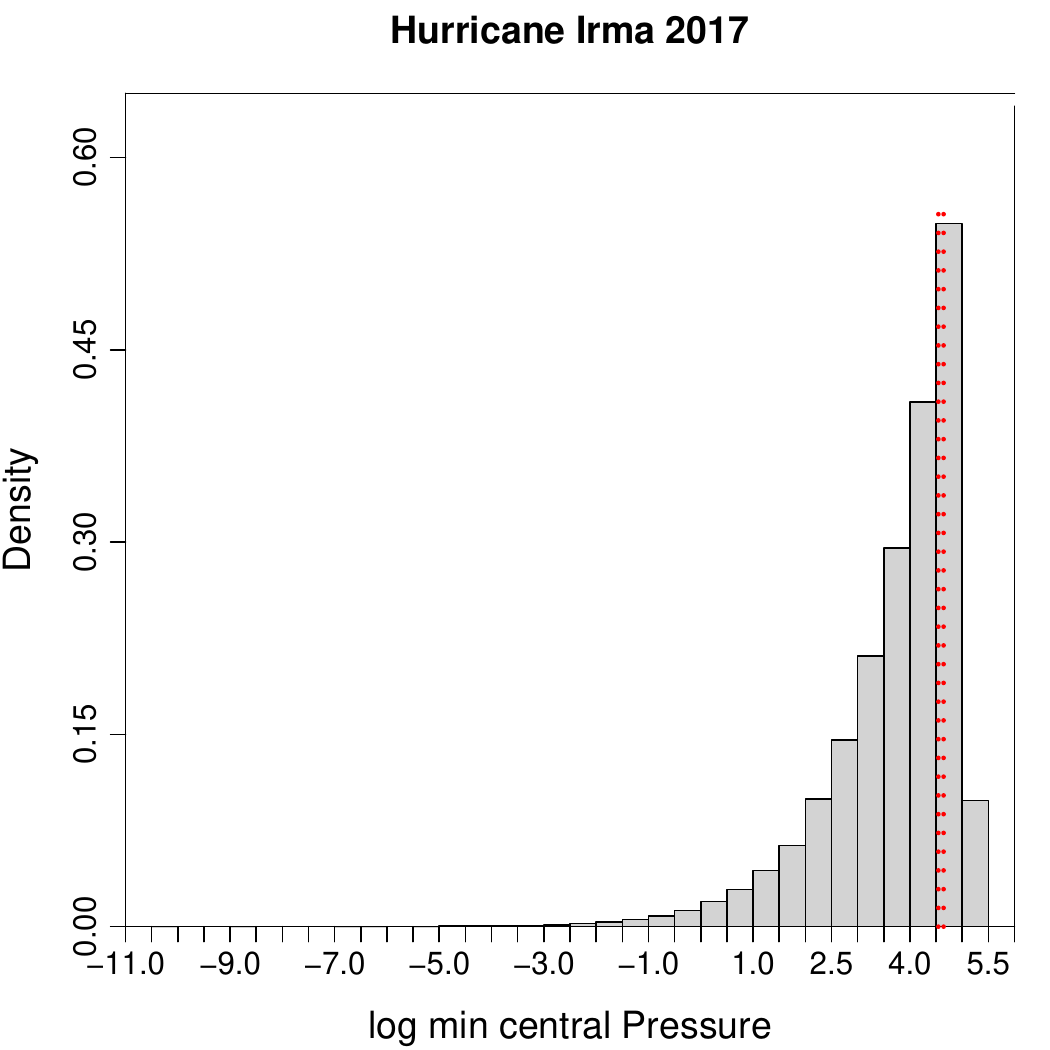}}
& 
{\includegraphics[width=0.30\textwidth]{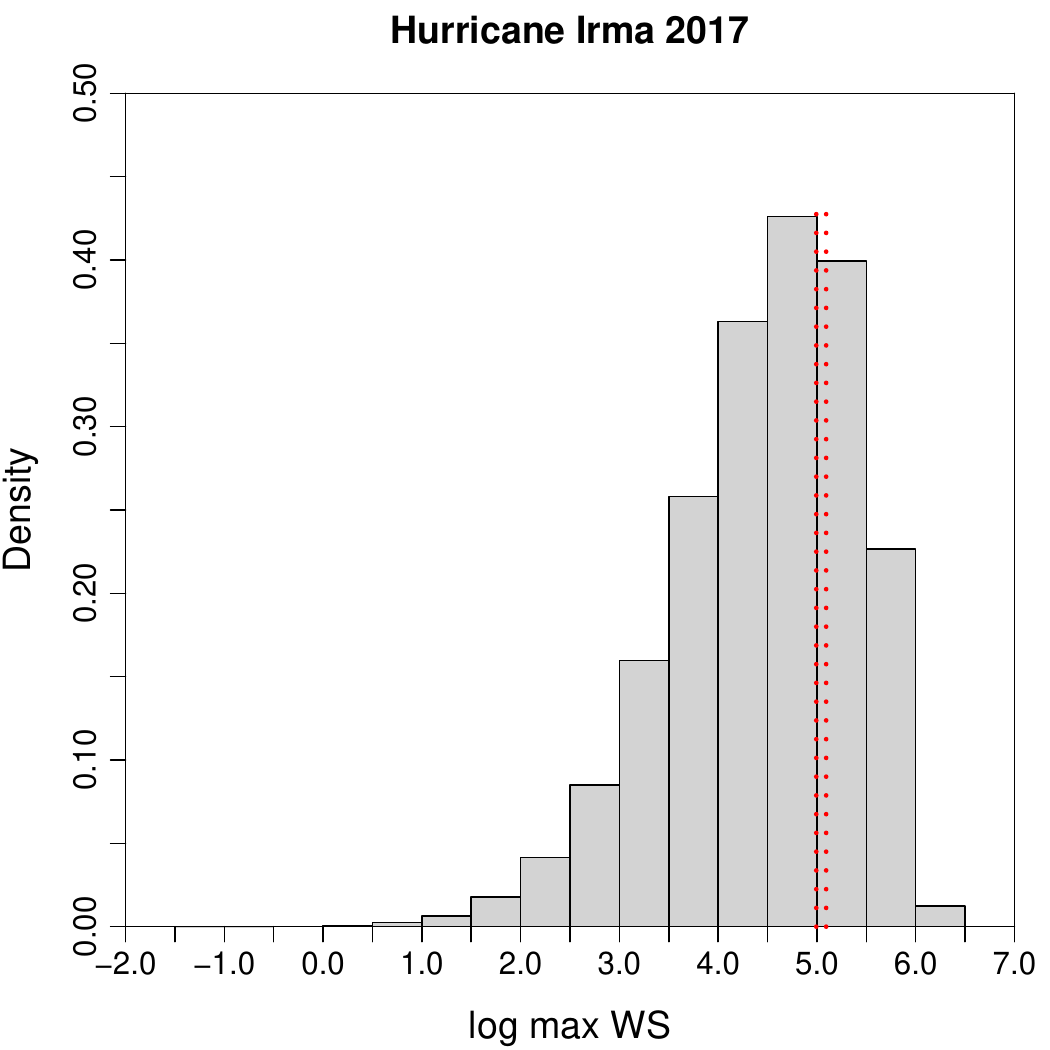}}
& 
{\includegraphics[width=0.30\textwidth]{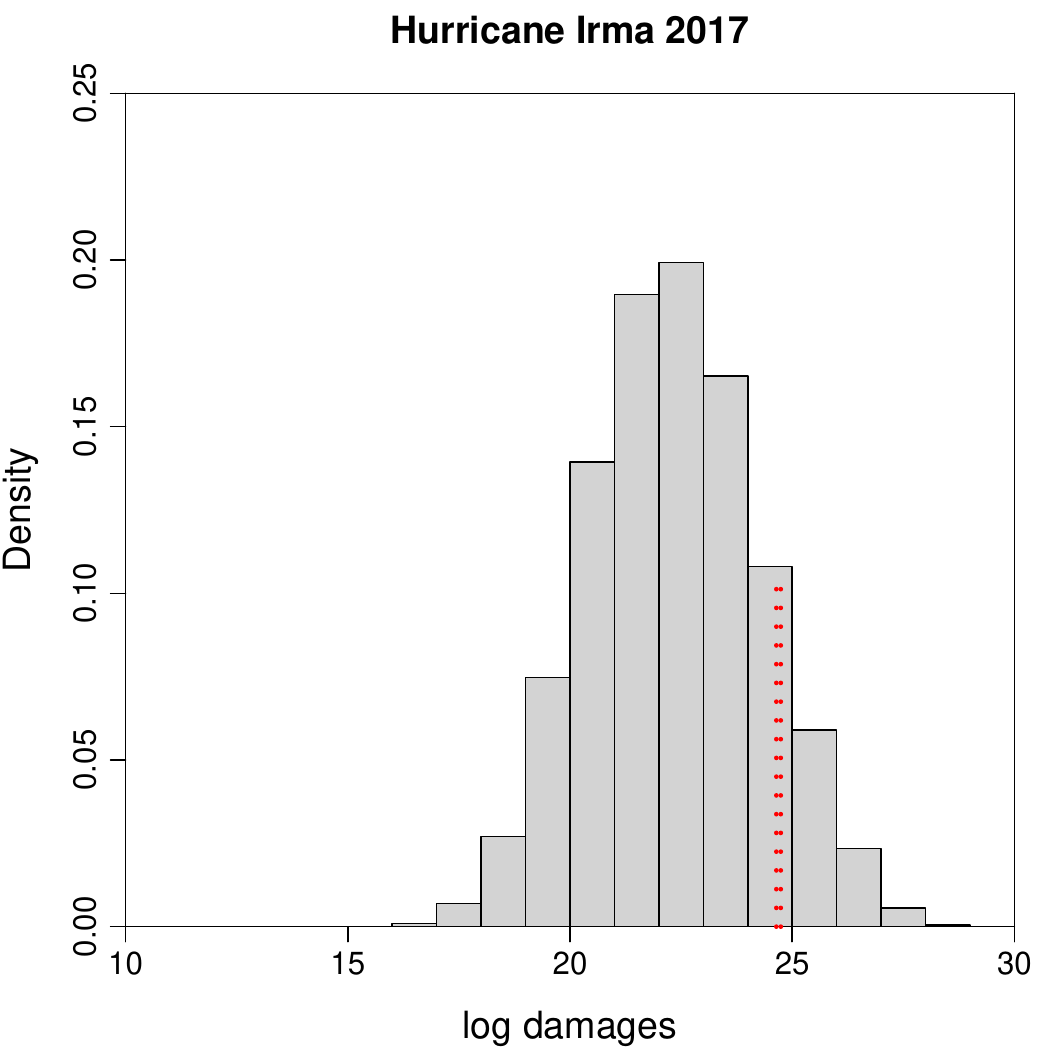}}
\end{tabular}
\end{figure}

\begin{figure}[h!] 
\centering
    \caption[Posterior Predictive Distributions for Storms in 2020]{\textbf{Posterior Predictive Distributions for Storms in 2020: }Posterior Predictive Distributions for minCP, maxWS and damages of tropical storms Delta (row a) and Eta (row b) and Sally (row c) in 2020 based on the hierarchical Bayesian GEV model with selected variables.  The actual values for the storms are displayed with red dashed lines.}
        \label{fig:2020hurricanes_hierGEV}
\begin{tabular}{ccc}
{(a)\includegraphics[width=0.30\textwidth]{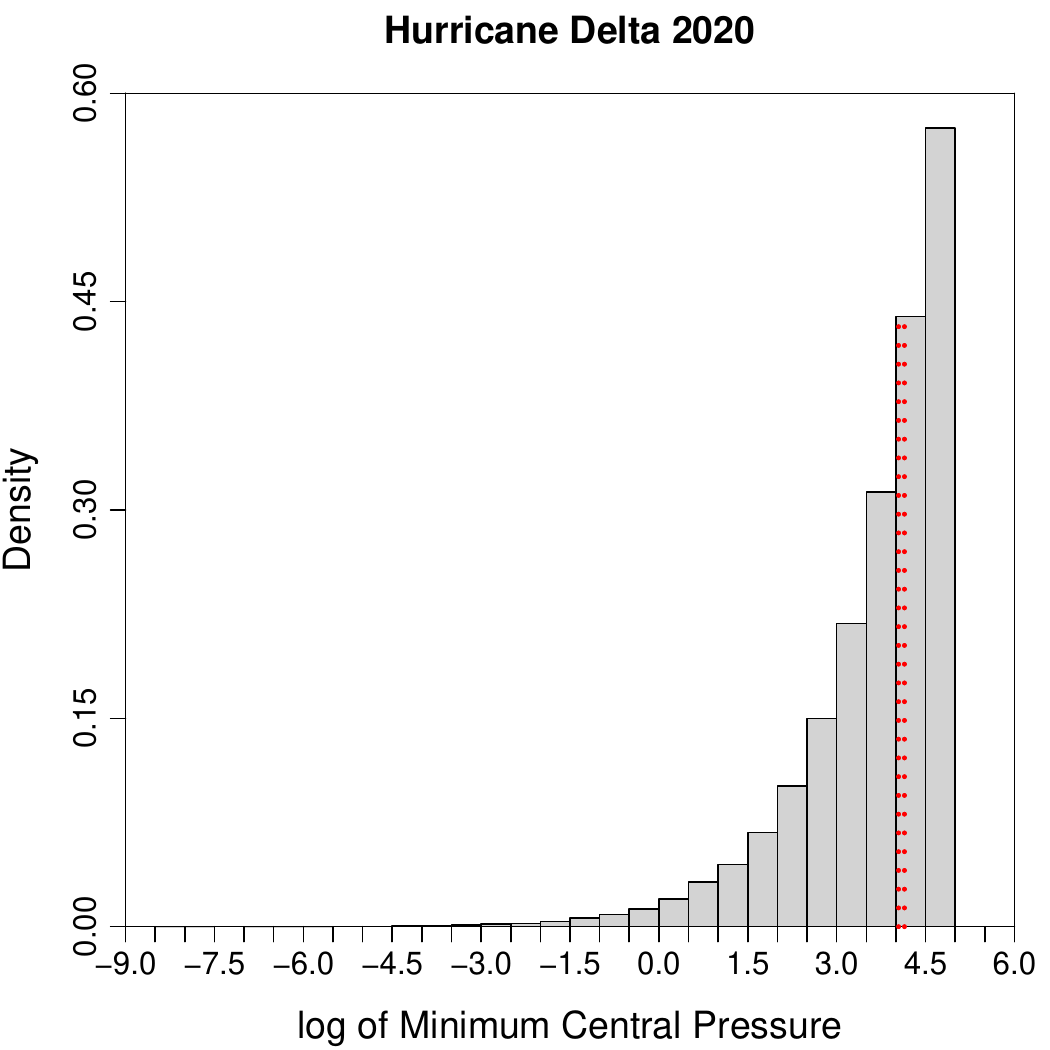}}
&
{\includegraphics[width=0.30\textwidth]{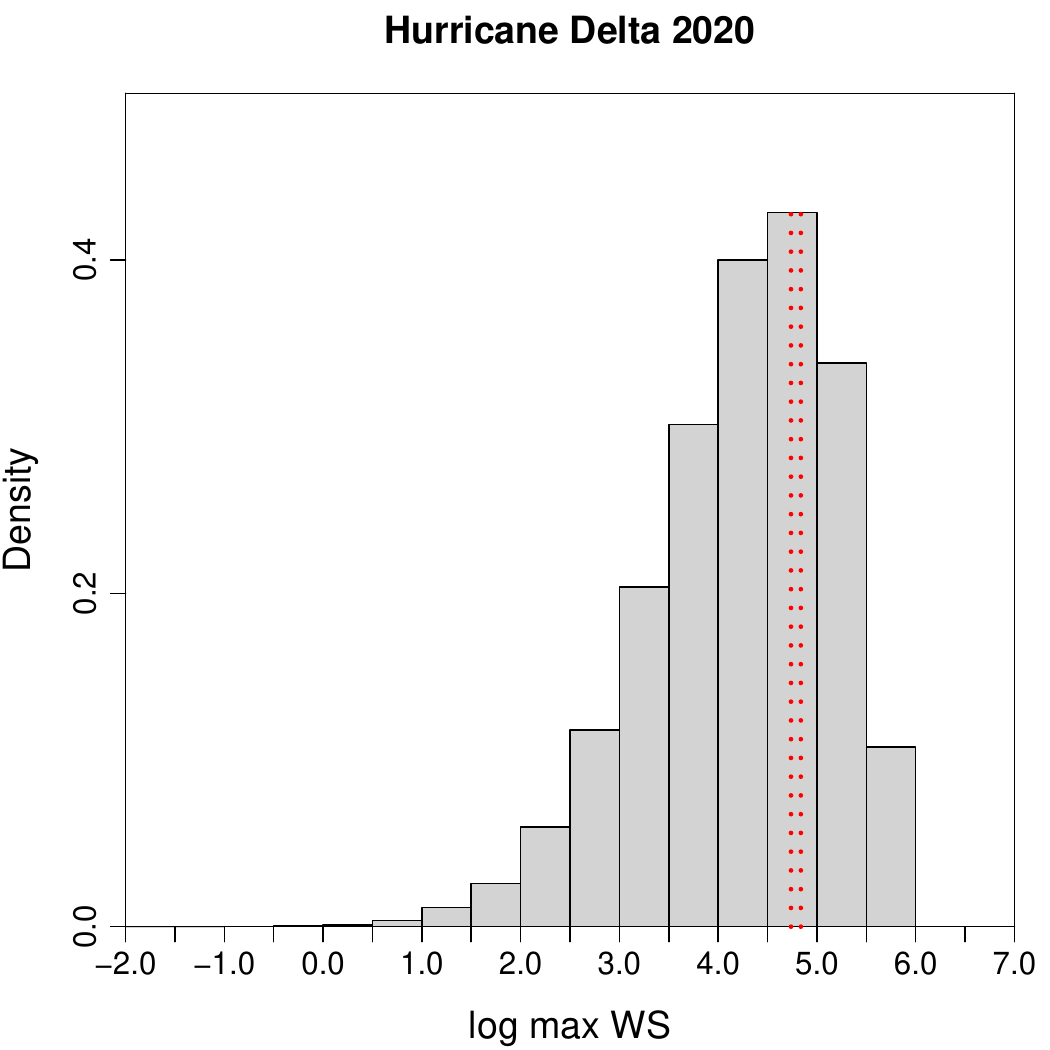}}
&
{\includegraphics[width=0.30\textwidth]{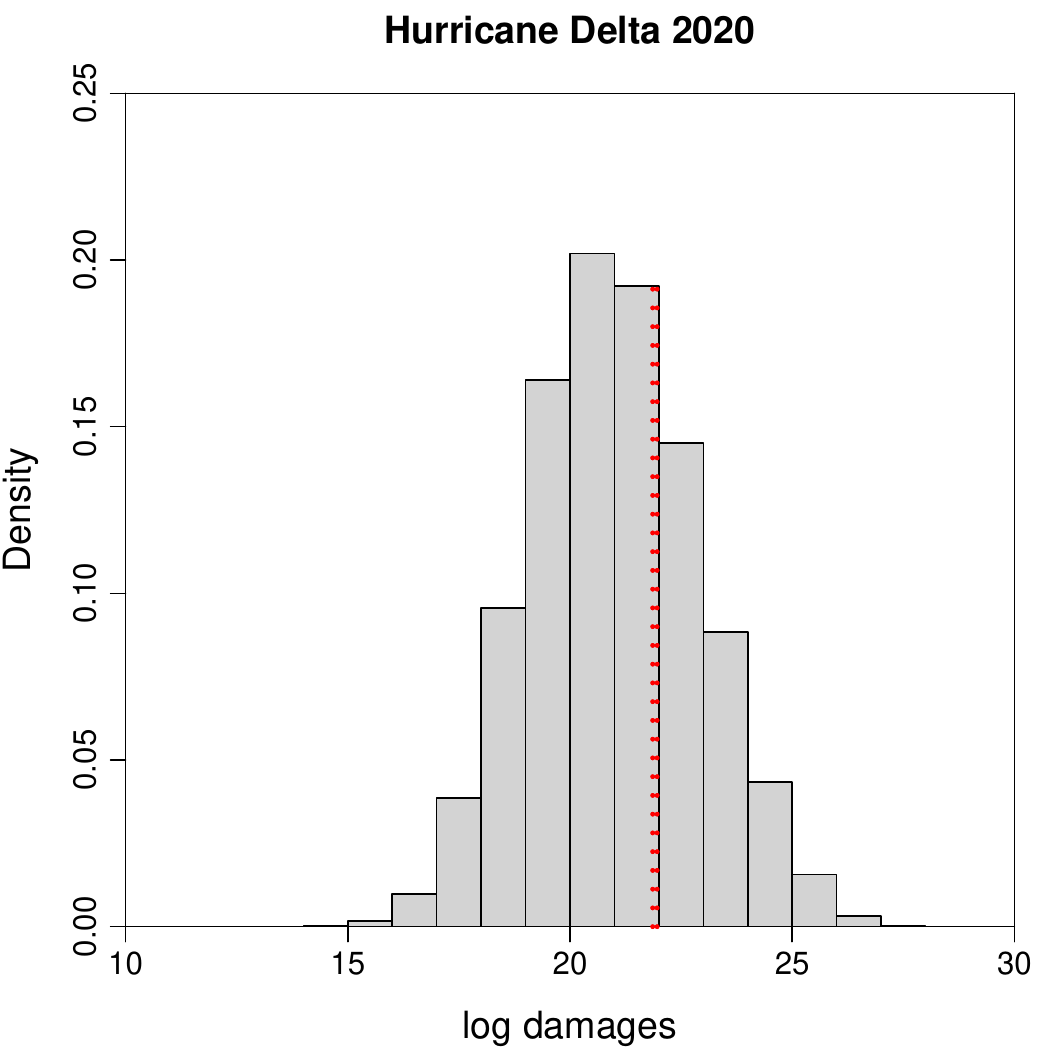}}
\\{(b)\includegraphics[width=0.30\textwidth]{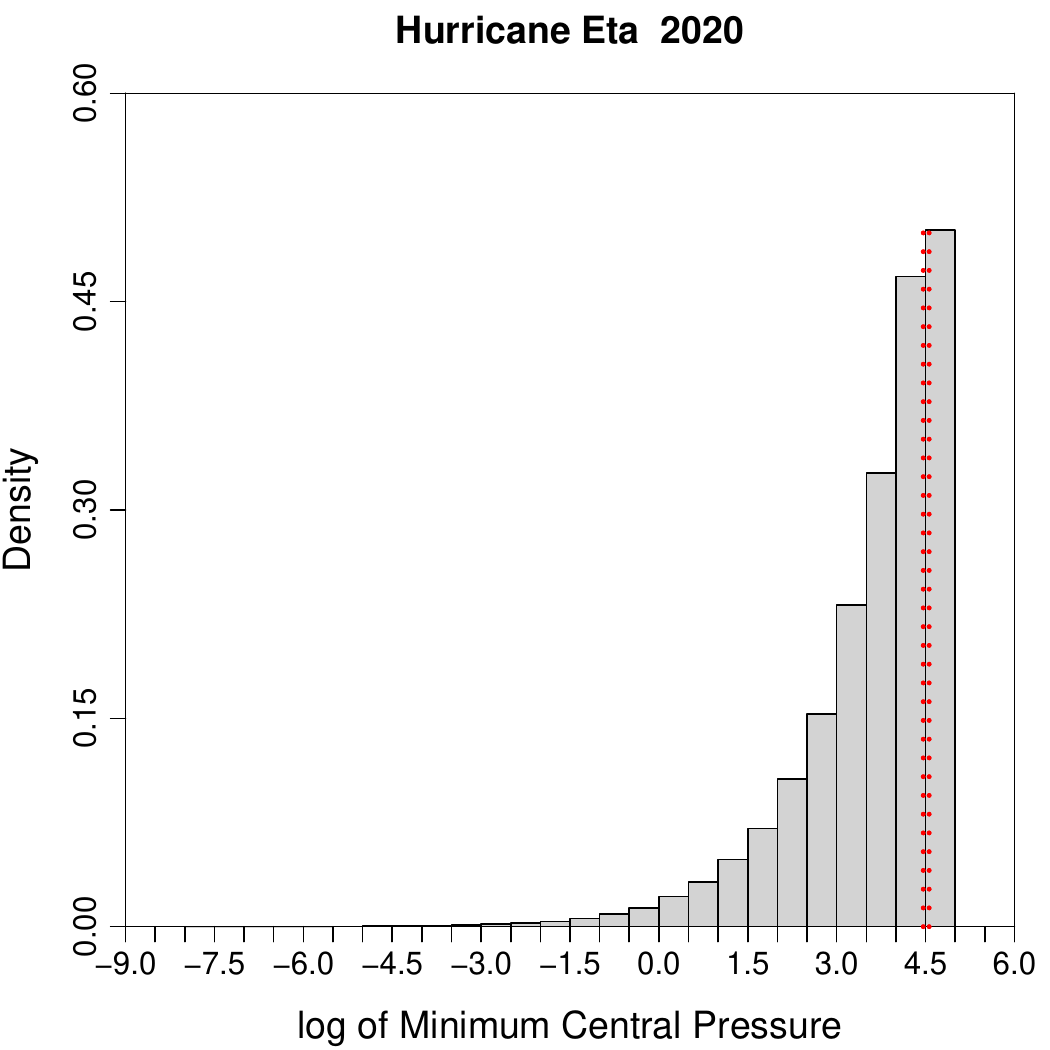}}
&
{\includegraphics[width=0.30\textwidth]{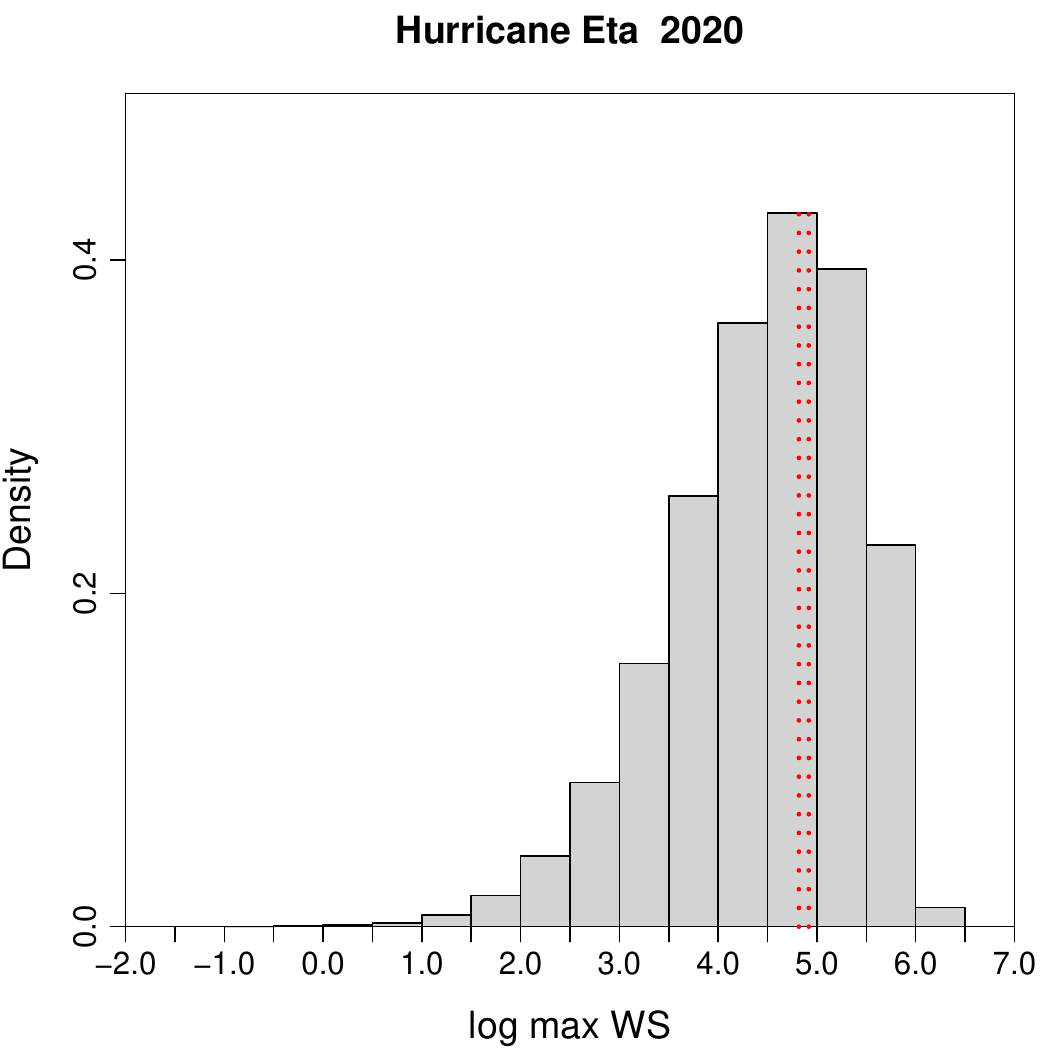}}
&
{\includegraphics[width=0.30\textwidth]{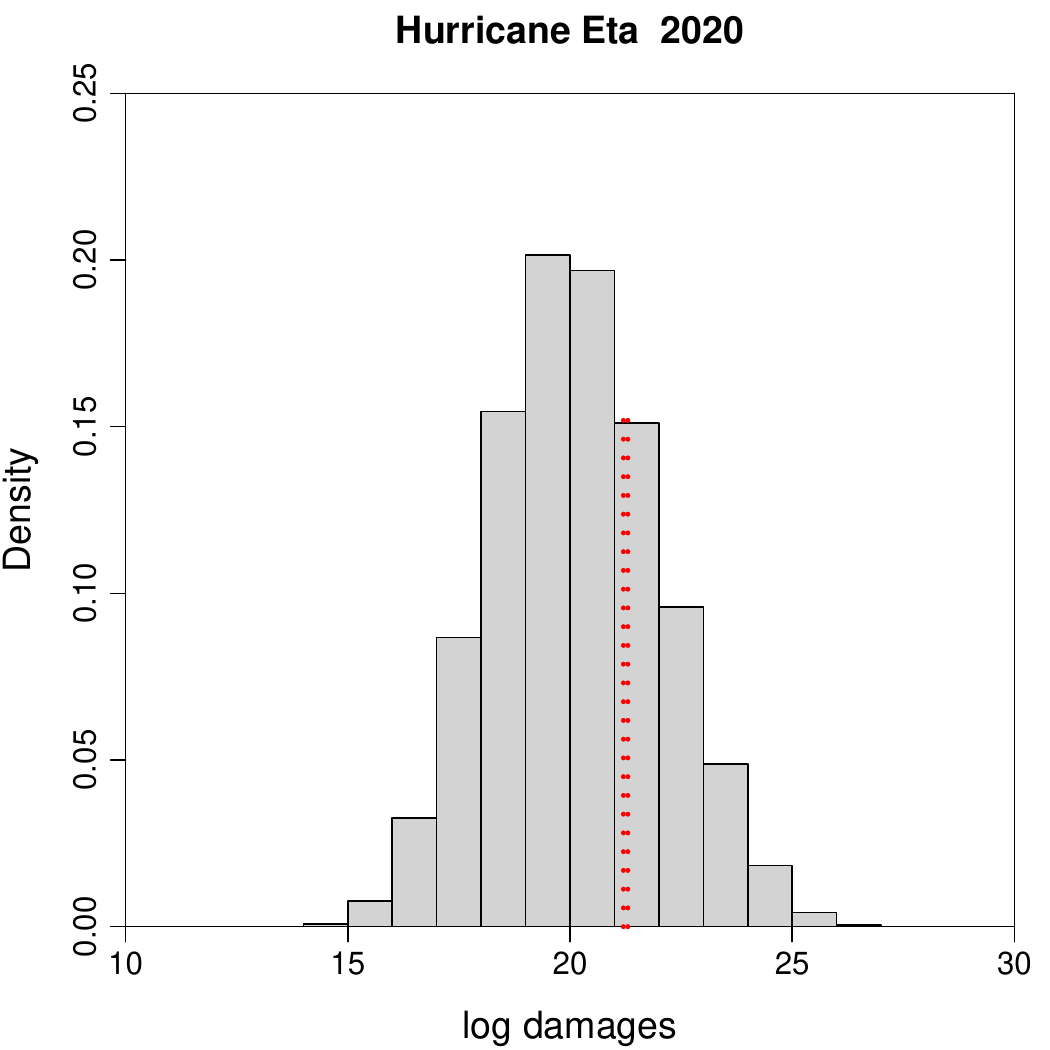}}\\
{(c)\includegraphics[width=0.30\textwidth]{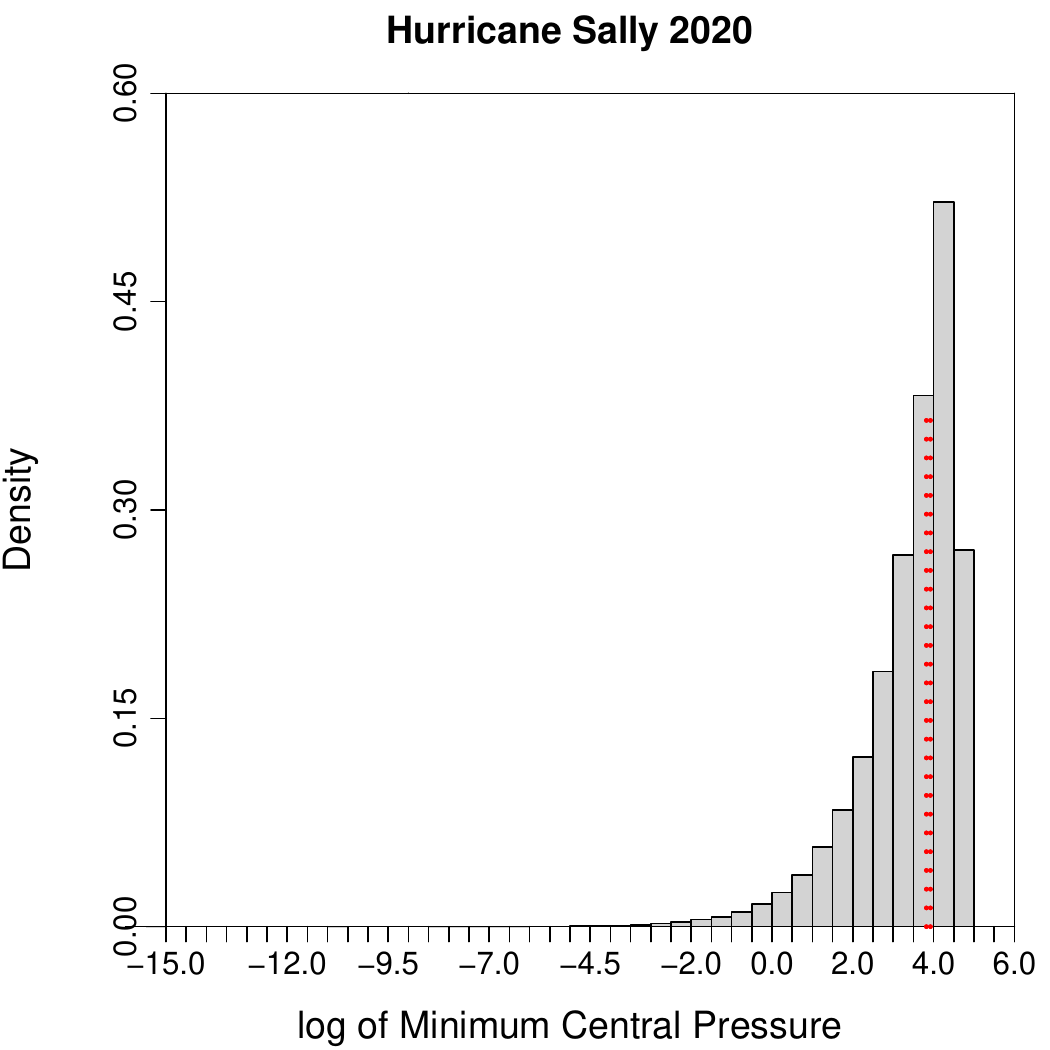}}
&
{\includegraphics[width=0.30\textwidth]{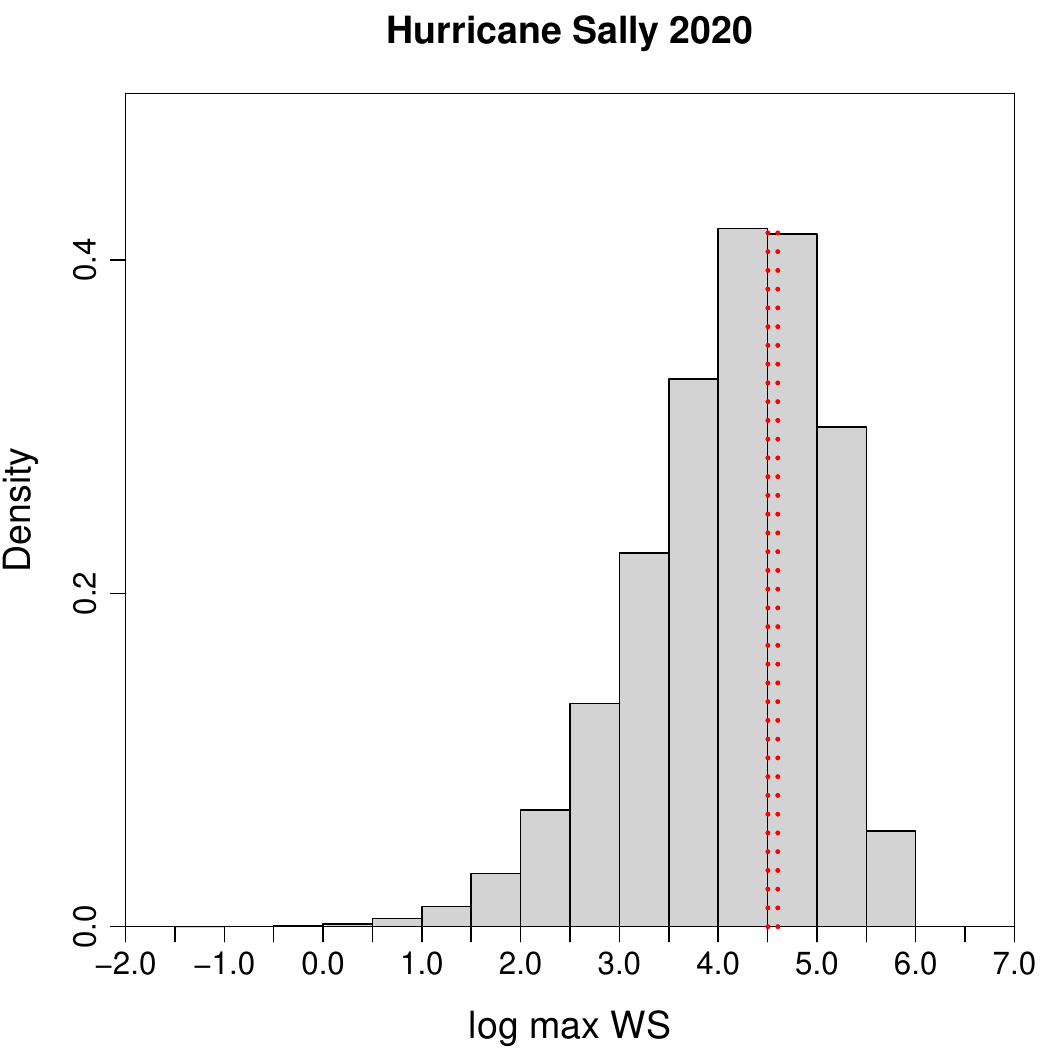}}
&
{\includegraphics[width=0.30\textwidth]{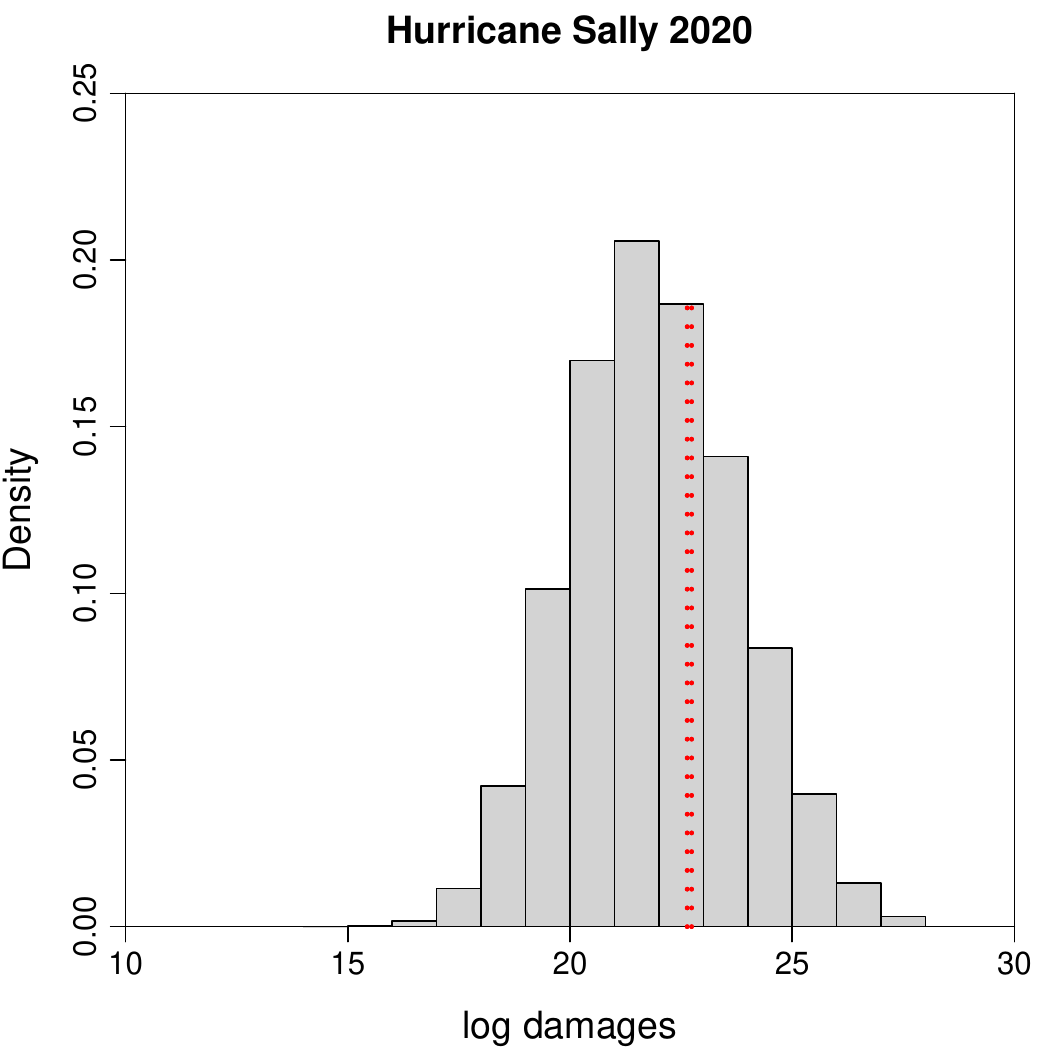}}
\end{tabular}
\end{figure}

\begin{figure}[h!] 
\centering
    \caption[Posterior Predictive Distributions for Storms in 2022]{\textbf{Posterior Predictive Distributions for Storms in 2022: }Posterior Predictive Distributions for minCP, maxWS and damages of tropical storms Ian (row a) and Nicole (row b) in 2022 based on the hierarchical Bayesian GEV model with selected variables.  The actual values for the storms are displayed with red dashed lines.}
        \label{fig:ian_nicole_hierGEV}
\begin{tabular}{ccc}
{(a)\includegraphics[width=0.30\textwidth]{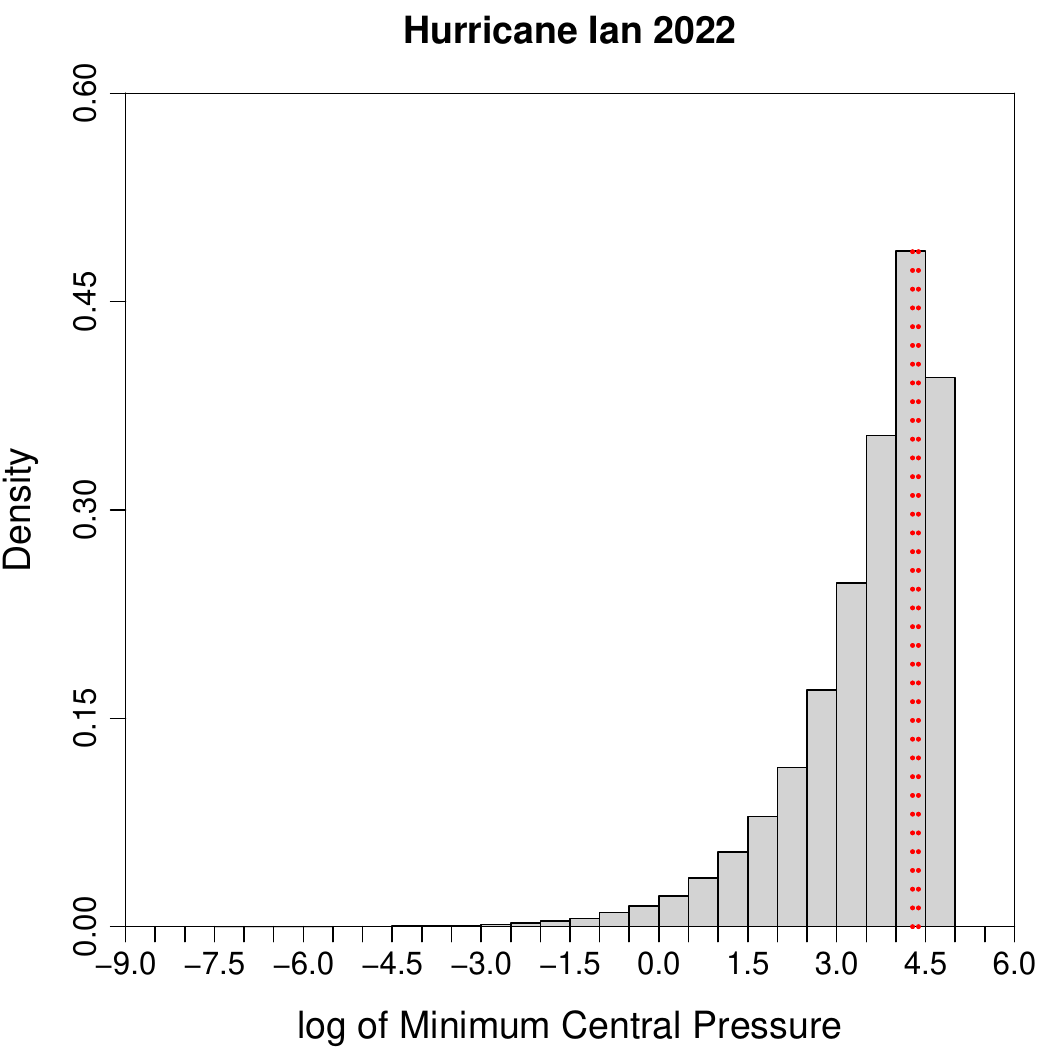}}
&
{\includegraphics[width=0.30\textwidth]{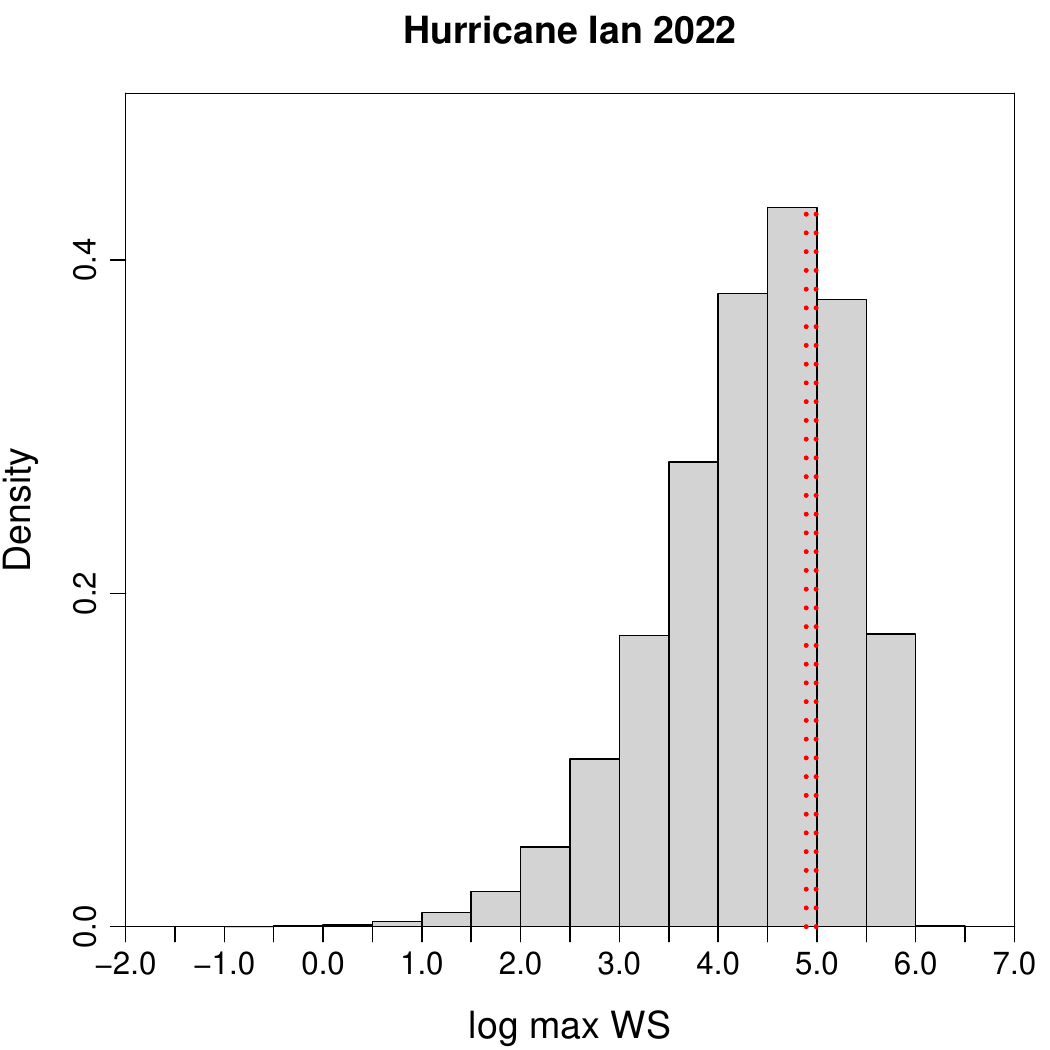}}
&
{\includegraphics[width=0.30\textwidth]{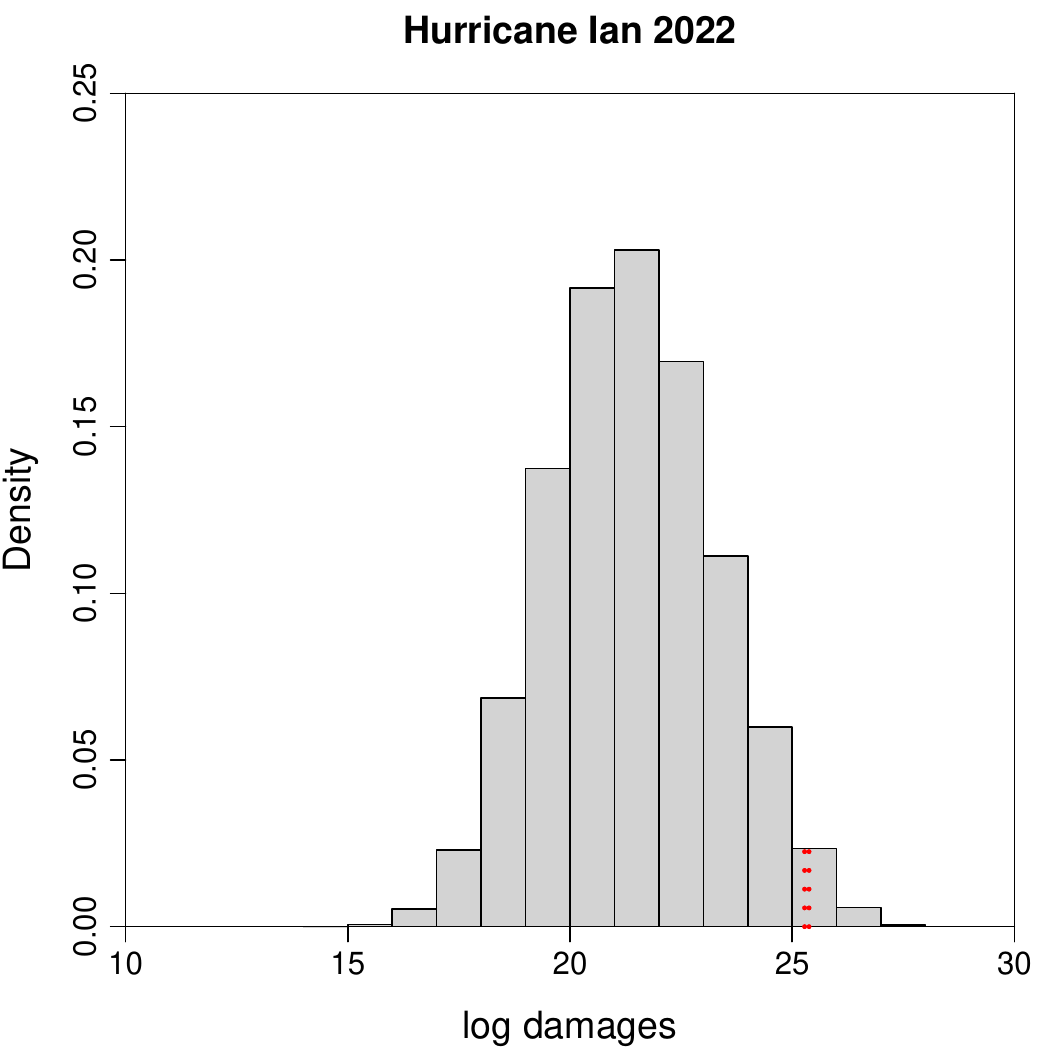}}
\\
{(b)\includegraphics[width=0.30\textwidth]{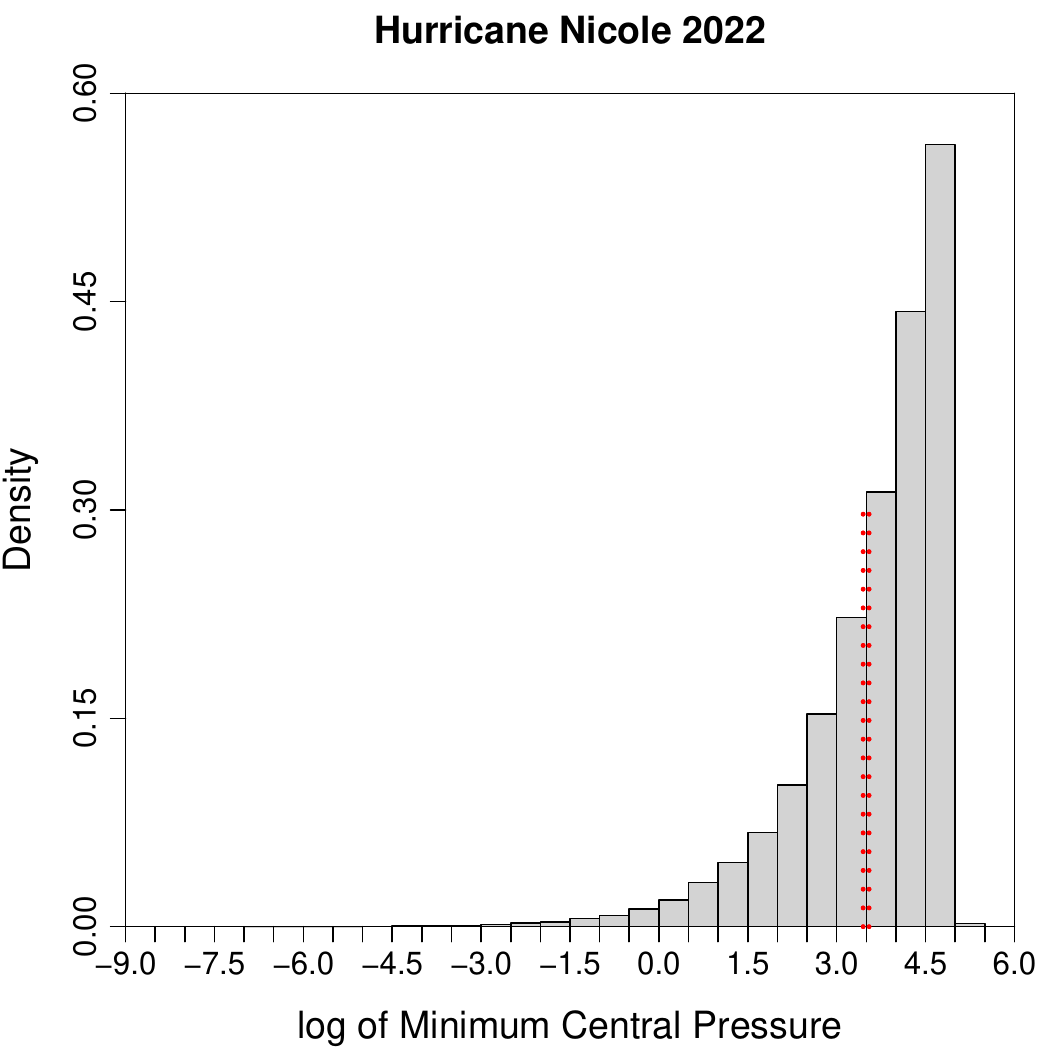}}
&
{\includegraphics[width=0.30\textwidth]{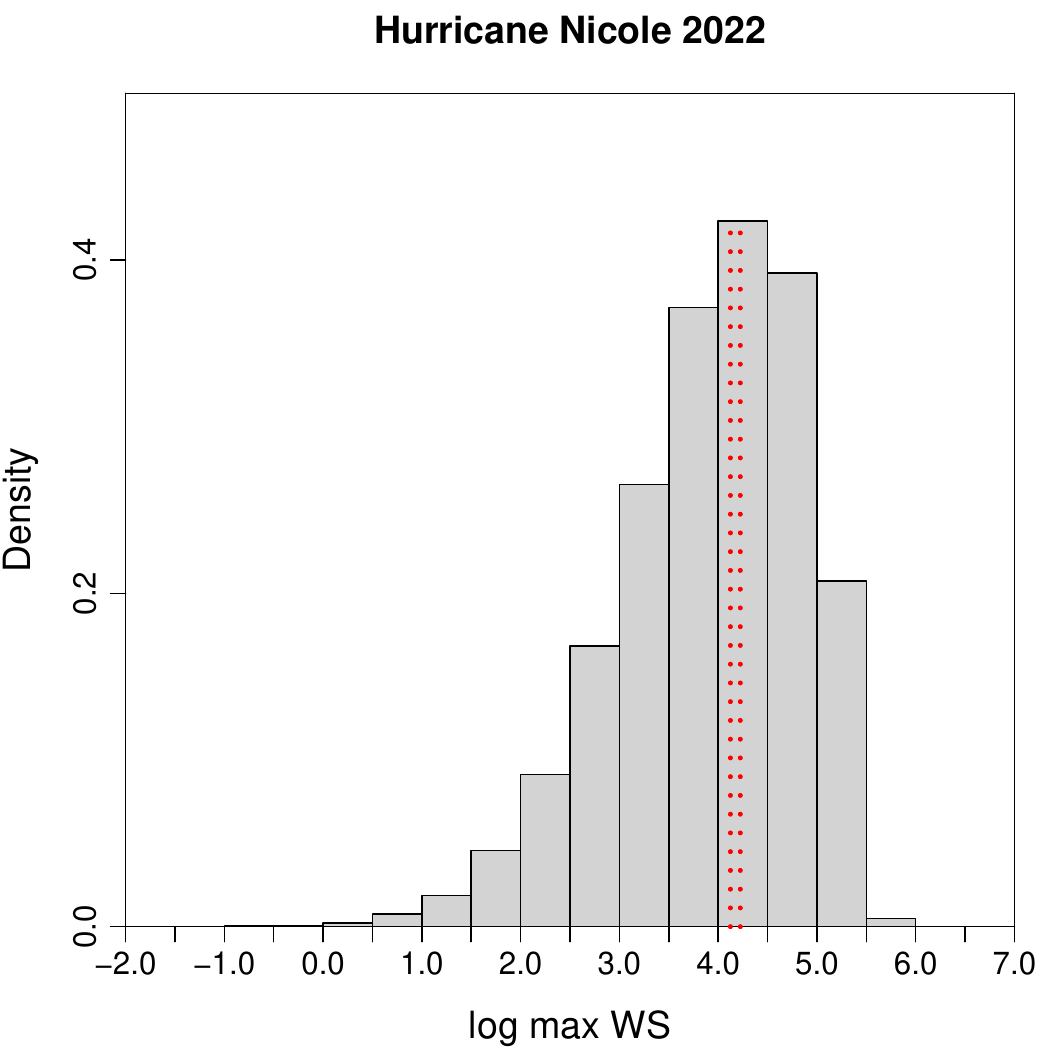}}
&
{\includegraphics[width=0.30\textwidth]{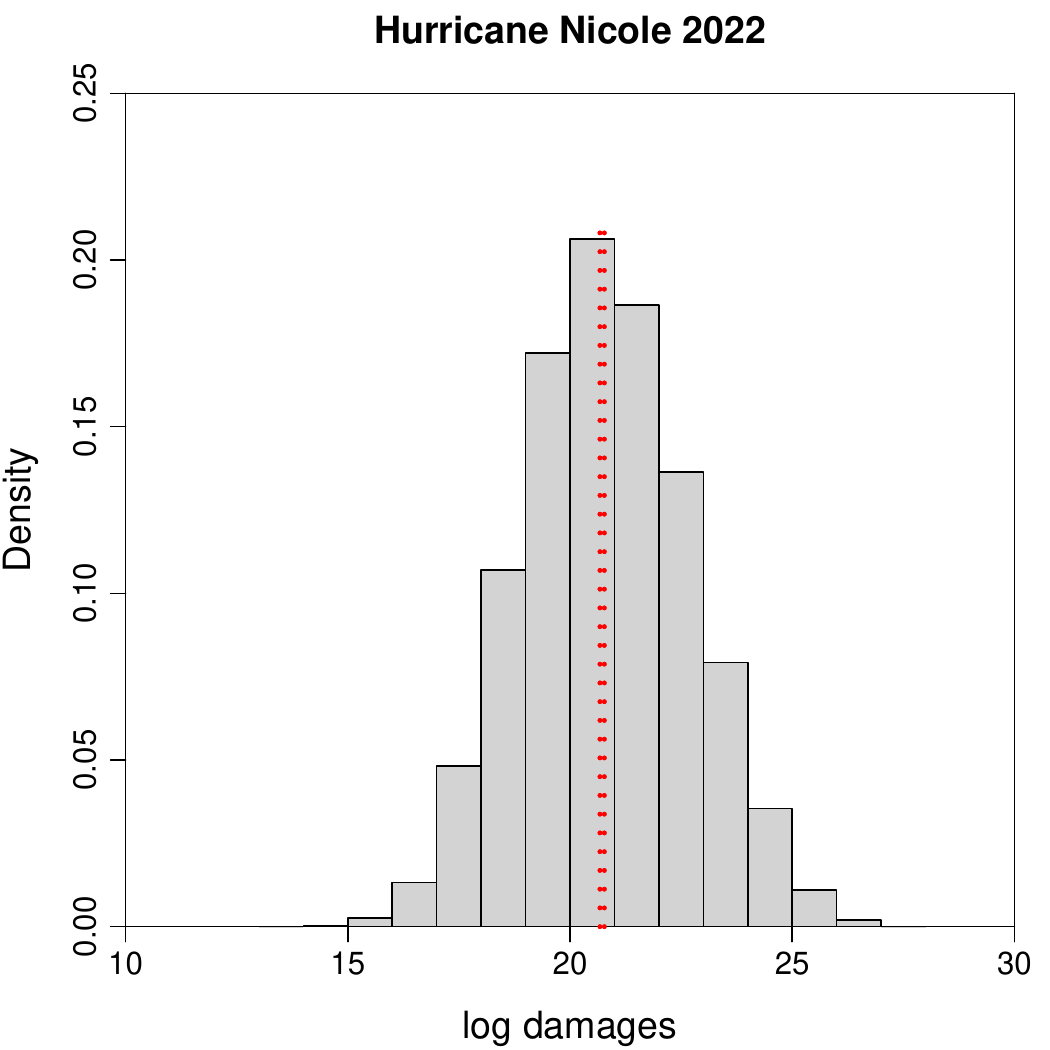}}
\end{tabular}
\end{figure}

The 2017 hurricane season was the costliest season since records began in 1851, in large part due to the devastation wrought by major hurricanes like Harvey, Irma, and Nate \citep{halverson2018costliest}. Natural disasters such as these highlight the need for quantitative estimates of the risk of such disasters.  
We display the posterior predictive densities for the  cyclones Harvey and Irma, two of the most historically damaging cyclones, in Figure~\ref{fig:harvey_irma_hierGEV}.  As can be seen from the figure, the actual amount of minimum central pressure and maximum wind speed for both the cyclones falls well within the predicted distribution. The actual amount of damage are within the limits of the predictive distribution for both Harvey and Irma, although towards the tails for damages. A similar trend can be noticed for Ian in 2022. For the other tropical cyclones of 2017, 2020, and 2022, the actual values were in substantially high posterior density regions, as can be observed in Figures~ \ref{fig:harvey_irma_hierGEV}, \ref{fig:2020hurricanes_hierGEV}, and \ref{fig:ian_nicole_hierGEV}, respectively. 

We conducted similar prediction analysis using a trivariate Bayesian GEV and a hierarchical Bayesian model with log-Normal damages, described in the supplementary materials (Appendix \ref{Sec:Storms_TrivariateGEV} and \ref{Sec:Storms_HierarchicalLogNormal}), and they enforce the narrative that our results are very robust against different data science techniques used.  Our predictive results also reflect that hurricanes Harvey and Ian were exceptional in terms of the damages, at the tails but within the range of the posterior distributions.  Our hierarchical Bayesian GEV model predicts that Harvey was one-in-a-thousand and Ian was a one-in-thirty event in terms of damages they inflicted.

\begin{table}[h!]
\centering
\begin{tabular}{|c c c c|}
\hline
Hurricanes & minCP & maxWS & Damage\\
\hline
Harvey (2017) &  0.45196     &   0.76586   &  0.00092   \\
Irma (2017) &    0.48402   &   0.62544     & 0.22868 \\
Nate (2017) &    0.76132   &  0.6869    &    0.70158  \\
Delta (2020) & 0.94194  & 0.62466 & 0.60472\\
Eta (2020) & 0.72962 &  0.75242 & 0.5528\\
Sally (2020) & 0.90424 & 0.72698 & 0.6722\\
Ian (2022) & 0.5744   &  0.60276  &   0.03816  \\
Nicole (2022) &  0.67722   &   0.88266  &   0.98188   \\
\hline
\end{tabular}
\caption{The $\delta = \delta = 2 \min\{ \alpha, 1-\alpha\}$ values with $\alpha$ denoting the empirical CDF for each of the three variables for the  tropical cyclones of 2017, 2020, and 2022, using the hierarchical GEV model. A $\delta$ value close to 1 reflects the truth to be close to the median of the posterior predictive distribution, and close to 0 reflects the truth lying in the tails of the distribution.}
\label{tab:delta_value_hierGEV}
\end{table}

\subsection*{{Data and data processing}}
\label{Sec:Data}

The National Hurricane Center (NHC) maintains the North Atlantic-basin hurricane database (HURDAT2, or Best Track), containing six-hourly information on the location, maximum winds, central pressure, and (beginning in 2004) size of all known tropical cyclones and subtropical cyclones since 1851 \citep{Landsea}. 
We use the data from the Atlantic tropical cyclone basin from 1960 up to 2022 in {HURDAT2}. 
Data before 1960 is not used in this study owing to possible inaccuracies.  
We define the maximum category a cyclone achieves by applying the Saffir-Simpson scale to the highest maximum wind speed\footnote{Maximum wind speed is defined as maximum wind speed value over 1 minute} over the cyclone's lifetime.  Tropical storms are included with tropical cyclones due to several damage events attributed to these less powerful cyclones.  Monetized damage estimates for all tropical cyclones since 1900 have been compiled \citep{Pielke}, and updated by the ICAT catastrophe insurance company {\url{https://public.emdat.be/}}  
The data is normalized to {2024} dollars to reflect changes in inflation, wealth, and population in the cyclone area \citep{Pielke}.

We use the Atlantic Multidecadal Oscillation {(AMO)}, the Southern Oscillation Index (SOI), the North Atlantic Oscillation (NAO), Ni\~{n}o 3.4 anomaly series, 
sea surface temperature  {SST}  and  sunspot activity {SSN} as covariates. 
The Atlantic Multidecadal Oscillation (AMO) is an ongoing series of long-duration changes in the sea surface temperature of the North Atlantic Ocean, with cool and warm phases that may last for 20-40 years at a time.  It is the ten-year running mean of detrended Atlantic  {SST} Anomalies north of the equator.  Data \citep{Enfield}  are retrieved from the National Oceanic and Atmospheric Administration (NOAA), Earth System Research Laboratories {(ESRL)} \url{http://www.esrl.noaa.gov/psd/data/timeseries/AMO/}. The relationship between AMO and hurricane frequency has been studied previously, with some attributing the increase in hurricane activity to increases in  AMO \citep{trenberth2006atlantic,zhang2006impact,li2009review,alexander2014climate,loehle2020hurricane}.

The Southern Oscillation Index ({SOI}) is defined as the normalized sea-level pressure difference between Tahiti and Darwin. Negative values of the  {SOI} indicate an El Ni{\~n}o event.  Monthly  {SOI} values are obtained from  National Centers for Environmental Prediction's {(NCEP)} Climate Prediction Center {(CPC)} (\url{ftp://ftp.cpc.ncep.noaa.gov/wd52dg/data/indices/soi}).   Annual averages of SOI over the months of August-October are used as indicators of shear upon North Atlantic hurricanes \citep{Elsner14,hodges2012spatial}.

The North Atlantic Oscillation {(NAO)} is characterized by fluctuations in sea level pressure differences.  Strong positive phases of the  {NAO} tend to be associated with above-average temperatures in the eastern United States and thus, provide a conducive environment for tropical cyclone development.  Index values for the  {NAO} are calculated as the difference in sea level pressure between Gibraltar and a station over southwest Iceland and are collected from  Physical Sciences Laboratory {(PSL)}, {NOAA} {(\url{https://psl.noaa.gov/data/correlation/nao.data})}.

The Ni{\~n}o 3.4 anomaly series is collected from  {NCEP}  {CPC} {(\url{http://www.cpc.ncep.noaa.gov/data/indices/ersst5.nino.mth.91-20.ascii})}.  This series is an average of the  {SST} from 5$^{\circ}$S-5$^{\circ}$N by 170$^{\circ}$W-120$^{\circ}$W with the 1951-2000 mean removed.  Other Ni{\~n}o indices exist but are highly correlated with Ni{\~n}o 3.4, and are less commonly used in literature.

Sea-surface temperatures (SST) are an important component of tropical intensification. Higher SST, all else being constant, is believed to provide a more conducive environment for tropical cyclone development \citep{fraza2015climatological,Dailey}.
Atlantic  {SST} averages gridded values over the region from 10-25$^{\circ}$N by 80$^{\circ}$W-20$^{\circ}$W.  Raw (unsmoothed and not detrended) monthly  {SST} values are obtained via the  {NOAA}  {PSL} {\url{https://psl.noaa.gov/data/gridded/tables/sst.html}}.  This is version 5 of the data known as  {ERSST} and was constructed using the most recently available  {ICOADS}  {SST} data \citep{Smith2008}.

Sunspots are magnetic disturbances  (SSN) of the sun's surface having both dark and brighter regions. Variations in solar activity are monitored by sunspots. These are visible disturbances in the photosphere of the sun.  The brighter regions increase the intensity of the ultraviolet emissions. Increased sunspot numbers correspond to more magnetic disturbances, which some studies predict leads to a reduction in the potential intensity of hurricanes \citep{hodges2014sun}.  {SSN} are obtained from World Data Center-Sunspot Index and Long-term Solar Observations at the Royal Observatory of Belgium {(\url{http://www.sidc.be/silso/datafiles})}.

We use the monthly time series available for each covariate.  Previous works \citep{Elsner28} suggest an average of the May and June values of the  {SOI},  {NAO}, and Atlantic  {SST} anomalies for prediction.  We operate under this same premise for  {AMO} and the Ni\~{n}o 3.4 anomaly.  However, with SSN we use the average of the monthly average sunspots for July to June of the predicting year, i.e., for 2019, we average July 2018 to June 2019 monthly values.

\section*{Conclusion and Discussion}
\label{Sec:Discussion}


In this paper, we propose a data-driven framework for predicting the monetary value of damages caused by Atlantic tropical cyclones and storms. The framework is developed to analyze data at two-time scales: for an entire season of cyclones and each individual cyclone or storm. The seasonal model predicts storm or cyclone frequency, the probability of causing any damage, and the amount of damage. The individual cyclone model predicts the minimum central pressure, maximum wind speed, and the amount of damage caused. Both the models exhibit excellent predictive power as evident from Figures~\ref{fig:2017_pred}~-~\ref{fig:harvey_irma_hierGEV} and other figures and tables in supplementary materials. The inference results for our proposed Bayesian hierarchical models are robust, as verified by replicating the studies with alternative data science models and posterior predictive checks. The model fits are also satisfactory as seen from the diagnostics of the Markov Chain Monte Carlo procedures.
Our estimates indicate that on average, the United States should prepare for approximately 12.597 billion dollars billion dollars worth of damages per year at current prices.

Our Bayesian hierarchical models can easily accommodate additional features and variables like exact location of tropical cyclone landing and degree of urbanization, monetized ecological and environmental losses \citep{pruitt2019call}, and so on. Minor computational extensions of our model can be used for variable or model selection also, for example, the exact landing spot of a tropical cyclone may not have substantial predictive value from a data science perspective. While using additional features can potentially lead to more precise predictions and narrower prediction intervals, we restrict our analysis to those features for which trustworthy and adequate data were available.
The Bayesian hierarchical models proposed here may also be useful for other basins of tropical cyclone activities. However, adequate data on cyclone damages seems to be available only for the Atlantic basin currently.

Our predictive models can be useful in many ways. Forecasting deadly tropical cyclones is a challenging task \citep{vecchi2014seasonal,camp2015seasonal,vecchi2014next}, and our proposed methodology and results can provide valuable insights here. Our models may be used by the insurance and reinsurance industry \citep{dlugolecki2000climate}, as well as the broader community. 
A data sciences framework like the one proposed in this paper can serve as a paradigm for using observable physical, chemical, biological, or other observable characteristics of natural or man-made events for predicting quantifiable gains and losses resulting from the event. More generally, this paradigm can be useful for guiding the effects of different kinds of interventions, or adaptation and mitigation strategies related to the event. 

\section*{Acknowledgment} 
The content is solely the responsibility of the authors. The authors declare no conflict of interest.

\section*{Disclosure Statement}
No potential conflict of interest was reported by the author(s).
\section*{Funding}


\bibliography{Pressure_Wind_Speed,BayesEVDmulti,bayesian_model,sample}
 \appendix
 \section{Additional Analysis for Individual cyclones}
 \subsection{Trivariate Extreme Value Model}
 \label{Sec:Storms_TrivariateGEV}
Here, we model $[\mathbf {maxWS},  \mathbf {minCP}, \mathbf {Damage}]$ using a tri-variate extreme values distribution (GEV) model. 
Using the same notation as before ($Y_1$ for maxWS, $Y_2$ for logarithm of damages and $Z_1$ for minCP). We also add non-stationarity in the location parameters to fully utilize all information in the covariates $X \in \mathbb{R}^{N \times p}$. We consider the same hierarchy as in Section \ref{Sec:Storm_Model} for modeling the location parameters. That is, for $\alpha \in \mathbb{R}^p, \beta \in \mathbb{R}^{p+1}$ and $\gamma \in \mathbb{R}^{p+2}$:
\begin{align}
\mu_{z_1}(X) &=  X \alpha \label{eq: loc_minCP}\\
\mu_{y_1}(Z_1, X) &= [Z_1, X]\beta \label{eq: loc_maxWS}\\
\mu_{y_2}(Z_1, Y_1, X) &= [Z_1, Y_1, X] \gamma. \label{eq: damages}
\end{align}
The location and scale parameters have subscripts indicating the variable they represent. The joint distribution of a logistic dependence model \cite{tawn1990modelling} for three variables $(Z_1, Y_1, Y_2)$ is given by,
\begin{align*}
G(z_1, y_1, y_2) &= \exp(-(t(z_1)^{1/r} + t(y_1)^{1/r} + t(y_2)^{1/r})^r),
\end{align*}
where $r$ ($0 < r \leq 1$) is the dependence parameter. We can differentiate this w.r.t. $z_1, y_1, y_2$ to derive the joint density.
\begin{align*}
& g(z_1, y_1, y_2) \\
&= \dfrac{t(z_1)^{\xi_{z_1} + 1}}{\sigma_{z_1}} \dfrac{t(y_1)^{\xi_{y_1} + 1}}{\sigma_{y_1}} \dfrac{t(y_2)^{\xi_{y_2} + 1}}{\sigma_{y_2}} t(z_1)^{\frac{1}{r} - 1}  t(y_1)^{\frac{1}{r} - 1} t(y_2)^{\frac{1}{r} - 1} 
\\ & \hspace{0.5cm} \times 
\exp(-(t(z_1)^{1/r} + t(y_1)^{1/r} + t(y_2)^{1/r})^r) \\
&\quad \quad \times \Bigg[\dfrac{(1-r)(2-r)}{r} - \left(\dfrac{1-r}{r} \right) (t(z_1)^{1/r} + t(y_1)^{1/r} + t(y_2)^{1/r})^r \\
&\quad \quad \quad + (t(z_1)^{1/r} + t(y_1)^{1/r} + t(y_2)^{1/r})^{2r} \Bigg] \\
& \quad \quad \quad \quad \quad \quad \times (t(z_1)^{1/r} + t(y_1)^{1/r} + t(y_2)^{1/r})^{r-3}.
\end{align*}
Then the log-likelihood is given by,
\begin{align*}
& l(\theta| Z_1, Y_2, Y_1; X) \\
&= \sum_{i=1}^n \Big((\xi_{z_1} + 1) \log(t(z_{1i})) - \log(\sigma_{z_1}) + (\xi_{y_1} + 1) \log(t(y_{1i})) - \log(\sigma_{y_1})  \\
&\quad  + (\xi_{y_2} + 1) \log(t(y_{2i})) - \log(\sigma_{y_2})  + \left(\frac{1}{r} -1\right) \log(t(z_{1i})) \\
& \quad \quad + \left(\frac{1}{r} -1\right) \log(t(y_{1i})) + \left(\frac{1}{r} -1\right) \log(t(y_{2i}))\\
& \quad \quad \quad  - (t(z_{1i})^{1/r} + t(y_{1i})^{1/r} + t(y_{2i})^{1/r})^r\\
&\quad \quad \quad\quad  + \log\Bigg[\dfrac{(1-r)(2-r)}{r} \\
& \quad \quad \quad \quad \quad  - \left(\dfrac{1-r}{r} \right) (t(z_{1i})^{1/r} + t(y_{1i})^{1/r} + t(y_{2i})^{1/r})^r   \\
&\quad \quad \quad \quad \quad \quad + (t(z_{1i})^{1/r} + t(y_{1i})^{1/r} + t(y_{2i})^{1/r})^{2r} \Bigg]\\
&\quad \quad \quad \quad \quad \quad \quad  + (r-3)\log(t(z_{1i})^{1/r} + t(y_{1i})^{1/r} + t(y_{2i})^{1/r}) \Big)
\end{align*}
Our choice of priors are non-informative with variances chosen to ensure proper coverage of the sample space and reasonanle good acceptance rates ($\approx 20\%$). 
\begin{align*}
\alpha_0, \alpha_1, \beta_0, \beta_1, \beta_2 &\overset{iid}{\sim} \text{N}(0,10^2)\\
\gamma_0, \gamma_1, \gamma_2, \gamma_3 & \overset{iid}{\sim} \text{N}(0,10^3)\\
\sigma_{z_1}, \sigma_{y_1},  \sigma_{y_2} &\overset{iid}{\sim} \text{IG}(\alpha = 1, \beta = 1)\\
 \xi_{z_1} &\sim \text{Unif}(-1,1)\\
 \xi_{y_1}, \xi_{y_2} &\overset{iid}{\sim} \text{Unif}(-0.5, 0.5)\\
 r & \sim \text{Unif}(-0.05,1)
\end{align*}
The standard regularity conditions for the likelihood of an GEV are satisfied for $\xi < 0.5$. In particular for the range of values $-0.5 < \xi < 0.5$ are most often encountered in practice \citep{hosking1984testing}. Hence, we model the shape parameters by a uniform prior between -0.5 and 0.5. Since, in our parameterization of the dependence parameter, $r$, $0 < r \leq 1$, we model it using a Uniform distribution but with a lower bound of 0.05 to avoid computational overflow. The Metropolis Hastings algorithm is run for $10^6$  MCMC steps. We present the posterior means and standard deviations for the selected parameters after fitting the Bayesian model for the entire dataset from 1960-2022 in Table \ref{tab: selectedcovs_estimatestrivar} and the estimates are very close to the ones in the Hierarchical Bayesian Extreme value model in Table \ref{tab: selectedcovs_estimatesGEV}.  Interestingly, the estimate  for dependence parameter $r$ is 0.9972 which is very close to 1 signifying independence between the three variables considered. We suspect that this is because the non-stationarity considered in location parameters already accounts for the dependence between the three variables. The results seem pretty stable on change on starting values and modifications to the prior specifications. 

\begin{table}[ht]
\centering
\begin{tabular}{rrr}
  \hline
 & Posterior means & Posterior standard deviation \\ 
  \hline
  \textcolor{blue}{Min CP} & \\
  \hline
Intercept & 3.2193 & 0.0751 \\ 
  Avg. Latitude & -0.1534 & 0.0240  \\ 
  Avg. Longitude & -0.1284 & 0.0364 \\ 
  Start month & 0.1424 & 0.0536 \\ 
  Year & -0.0569 & 0.0372 \\ 
  AMO & -0.1382 & 0.0501 \\ 
  Atlantic SST & 0.2254 & 0.0357  \\
  Sunspots & -0.0426 & 0.0392\\
  $\xi$  & -0.9336 & 0.0460 \\
  $\sigma$ & 1.2905 & 0.0618 \\  
  \hline
  \textcolor{blue}{Max WS} & \\
  \hline
  Intercept & 3.8126 & 0.0819 \\ 
  Min CP (scaled) & 0.3621 & 0.0760 \\ 
  $\xi$ & -0.5414 & 0.0109 \\ 
  $\sigma$  & 1.0106 & 0.0471 \\
  \hline
  \textcolor{blue}{Damages} & \\
  \hline
  Intercept & 19.7481 & 0.1593 \\ 
  Max WS (scaled) & 0.5890 & 0.2009 \\ 
  Min CP (scaled) & 0.7217 & 0.2027 \\ 
  Avg. Latitude & -0.3533 & 0.1524\\ 
  Year & 0.2892 & 0.1776 \\ 
  SOI & 0.1660 & 0.1955 \\ 
  ANOM.3.4 & 0.4119 & 0.1875 \\  
  Atlantic SST & -0.4051 & 0.1831 \\ 
  Sunspots & -0.1670 & 0.1671 \\ 
  $\xi$ & -0.2464 & 0.0335 \\  
  $\sigma$ &  1.8364 & 0.0713 \\ 
   \hline
   $r$ & 0.9972 & 0.0043 \\ 
   \hline
\end{tabular}
\caption{Posterior mean and standard deviation estimates for models \eqref{mod1_main},\eqref{mod2_main}, \eqref{mod3_main} for the trivariate Bayesian Generalized Extreme Value (GEV) model with selected covariates.}
\label{tab: selectedcovs_estimatestrivar}
\end{table}

\subsection{Hierarchical Bayesian Extreme Value model with $\log$-Normal Damages}
\label{Sec:Storms_HierarchicalLogNormal}

Since it is possible that storm  damages follow a heavy-tailed but not necessarily extreme valued distribution, we model it has a log-normal distribution in this section. We consider the following model: \begin{align}
Z_1|X &\sim \text{GEV}(\mu_{z_1}(X), \sigma_{z_1}, \xi_{z_1}) \label{mod1}\\
Y_1|(Z_1, X) &\sim \text{GEV}(\mu_{y_1}(Z_1,X), \sigma_{y_1}, \xi_{y_1}) \label{mod2}\\
Y_2|(Y_1, Z_1, X) &\sim \text{Lognormal}(\mu_{y_2}(Y_1, Z_1, X), \sigma_{y_2}) \label{mod3},
\end{align}
where the hierarchy essentially comes in the location parameters as Sections~\ref{Sec:Storm_Model} and \ref{Sec:Storms_TrivariateGEV}.
The joint density can then be written as,
\begin{align*}
& f(Z_1, Y_1, Y_2| X, \theta) \\
&=  f(Y_2|Y_1, Z_1, X) f(Y_1|Z_1,X) f(Z_1|X)\\
&= \dfrac{1}{y_2 \sigma_{y_2} \sqrt{2\pi}} \exp\left(- \dfrac{(\log(y_2) - \mu_{y_2}(y_1, z_1, x))^2}{2\sigma_{y_2}^2} \right) \\
& \quad \quad \times \dfrac{1}{\sigma_{y_1}} (t(y_1))^{\xi_{y_1} + 1} \exp(-t(y_1))\\
&\quad \quad \quad \times \dfrac{1}{\sigma_{z_1}} (t(z_1))^{\xi_{z_1} + 1} \exp(-t(Z_1)),
\end{align*}
where,
\begin{align*}
t(x) &= 
\begin{cases}
(1 + \xi (\frac{x - \mu}{\sigma}))^{-1/\xi} & \xi \neq 0\\
\exp\left(- \frac{x - \mu}{\sigma} \right) & \xi = 0.
\end{cases}
\end{align*}
Then the log-likelihood can be given by,
\begin{align*}
l(\theta|Y_1, Y_2, Z_1, X) &= \sum_{i=1}^n \Big[-\log(y_{2i} \sigma_{y_2} \sqrt{2\pi}) - \dfrac{(\log(y_{2i}) - \mu_{y_2})^2}{2\sigma_{y_2}^2}\\ & \hspace{0.5cm}
 - \log(\sigma_{y_{1i}}) + 
(\xi_{y_{1i}} + 1) \log(t(y_{1i})) - t(y_{1i}) \\
&\quad \quad - \log(\sigma_{z_{1i}}) + (\xi_{z_{1i}} + 1) \log(t(z_{1i})) - t(z_{1i})\Big].
\end{align*}

\textbf{Priors: } 
The location parameter priors for both the GEV models (minCP and maxWS) and the log-normal model (damages) are non-informative with variances chosen to be drawn for Inverse-Gamma(1,1) to ensure proper coverage of the sample space leading to good acceptance rates. 
\begin{align*}
\alpha_0, \alpha_1 &\overset{iid}{\sim} \text{N}(0,10^3)\\
\beta_0, \beta_1, \beta_2 &\overset{iid}{\sim} \text{N}(0,10^2)\\
\gamma_0, \gamma_1, \gamma_2, \gamma_3 & \overset{iid}{\sim} \text{N}(0,10^3)\\
\sigma_{z_1}, \sigma_{y_1}, \sigma_{y_2} &\overset{iid}{\sim} \text{IG}(\alpha = 1, \beta = 1)\\
 \xi_{z_1} &\sim \text{Unif}(-1,1)\\
 \xi_{y_1} &\overset{iid}{\sim} \text{Unif}(-0.5, 0.5).
\end{align*}

We use Metropolis-Hastings algorithm to sample from the posterior distribution.
The chain is run for $N= 10^6$ MCMC sample size and the step-sizes are chosen to achieve about 20\% acceptance rate. 
The results from this model fitting is given in Table~\ref{tab: selectedcovs_estimates_hierlogNormalD}.
The results of this are very close to those in Table~\ref{tab: selectedcovs_estimatestrivar} and Table~\ref{tab: selectedcovs_estimatesGEV}, where we reported the results using the trivariate Bayesian GEV model and the hierarchical Bayesian GEV model respectively. Thus, the conclusions from the data are quite robust to the data science framework used for analysis.

Note that the frequentist estimates for the location parameters are somewhat similar to the posterior means for the location parameters in Table \ref{tab: selectedcovs_estimates_hierlogNormalD}. In terms of significance of estimates, coefficient of average latitude, average longitude, starting month, AMO, Atlantic SST are significant in modeling location parameter for minCP. Similarly, the effect of minCP ($\beta_1$) in modeling location parameter for maxWS seems to be significant. For damages, we notice significant effect corresponding to minCP, average latitude, SOI, ANOM 3.4, Atlantic SST on the location parameter of damages, with a highly significant intercept term. The scale parameter estimates are significant across models, and the first two models in the hierarchy have a negative estimate for the shape parameters signifying Reverse Wiebull distributions for the marginals of each of log(minCP) and log(maxWS) respectively.

\begin{table}[ht]
\centering
{\scriptsize
\begin{tabular}{rrrrr}
  \hline
 & Posterior means & Posterior standard deviation  & Frequentist est & Freq se\\ 
  \hline
  \textcolor{blue}{Min CP} & \\
  \hline
Intercept & 3.2099 & 0.0717 & 3.5613 & 0.0525 \\ 
  Avg. Latitude & -0.1522 & 0.0252 & -0.2110 & 0.0362  \\ 
  Avg. Longitude & -0.1132 & 0.0398 & -0.2573 & 0.0553 \\ 
  Start month & 0.1589 & 0.0508 & 0.1253 & 0.0466\\ 
  Year & -0.0235 & 0.0241 & -0.0512 & 0.0316\\ 
  AMO & -0.1527 & 0.0508 & -0.0413 & 0.0716\\ 
  Atlantic SST & 0.2245 & 0.0399  & 0.1724 & 0.0554\\
  $\xi$  & -0.9220 & 0.0473 & 0.3449 & 0.2068 \\
  $\sigma$ &  1.2900 & 0.0630 & 0.6279 & 0.2230  \\  
  \hline
  \textcolor{blue}{Max WS} & \\
  \hline
  Intercept & 3.8045 & 0.0847 & 0.6076 & 0.0417\\ 
  Min CP (scaled) & 0.3741 & 0.0755 & -0.5210 & 0.0660 \\ 
  $\xi$ &  -0.5416 & 0.0104 & -0.4362 & 0.1964 \\ 
  $\sigma$  & 1.0156 & 0.0464 & -0.2865 & 0.1864\\
  \hline
  \textcolor{blue}{Damages} & \\
  \hline
  Intercept & 20.4610 & 0.1450 & 4.3972 & 0.0117\\ 
  Max WS (scaled) & 0.4091 & 0.2807 & 0.3628 & 0.0128  \\ 
  Min CP (scaled) & 1.0026 & 0.2798  & 0.1402 & 0.0076\\ 
  Avg. Latitude &  -0.3760 & 0.1446 &-0.2978 & 0.0316\\ 
  Year & 0.2191 & 0.1736 & 20.5228 & 0.1608 \\ 
  SOI & 0.3407 & 0.1629 & 0.3315 & 0.5400\\ 
  ANOM.3.4 & 0.6013 & 0.1753  & 1.1006 & 0.5230\\  
  Atlantic SST &  -0.4038 & 0.1636 & -0.4025 & 0.1832 \\ 
  Sunspots & -0.2805 & 0.1566 & 0.2408 & 0.2033\\  
  $\sigma$ & 1.7822 & 0.0728 & 2.0535 & \\ 
   \hline
\end{tabular}
\scriptsize}
\caption{Posterior mean and standard deviation estimates for models \eqref{mod1_main},\eqref{mod2_main}, \eqref{mod3_main} for the Hierarchical Bayesian Generalized Extreme Value (GEV) model with $log$-Normal damages and selected covariates.}
\label{tab: selectedcovs_estimates_hierlogNormalD}
\end{table}

\subsection{Individual Storms: Additional Predictions}
\label{Storms_Prediction_Appendix}

Similar to the analysis presented in Section~\ref{Sec:Storm_Prediction}, we now 
present the prediction results for the Atlantic propical storms of 2017, 2020, and 2022 using the statistical models described in Section~\ref{Sec:Storms_TrivariateGEV} (trivariate Bayesian GEV) and in Section~\ref{Sec:Storms_HierarchicalLogNormal} (hierarchical Bayesian model with log-Normal distribution for damages).

We consider the predictions from these models for the years 2017, 2020 and 2022.  The details are same as those given in Section~\ref{Sec:Storm_Prediction}.
We note from Table~\ref{tab:credible_int_count} that all the true minimum central pressure values were always within the 95\% credible interval for all five storms. Even the true maximum wind speed fell within the intervals for all the three models, except it missed the mark very closely for hurricane Nate in hierarchical Bayesian model with log-Normal for damages. More specifically, the 95\% credible interval for Nate was $(3.1304, 4.3807)$ and the true value was $4.3820$, on the log-scale. For damages, none of the credible intervals for the three models (hierarchical GEV in Section~\ref{Sec:Storm_Prediction}, trivariate GEV in Section~\ref{Sec:Storms_TrivariateGEV} and hierarchical model with log-Normal loss in Section~\ref{Sec:Storms_HierarchicalLogNormal}) could capture the observed value for Harvey in the 95\% interval. However the damages due to Harvey is within the range of the posterior for the  hierarchical GEV method of  Section~\ref{Sec:Storm_Prediction}.

Similarly, the true damage value for Irma only fell in the 95\% credible interval when the hierarchical GEV  model of in Section~\ref{Sec:Storm_Prediction} was fit, but was missed in the other two methods. However, the truth was not too far away in the tails of the predictive distribution. Given that Harvey and Irma both are in the top-five most damage causing Atlantic storms, these results are not surprising. 
We present the $\delta$-values using the models described in Section~\ref{Sec:Storms_TrivariateGEV}
and in Section~\ref{Sec:Storms_HierarchicalLogNormal}
in Tables \ref{tab:delta_value_triGEV} and \ref{tab:delta_value_logNormal}
respectively. These values are not substantially different from those of 
Table~\ref{tab:delta_value_hierGEV}.

\begin{table}[h!]
\centering
\begin{tabular}{|c c c c|}
\hline
& minCP & maxWS & damages\\
\hline
\hline
Hierarchical GEV & 100\% (13/13) & 100\% (13/13) & 85\% (11/13)\\
Trivariate GEV & 100\% (13/13)  & 100 \% (13/13) & 85\% (11/13)\\
Hierarchical log-Normal damages & 100\% (13/13) & 100\% (13/13) & 85\% (11/13)\\
\hline
\end{tabular}
\caption{The proportion of the 13 selected storms with three from 2017, 8 randomly selected one from 2020 and two from 2022 where the truth was contained in the 95\% credible interval.}
\label{tab:credible_int_count}
\end{table}

\begin{table}[h!]
\centering
\begin{tabular}{|c c c c|}
\hline
Hurricanes & minCP & maxWS & Damage\\
\hline
Harvey (2017) &  0.1434     &   0.50638  &  0.02896  \\
Irma (2017) &    0.64704   &   0.8257     & 0.1295 \\
Nate (2017) &    0.76008   & 0.68278    &     0.83464  \\
Delta (2020) &  0.95046  & 0.62686 & 0.12042\\
Eta (2020) & 0.72122 &  0.7524 & 0.32816\\
Sally (2020) & 0.93118 & 0.72956 & 0.54966\\
Ian (2022) & 0.5781  &  0.59854  &   0.0173 \\
Nicole (2022) & 0.67764   &  0.88466 &   0.69672 \\
\hline
\end{tabular}
\caption{For the trivariate GEV model in Section~\ref{Sec:Storms_TrivariateGEV}, $\delta$ values for each of the hurricanes, closer to 1 reflects the truth to be close to the median of the posterior predictive distribution and closer to 0 reflects the truth lying in the tails of the distribution.}
\label{tab:delta_value_triGEV}
\end{table}

\begin{table}[h!]
\centering
\begin{tabular}{|c c c c|}
\hline
Hurricanes & minCP & maxWS & Damage\\
\hline
Harvey (2017) &  0.49068     &   0.76494  &  0.00204   \\
Irma (2017) &    0.56628   &  0.63282    & 0.84256 \\
Nate (2017) &    0.77438   &  0.68424    &   0.96714  \\
Delta (2020) & 0.93806  & 0.62622 & 0.33648\\
Eta (2020) & 0.72088 & 0.7449 & 0.35948\\
Sally (2020) & 0.89708 & 0.72556 & 0.06786\\
Ian (2022) & 0.57756   & 0.6035 &   0.00016  \\
Nicole (2022) &  0.67716   &   0.88634  &    0.4203  \\
\hline
\end{tabular}
\caption{For the hierarchical GEV model with log-Normal damages in Section~\ref{Sec:Storms_HierarchicalLogNormal}, $\delta$ values for each of the hurricanes, closer to 1 reflects the truth to be close to the median of the posterior predictive distribution and closer to 0 reflects the truth lying in the tails of the distribution.}
\label{tab:delta_value_logNormal}
\end{table}

\section{Additional details for Seasonal Analysis of Tropical Cyclones}
\subsection{Additional Analysis: Empirical Bayesian Modeling Of Seasonal Data}
\label{Sec:Seasonal_EB}

To corroborate the full hierarchical Bayesian model presented in Section~\ref{Sec:Seasonal_Model}, we carried out multiple studies. One is an empirical Bayesian study, where we used the maximum likelihood estimators to inform the prior. The technical details are given below. 
As can be seen from Tables \ref{LowPosteriorResults_new} and \ref{HighPosteriorResults_new}, the results from empirical Bayes are very close to those obtained using the hierarchical Bayesian model.

{\bf Low Intensity Storm Bayesian Specification (EB): }
\begin{align}
[N_{1,i}|X_i,r,p_i,\boldsymbol{\beta}_1] &\sim \text{NegBinom}(r, p_i) \label{lowintensityfreq_new_EB}\\
p_i &= \frac{r}{r + \lambda(x_i)} \nonumber \\
\log(\lambda(x_i)) &=x_i\boldsymbol{\beta}_1; \boldsymbol{\beta}_1 \in \mathbb{R}^{q_1} \nonumber \\
[L_{1,i}|N_{1,i},X_i,\phi,\boldsymbol{\beta}_1,\theta_1] &\sim \text{Binomial}(n_{1,i},\theta_1)\label{lowintensityland_new_EB} \\
[D_{1,i}|L_{1,i},N_{1,i},X_i,\phi,\boldsymbol{\beta}_1,\theta_1,\mu_{1},\sigma_{1}]&\sim \nonumber \\
(1-(1-\theta_1)^{n_{i1}}) * \text{Lognormal}&(\mu_{1},\sigma_{1}) + (1-\theta_1)^{n_{i1}} * 0 \label{lowintensitydam_new_EB} \\
[{\boldsymbol\beta_{1}}] & \sim \mathcal{N}( ({\hat \beta}_{11}, \hdots,{\hat \beta}_{1q_1} ), 10^{4} I_{q_1})\label{lowintensitynorm1_new_EB}\\
[\log(\phi)]&\sim \mathcal{N}(\log(\hat\phi),1)\label{lowintensitynorm2_new_EB}\\
[\theta_{1}]&\sim \text{Beta}(\alpha(\hat{p}_1),\beta(\hat{p}_1))\label{lowintensitybeta_new_EB}\\
[\mu_1]&\sim \mathcal{N}(\bar{x}_1,10^{4})\label{lowintensitynorm3_new_EB}\\
[r] &\sim \text{Unif}(0,70)\\
\left[\frac{1}{\sigma_1^2}\right] &\sim \text{Gamma}(v_{11}(\hat{s}_1),v_{12}(\hat{s}_1))\label{lowintensitygamma_new_EB}
\end{align}

{\bf High Intensity Storm Bayesian Specification (EB): }
\begin{align}
[N_{2,i}|X_i,\boldsymbol{\beta}_2] &\sim \text{Poisson}(\lambda(x_i)) \label{highintensityfreq_new_EB}\\
&\log(\lambda(x_i))=x_i\boldsymbol{\beta}_2; \boldsymbol{\beta}_2 \in \mathbb{R}^{q_2} \nonumber \\
[L_{2,i}|N_{2,i},X_i,\boldsymbol{\beta}_2,\theta_2]& \sim \text{Binomial}(n_{2,i},\theta_2) \label{highintensityland_new_EB}\\
[D_{2,i}|L_{2,i},N_{i2},X_i,\boldsymbol{\beta}_2,\theta_2,\mu_{2},\sigma_{2}]&\sim \nonumber \\
(1-(1-\theta_2)^{n_{i2}}) * &\text{Lognormal}(\mu_{2},\sigma_{2}) + (1-\theta_2)^{n_{i2}} * 0 \label{highintensitydam_new_EB}\\
[{\boldsymbol\beta}_{2}] & \sim \mathcal{N}_2([{\hat \beta}_{21}, \hdots,{\hat \beta}_{2q_2}]^T , 10^4 I_{q_2}) \label{highintensitynorm1_new_EB}\\
[\theta_{2}]&\sim \text{Beta}(\alpha(\hat{p}_2),\beta(\hat{p}_2))\label{highintensitybeta_new_EB}\\
[\mu_2]&\sim \mathcal{N}(\bar{x}_2,10^{4}) \label{highintensitynorm2_new_EB}\\
\left[\frac{1}{\sigma_2^2}\right] &\sim \text{Gamma}(v_{21}(\hat{s}_2),v_{22}(\hat{s}_2))\label{highintensitygamma_new_EB}
\end{align}

\subsection{Additional Analysis: Data Clone Computations On Seasonal Data}
\label{Sec:Seasonal_DClone}

In addition to the hierarchical Bayes and empirical Bayes computations, we also used \verb#dclone# \citep{lele2007data} method to analyze the data, and provide further support and justification about the robustness of our findings. 

Data cloning was implemented using 1, 2, and 5 clones.  We report the estimates of the the 5 clone chain.  Each run consisted of running 3 chains of length 100,000 with adaptation on the first 100 iterations.   The Markov Chain Monte Carlo (MCMC) procedure with Gibbs sampling was also implemented using 3 chains, each with 100,000 iterations.  The first 100 iterations of each chain were discarded.

We report some comparision results across different statistical methodologies in Tables \ref{LowPosteriorResults_new} and \ref{HighPosteriorResults_new}. Here, we either used all the covariates, or used a select few depending on a  statistical model selection criterion  (in these cases, both Akaike and Bayesian model selection suggested the same model, where a few covariates are dropped). 
We compare the empirical Bayesian approach with the full hierarchical Bayesian approach for both these choices. Model fitting was done either by MCMC, or by adapting the non-Bayesian data cloning approach. The tables report the posterior mean and standard deviation for the different parameter values, for these various combinations. Both tables demonstrate that there is little difference between the results from different techniques, and that we have excellent robustness against a choice of statistical methodology. Additional robustness studies were also conducted. We have used the hierarchical bayesian approach with selected covariates as the main approach for the results reported in the paper.

\begin{table}[h!]
\caption{{Comparison of different statistical methodologies for analyzing  the low intensity tropical cyclones data. Here, \textit{all} refers to the model where all the covariates were used, while \textit{selected} refers to where a few selected covariates were used. These were selected using a model selection criterion on a model with all covariates. The abbreviation  EB refers empirical Bayes approach, while HB refers to hierarchical Bayesian approach that we use in the paper. The MCMC and dclone are two different computational approaches, the latter is non-Bayesian, but may be used in conjunction with empirical or hierarchical Bayesian techniques as well. Each entry is an expected value, with standard deviation in brackets. Results show excellent robustness across statistical methodologies.}}
\centering
\label{LowPosteriorResults_new}
{\scriptsize

\begin{tabular}{| l | l | l l | l l |}
\hline
&  & \multicolumn{2}{c}{All} & \multicolumn{2}{|c|}{Selected} \\
\hline
Parameter & Method      &      EB  & HB  & EB   &   HB   \\
\hline
{$\beta_{11}$ }& \tt MCMC &   -0.068 (0.059)  &  -0.068 (0.059) &  -0.060 (0.059) & -0.060 (0.059) \\
 (NAO)  & dclone  &  -0.068 (0.026) & -0.068 (0.026) & -0.060 (0.026) & -0.060 (0.026)\\
\hline
{$\beta_{12}$ }& \tt MCMC &  -0.016 (0.031)  &-0.016 (0.031) & -& - \\
 (SOI)  & dclone  &   -0.016 (0.014) &-0.016 (0.014) &- &- \\
\hline
{$\beta_{13}$ }& \tt MCMC &   -0.429 (0.211)  & -0.429 (0.211) & -0.456 (0.208) & -0.456 (0.208) \\
 (AMO)  & dclone  &   -0.429(0.092) & -0.429 (0.092)  & -0.456 (0.092) & -0.456 (0.092)\\
\hline
{$\beta_{14}$ }& \tt MCMC &  -0.253 (0.103)  & -0.254 (0.102) & -0.207 (0.075)& -0.207 (0.075)  \\
 Nino-3.4  & dclone  &   -0.253 (0.045) & -0.253 (0.045) & -0.207 (0.033) &-0.207 (0.033)\\
\hline
{$\beta_{15}$ }& \tt MCMC &   0.098 (0.003)  & 0.098 (0.003) & 0.100 (0.002)  & 0.100 (0.002)\\
 (SST)  & dclone &   0.098 (0.001) & 0.098 (0.001) &0.100 (0.001)& 0.100 (0.001)\\
\hline
{$\beta_{16}$ }& \tt MCMC &   0.001 (0.001)  & 0.001 (0.001) & -& -  \\
(Sunspots)   & dclone  &    0.001 (0.000) &  0.001 (0.000) & -&- \\
\hline
\multirow{2}{*}{$r$} & \tt MCMC    &   36.087 (15.590) & 36.065 (15.540) & 36.730 (15.474) & 36.807 (15.512)  \\
 & dclone &37.118(11.046) & 37.133 (11.064)  & 34.932 (10.447) & 34.934 (10.439) \\
\hline
\multirow{2}{*}{$\theta_1$}& \tt MCMC   &   0.142 (0.011) & 0.142 (0.011) & 0.142 (0.011)&0.142 (0.011)\\
 & dclone        & 0.142 (0.005) &0.142 (0.005) & 0.142 (0.005) & 0.142 (0.005) \\
\hline
\multirow{2}{*}{$\mu_1$}& \tt MCMC  &  18.297 (0.344) & 18.296 (0.337) & 18.295 (0.344)& 18.296 (0.337)   \\
 & dclone     &18.296 (0.149) & 18.295 (0.149) &18.296 (0.149)  &18.296 (0.149)\\
\hline
\multirow{2}{*}{$\sigma_1^2$}& \tt MCMC &  5.074 (1.155) &4.882 (1.091) & 5.077(1.161) & 4.884 (1.093)  \\
 & dclone &4.790 (0.467) & 4.756 (0.461)& 4.789 (0.466) & 4.756 (0.463)  \\
\hline
\end{tabular}

\scriptsize}
\end{table}

\begin{table}[h!]
\caption{{Comparison of different statistical methodologies for analyzing  the high intensity tropical cyclones data. Here, \textit{all} refers to the model where all the covariates were used, while \textit{selected} refers to where a few selected covariates were used. These were selected using a model selection criterion on a model with all covariates. The abbreviation  EB refers empirical Bayes approach, while HB refers to hierarchical Bayesian approach that we use in the paper. The MCMC and dclone are two different computational approaches, the latter is non-Bayesian, but may be used in conjunction with empirical or hierarchical Bayesian techniques as well. Each entry is an expected value, with standard deviation in brackets. Results show excellent robustness across statistical methodologies.}}
\centering
\label{HighPosteriorResults_new}

{\scriptsize

\begin{tabular}{| l | l | l l | l l |}
\hline
&  & \multicolumn{2}{c}{All} & \multicolumn{2}{|c|}{Selected} \\
\hline
Parameter & Method      &      EB  & HB  & EB   &   HB   \\
\hline
{$\beta_{21}$ }& \tt MCMC &  0.140 (0.118) & 0.140(0.118) & - & -\\
(NAO) & dclone & 0.137(0.053)  & 0.137 (0.053) & - & -  \\
\hline
{$\beta_{22}$ }& \tt MCMC &  0.025 (0.061) & 0.025 (0.061) & - & -\\
(SOI) & dclone & 0.024 (0.027)  &0.024 (0.027) & - &-   \\
\hline
{$\beta_{23}$ }& \tt MCMC &  1.908 (0.434) & 1.907 (0.435) & 1.738 (0.385) & 1.740 (0.386)\\
(AMO) & dclone &1.893 (0.193)   & 1.893 (0.194) & 1.733 (0.172) &  1.732 (0.173) \\
\hline
{$\beta_{24}$ }& \tt MCMC &  -0.223 (0.191) & -0.224 (0.191) & -0.321 (0.147) &  -0.321 (0.147)\\
Nino-3.4 & dclone & -0.224 (0.085)  &-0.224 (0.085)  &-0.321 (0.066) & -0.321 (0.066)\\
\hline
{$\beta_{25}$ }& \tt MCMC &  0.033 (0.006) &0.033 (0.006) & 0.031 (0.004) & 0.031 (0.004)\\
(SST) & dclone & 0.034 (0.003)   &0.034 (0.003)  & 0.031 (0.002) & 0.031 (0.002) \\
\hline
{$\beta_{26}$ }& \tt MCMC &  -0.001(0.001) & -0.001(0.001) & - & -\\
(Sunspots) & dclone & -0.001(0.001)   & -0.001(0.001) &- & - \\
\hline
\multirow{2}{*}{$\theta_2$} & \tt MCMC   &  0.390(0.040) & 0.391(0.040) & 0.390 (0.040) & 0.391(0.040)\\
& dclone        & 0.389 (0.018) & 0.390 (0.018)  & 0.389 (0.018) & 0.390 (0.018)\\
\hline
\multirow{2}{*}{$\mu_2$}& \tt MCMC  &  21.121 (0.447) & 21.122(0.436) & 21.120 (0.448)  & 21.121(0.435)\\
& dclone       &21.121 (0.193) & 21.121 (0.192) & 21.121(0.193)& 21.121 (0.192)\\
\hline
\multirow{2}{*}{$\sigma_2^2$}& \tt MCMC & 6.997 (1.794) & 6.644(1.662) & 6.993(1.798) & 6.649(1.668)\\
& dclone   & 6.508 (0.705) & 6.445(0.695) &  6.510(0.705) & 6.447 (0.695)\\
\hline
\end{tabular}

\scriptsize}
\end{table}
\section{Prediction for other years}
\label{sec: other_years_seasonal}
We present a detailed discussion, and additional figures, for the Bayesian predictive analysis. We discuss the results in detail for 2016 and 2019 below, to illustrate how the probability density or mass functions depicted in these figures reflect the data. In each analysis, we use the hierarchical Bayesian model, described in Section~\ref{Sec:Storm_Model}, the empirical Bayesian prediction
and the fast computation Bayesian prediction, both  described in Section~\ref{Sec:Seasonal_PosteriorPredictiveDistribution}. 

 \subsection{{2016 and 2019 Posterior Predictive Bayesian Seasonal Analysis}}
\begin{figure}[ht]
\centering
    \caption[Posterior Predictive Distributions for 2016 Low Intensity Storms]{Posterior predictive distributions for 2016 low intensity storms. 
   The top row is for empirical Bayes prediction, the middle row is for fast Bayesian prediction and the bottom row is for hierarchical Bayesian  prediction. 
   The left column is the probability mass function for cyclone frequency, middle column is for damage-inflicting probability, and right column is density for logarithm of damages.
    The actual values  are displayed with red dashed lines.}
        \label{pp12016}
\begin{tabular}{ccc}        
{\includegraphics[width=0.32\textwidth]{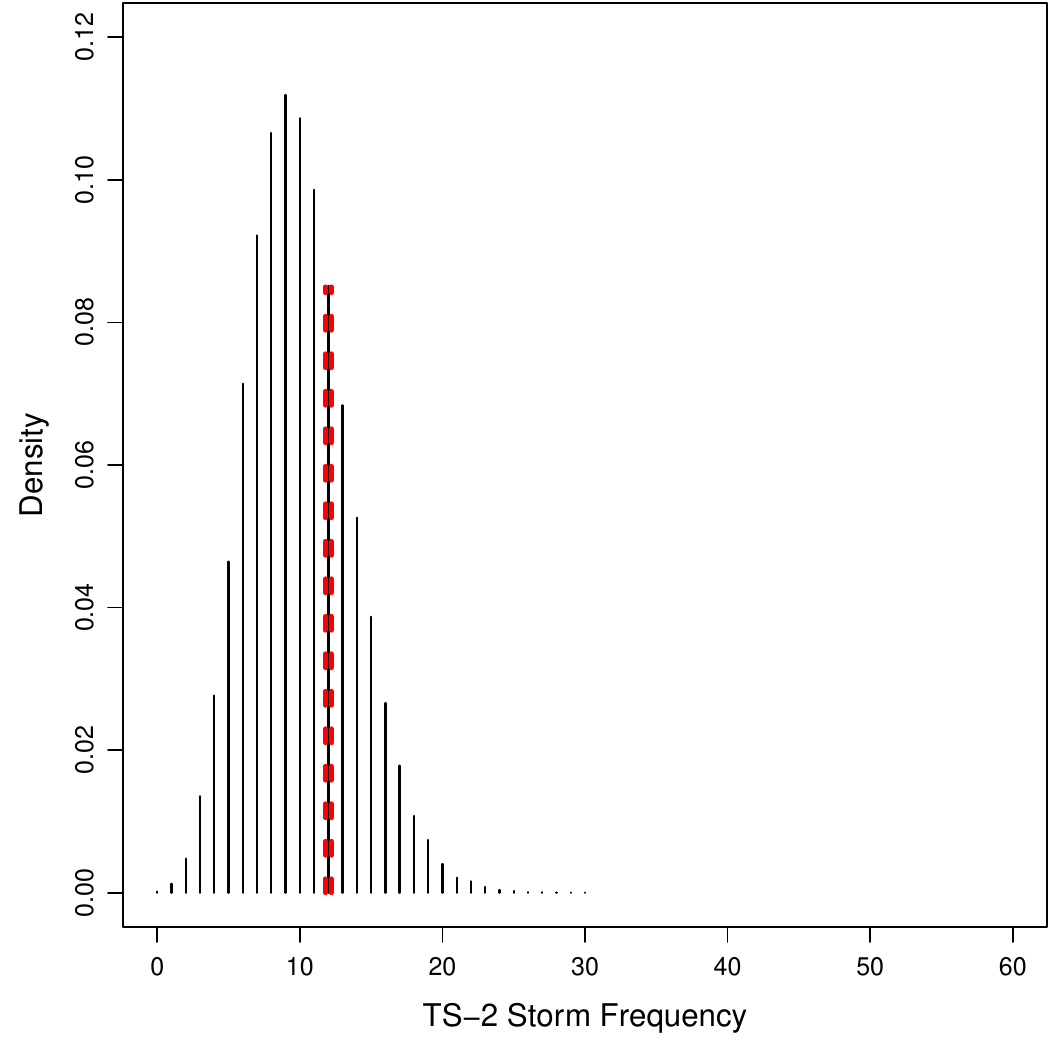}}
&
{\includegraphics[width=0.32\textwidth]{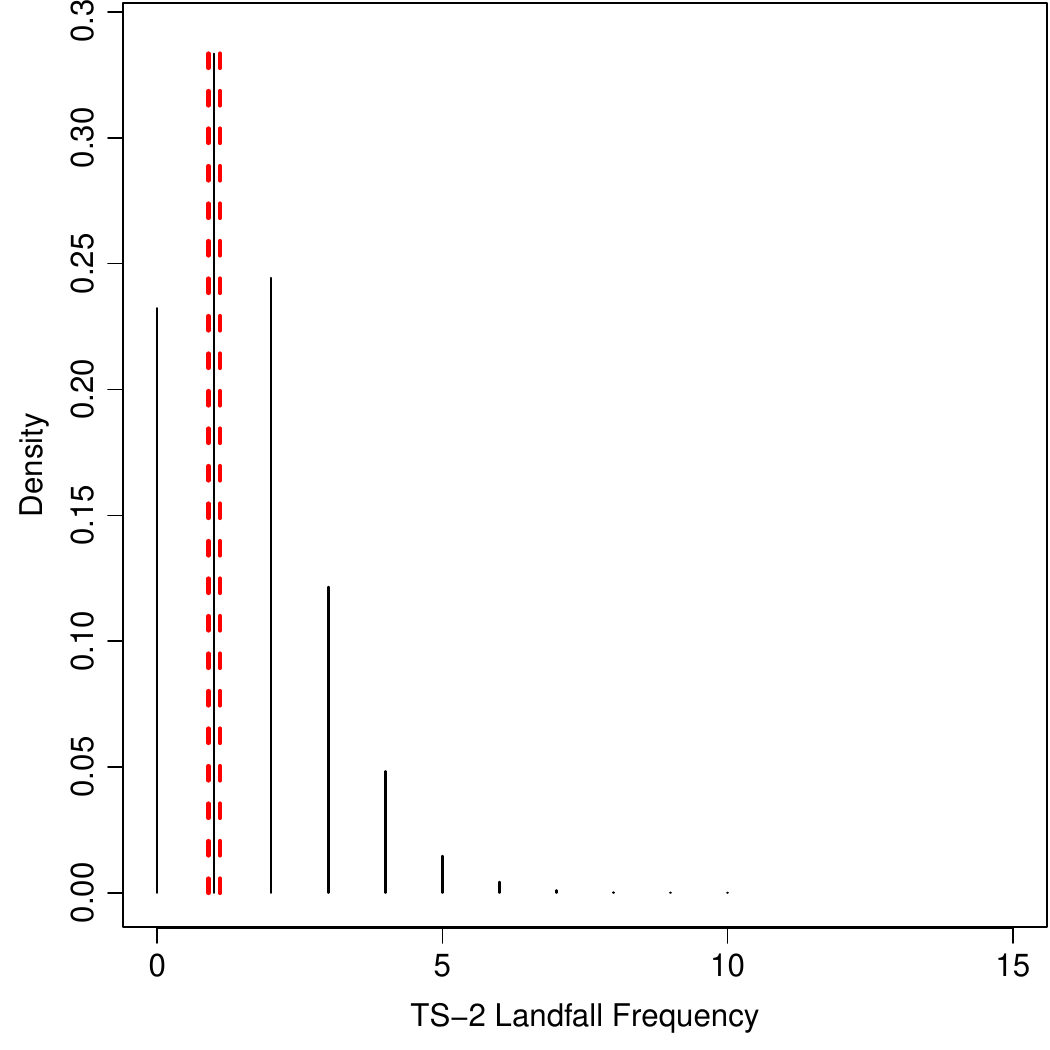}}
&
{\includegraphics[width=0.32\textwidth]{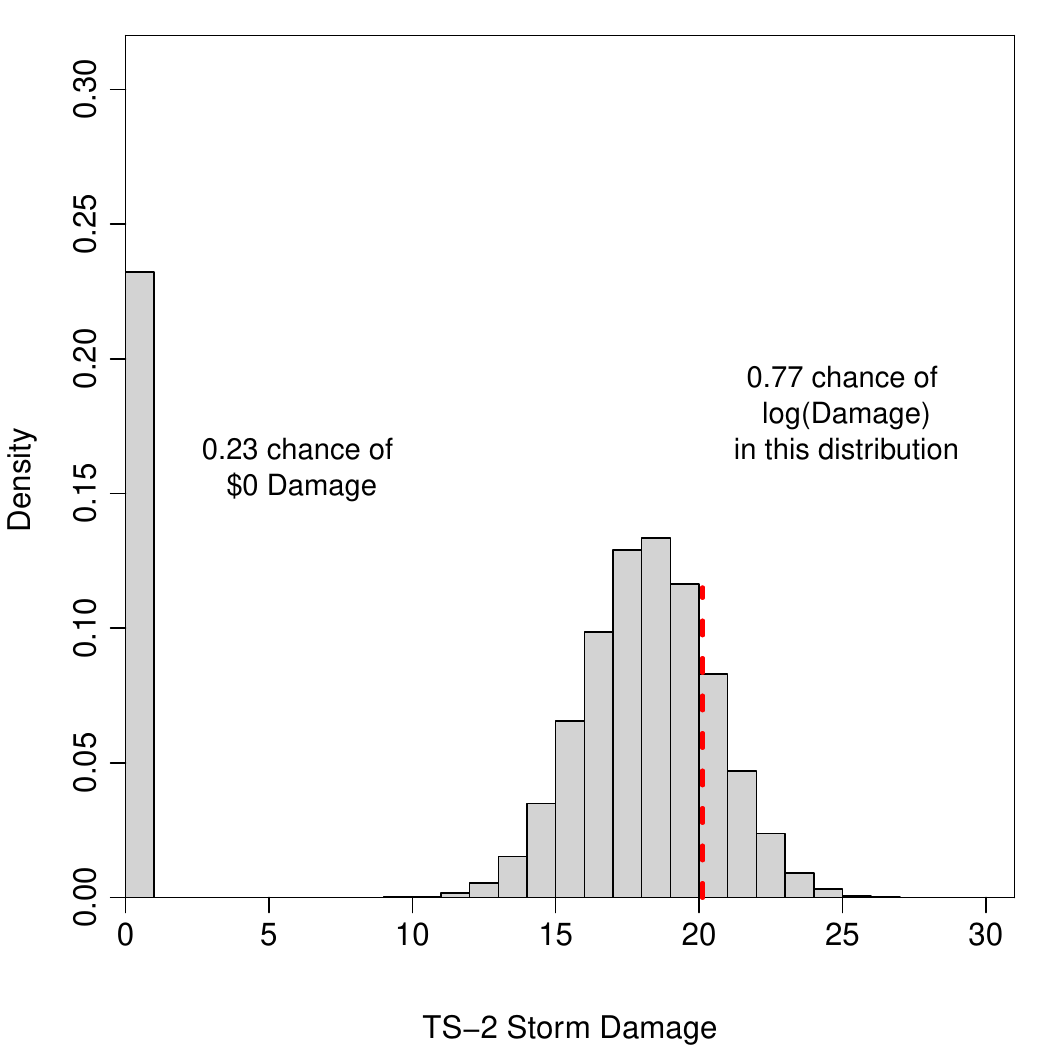}}\\
{\includegraphics[width=0.32\textwidth]{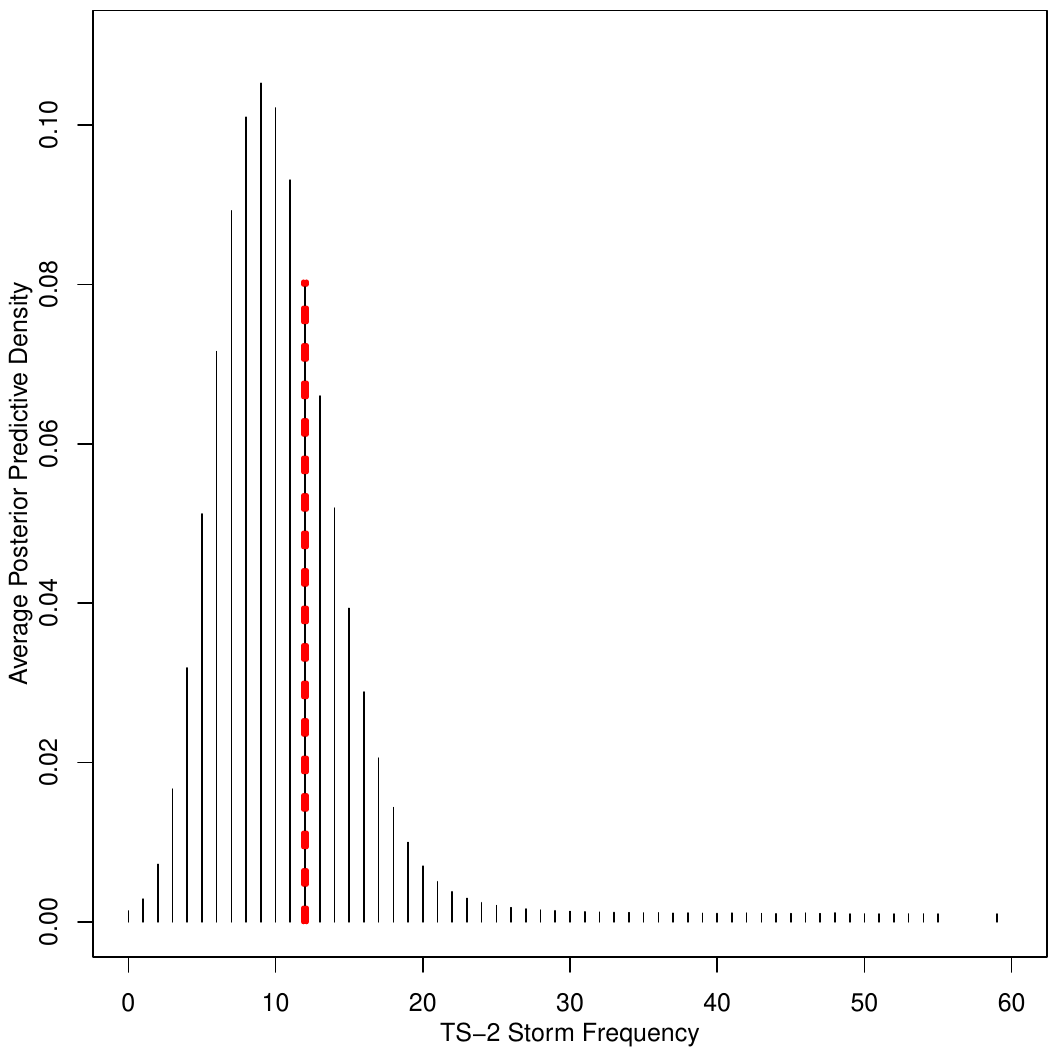}}
&
{\includegraphics[width=0.32\textwidth]{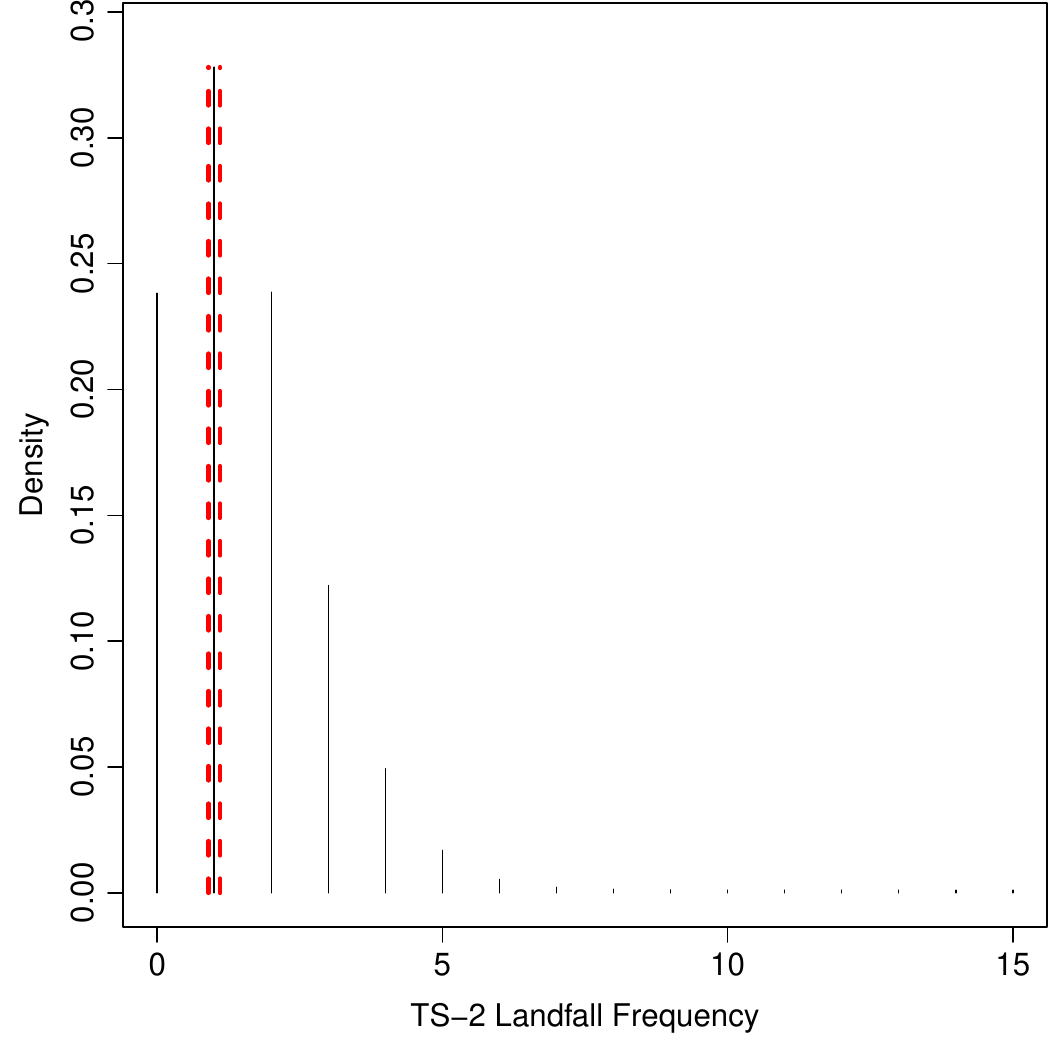}}
&{\includegraphics[width=0.32\textwidth]{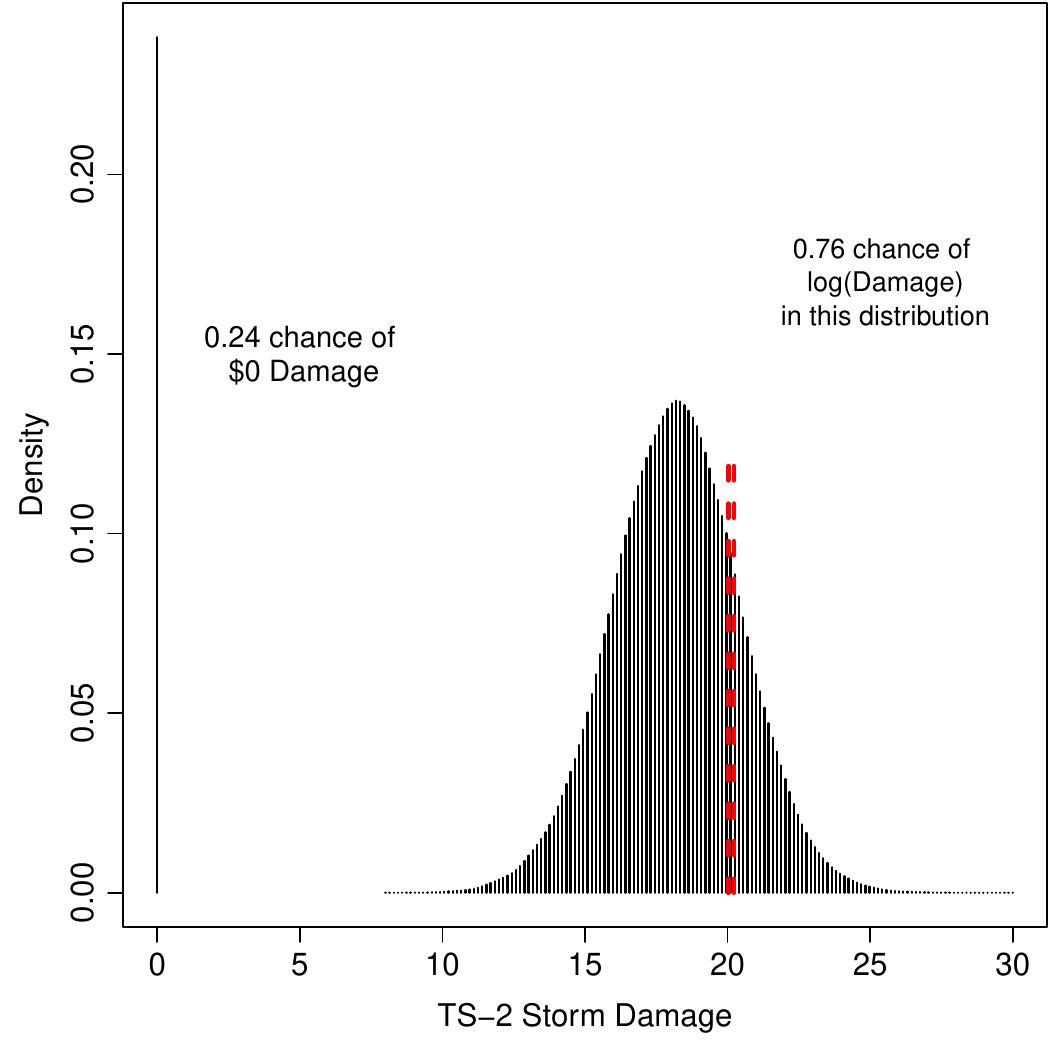}}
\\
{\includegraphics[width=0.30\textwidth]{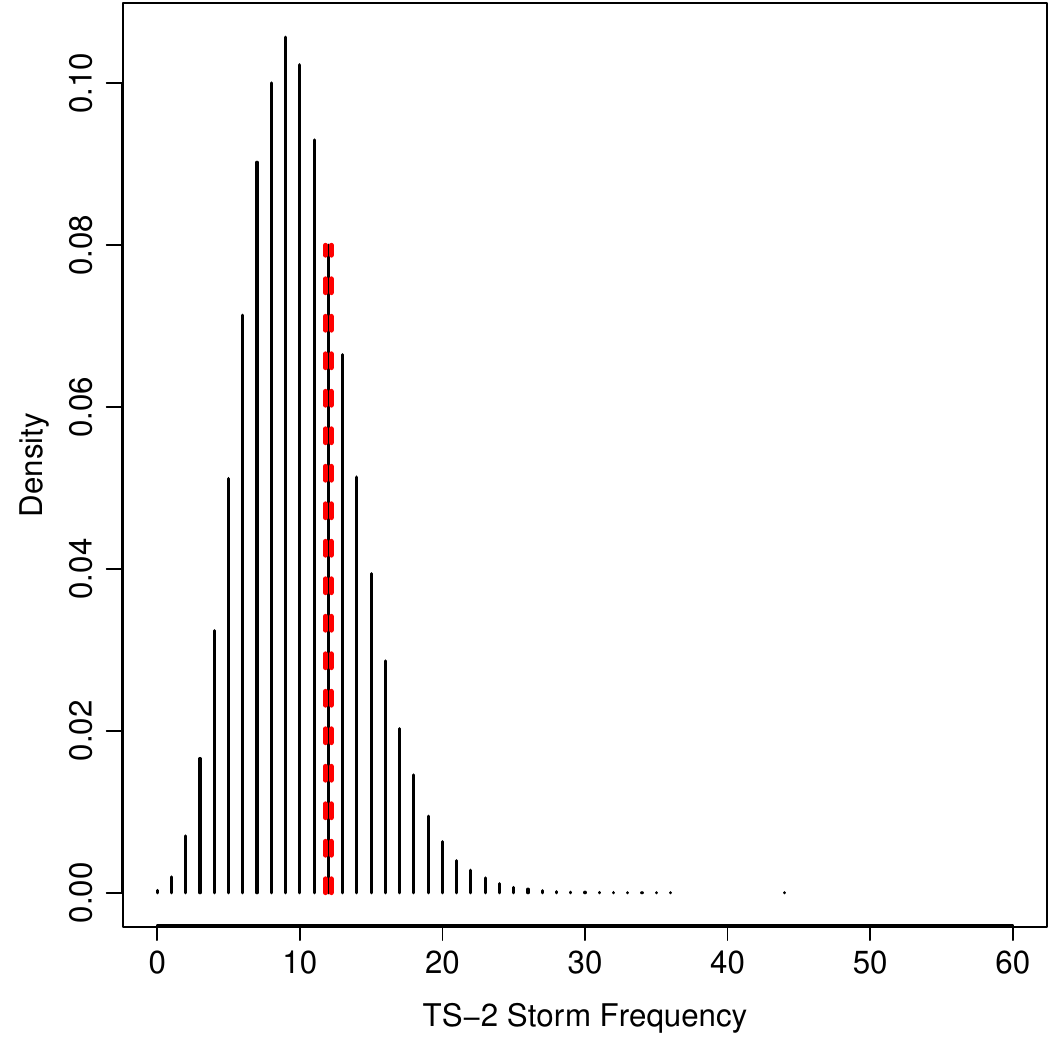}}
&
{\includegraphics[width=0.30\textwidth]{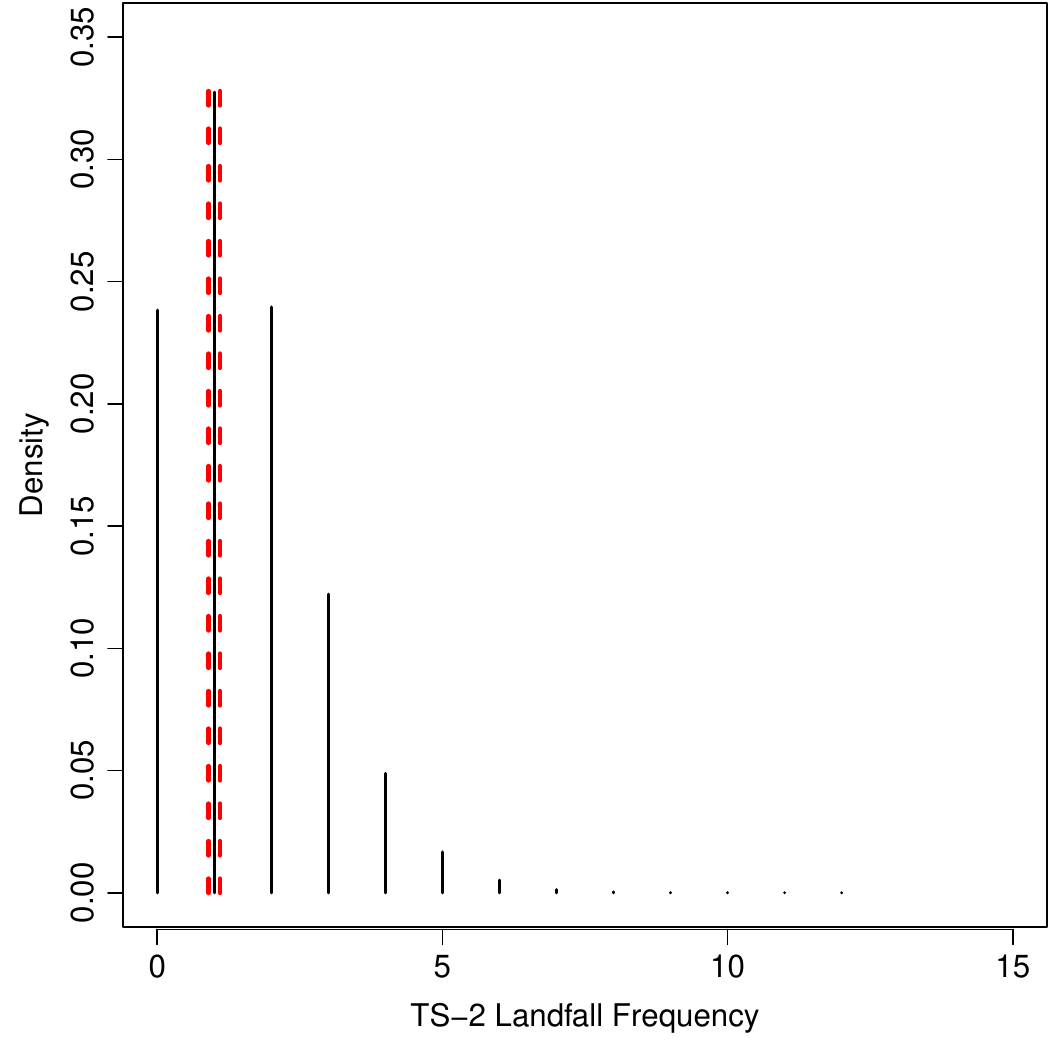}}
&
{\includegraphics[width=0.32\textwidth]{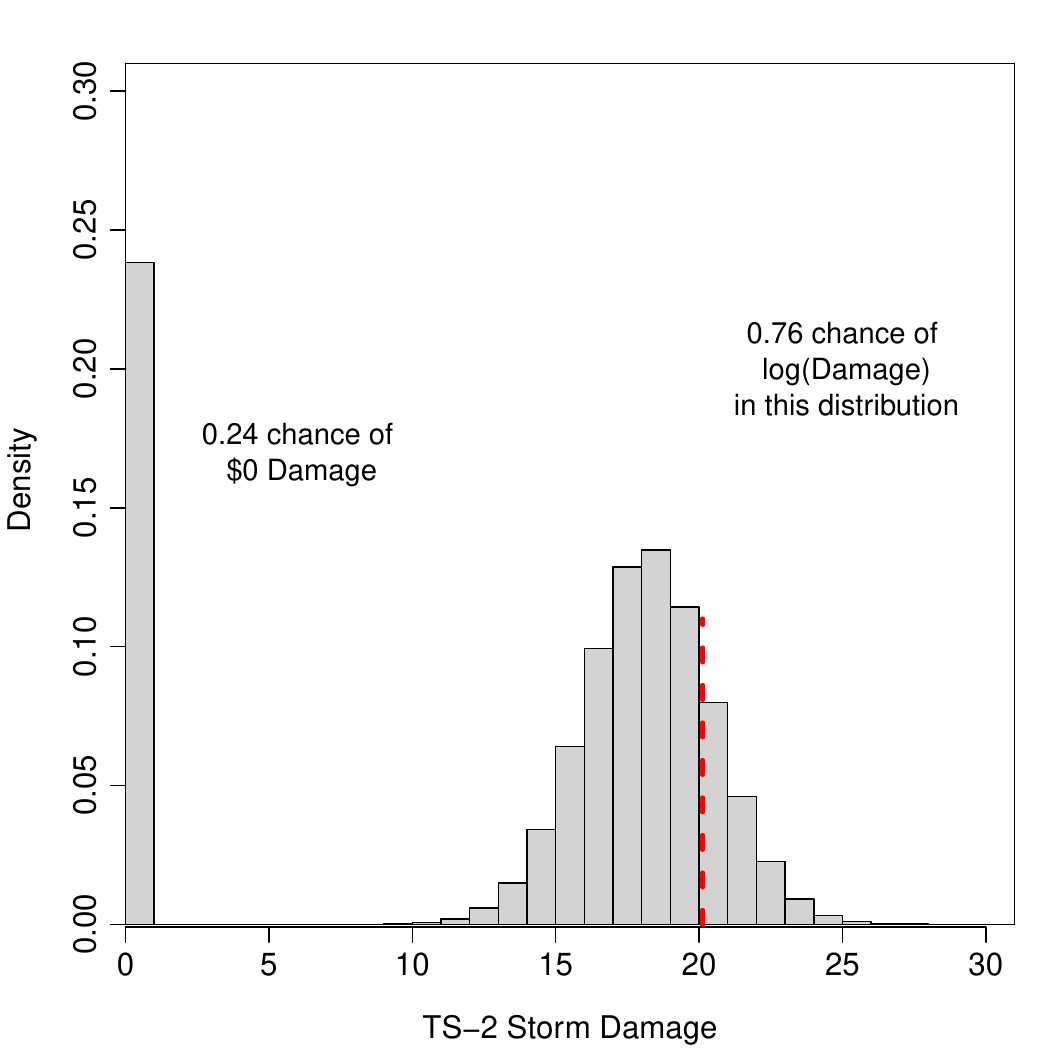}}
\end{tabular}
 \end{figure}

\begin{figure}[ht]
\centering
    \caption[Posterior Predictive Distributions for 2016 High Intensity Storms]{Posterior Predictive Distributions for 2016 High Intensity Storms. 
   The top row is for empirical Bayes prediction, the middle row is for fast Bayesian prediction and the bottom row is for hierarchical Bayesian  prediction. 
   The left column is the probability mass function for cyclone frequency, middle column is for damage-inflicting probability, and right column is density for logarithm of damages.
    The actual values  are displayed with red dashed lines.}
        \label{pp22016}
      \begin{tabular}{ccc}  
{\includegraphics[width=0.32\textwidth]{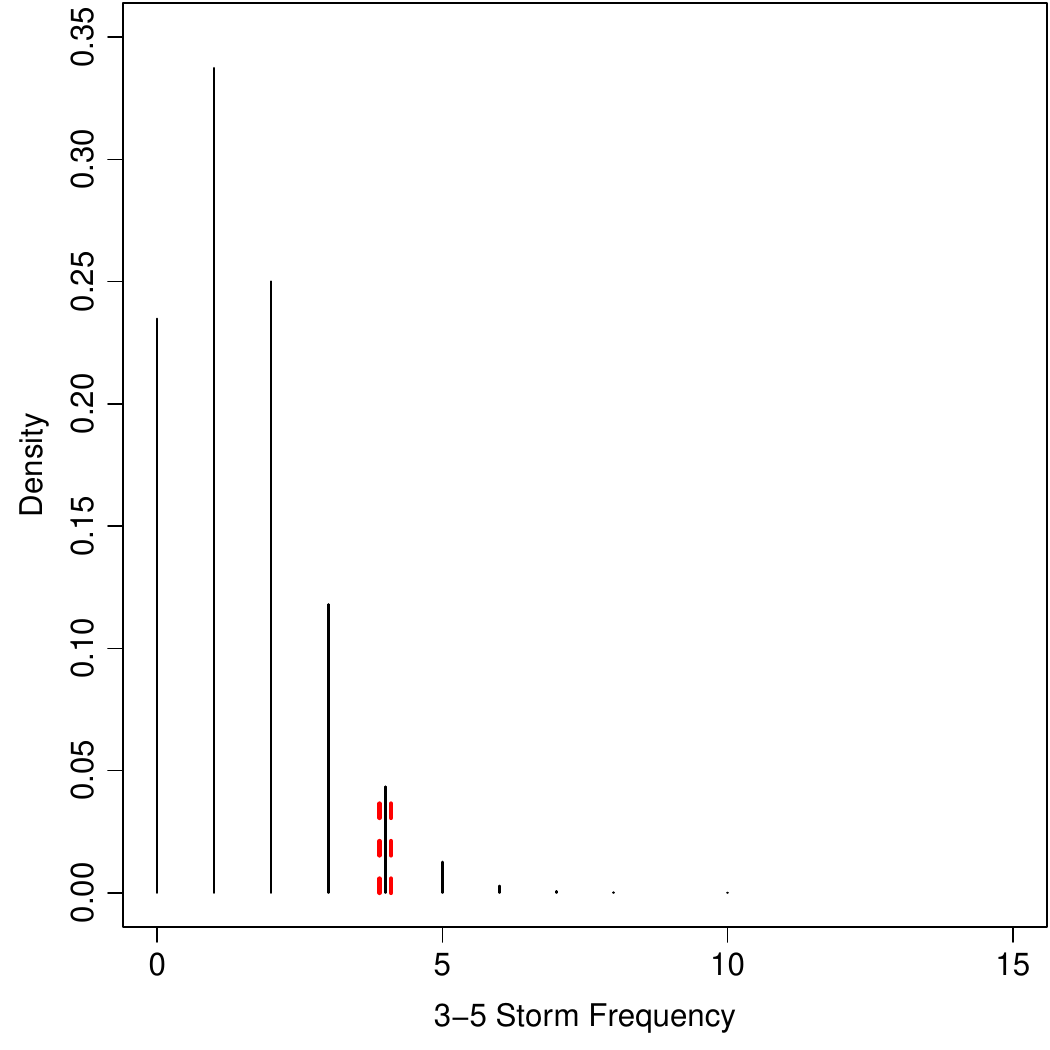}}
&
{\includegraphics[width=0.32\textwidth]{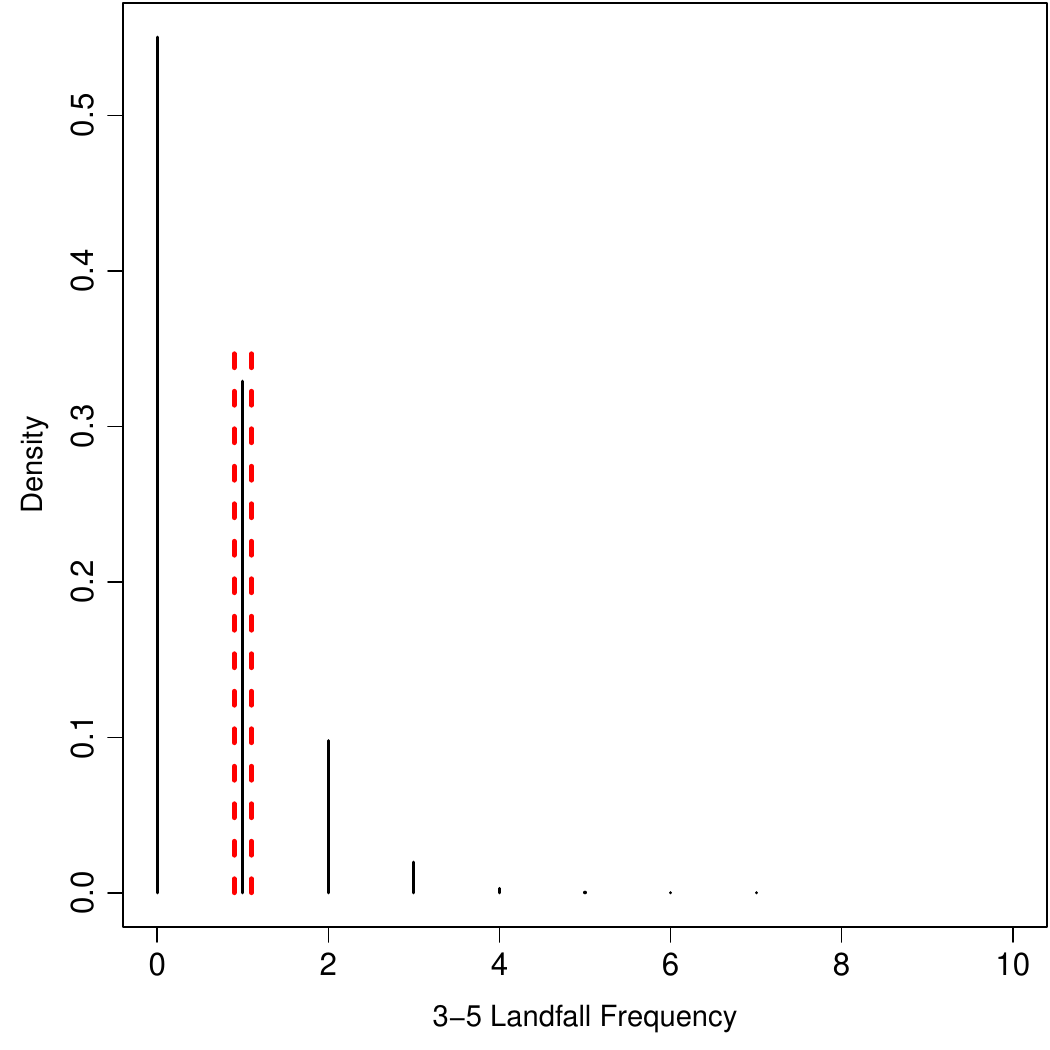}}
&
{\includegraphics[width=0.32\textwidth]{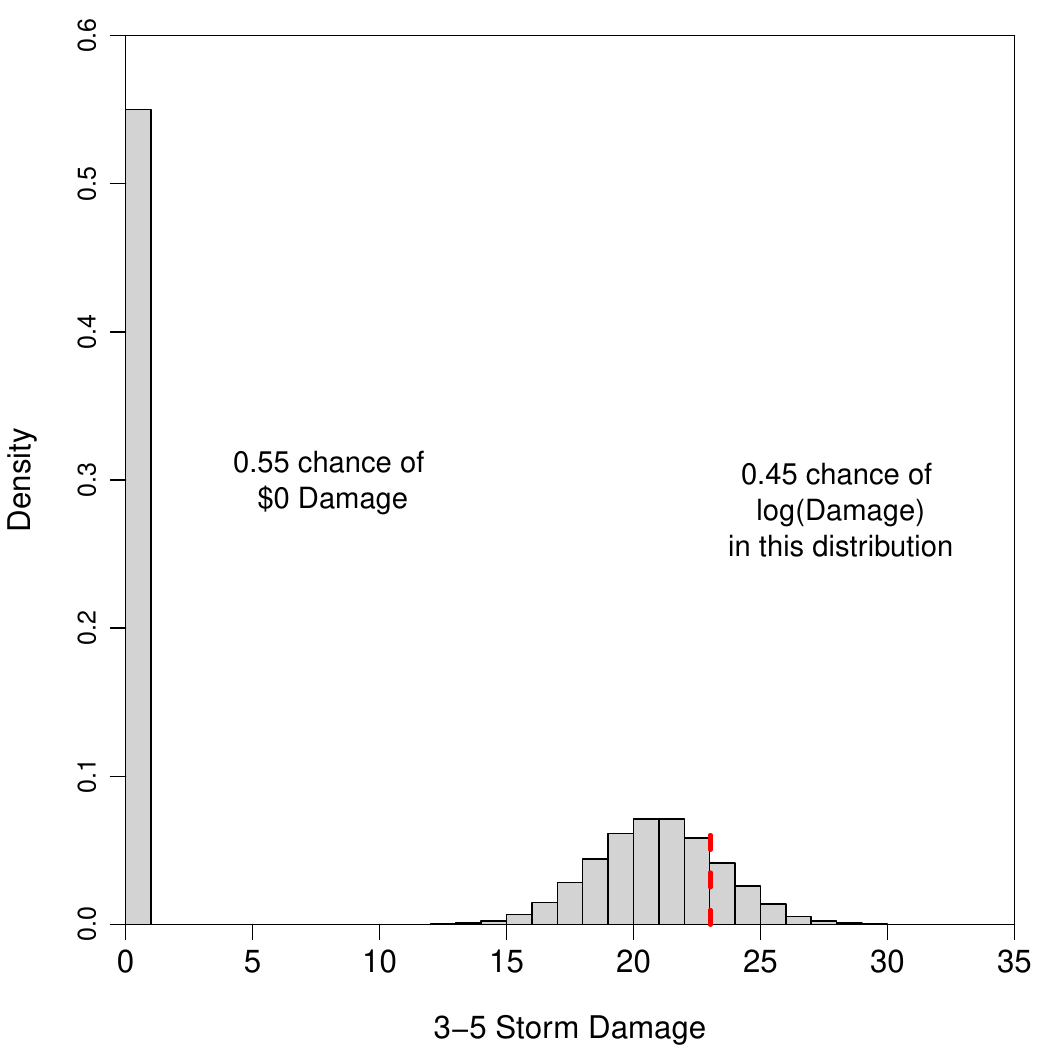}}\\
{\includegraphics[width=0.32\textwidth]{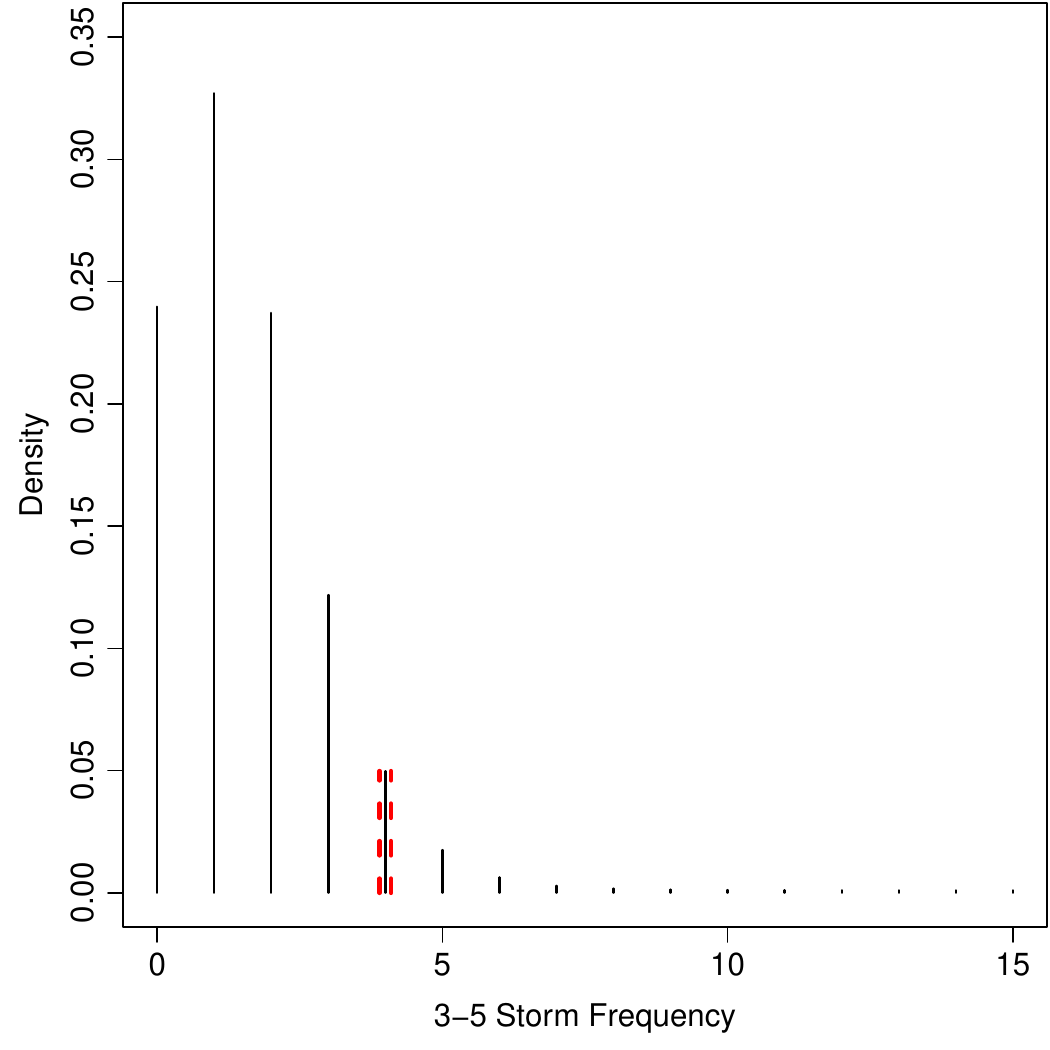}}
&{\includegraphics[width=0.32\textwidth]{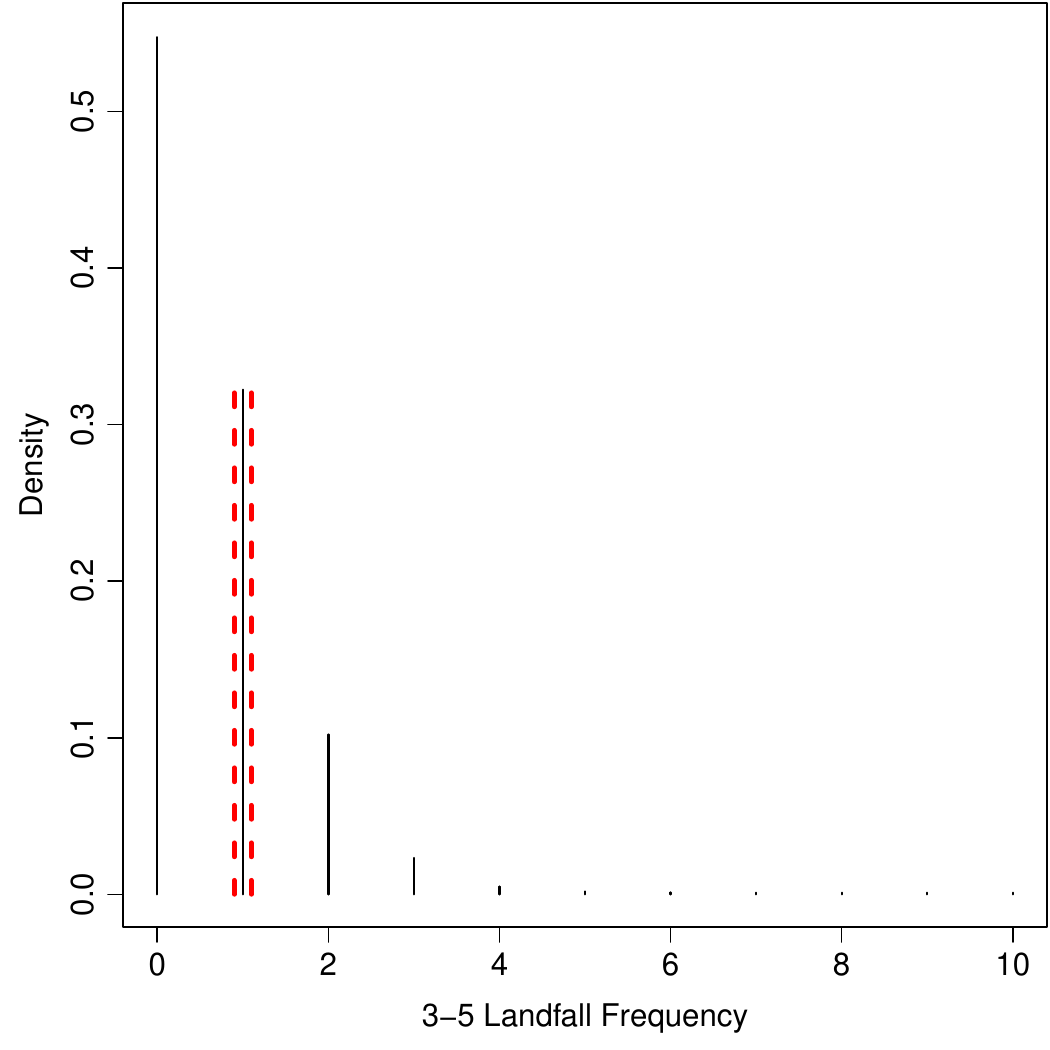}}
&{\includegraphics[width=0.32\textwidth]{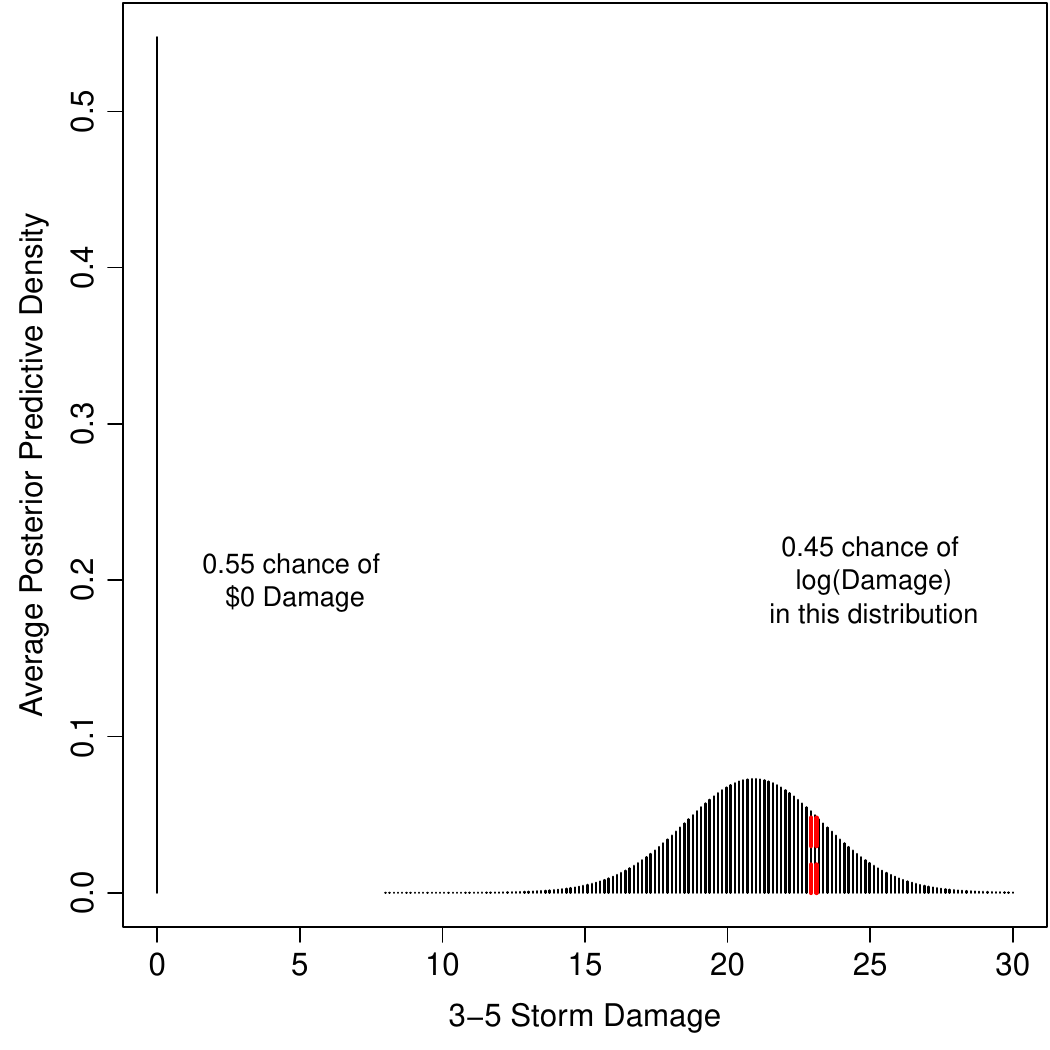}}\\
{\includegraphics[width=0.32\textwidth]{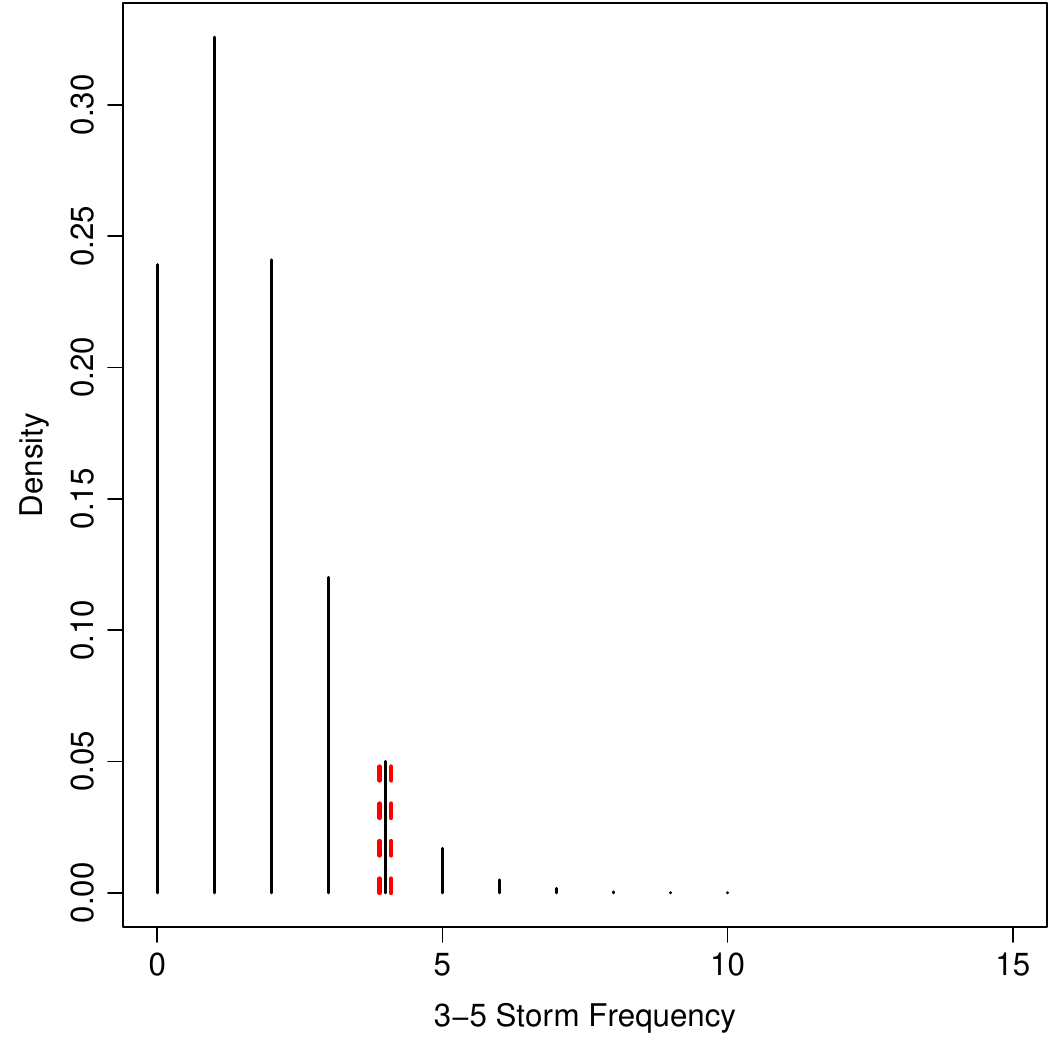}}
&{\includegraphics[width=0.32\textwidth]{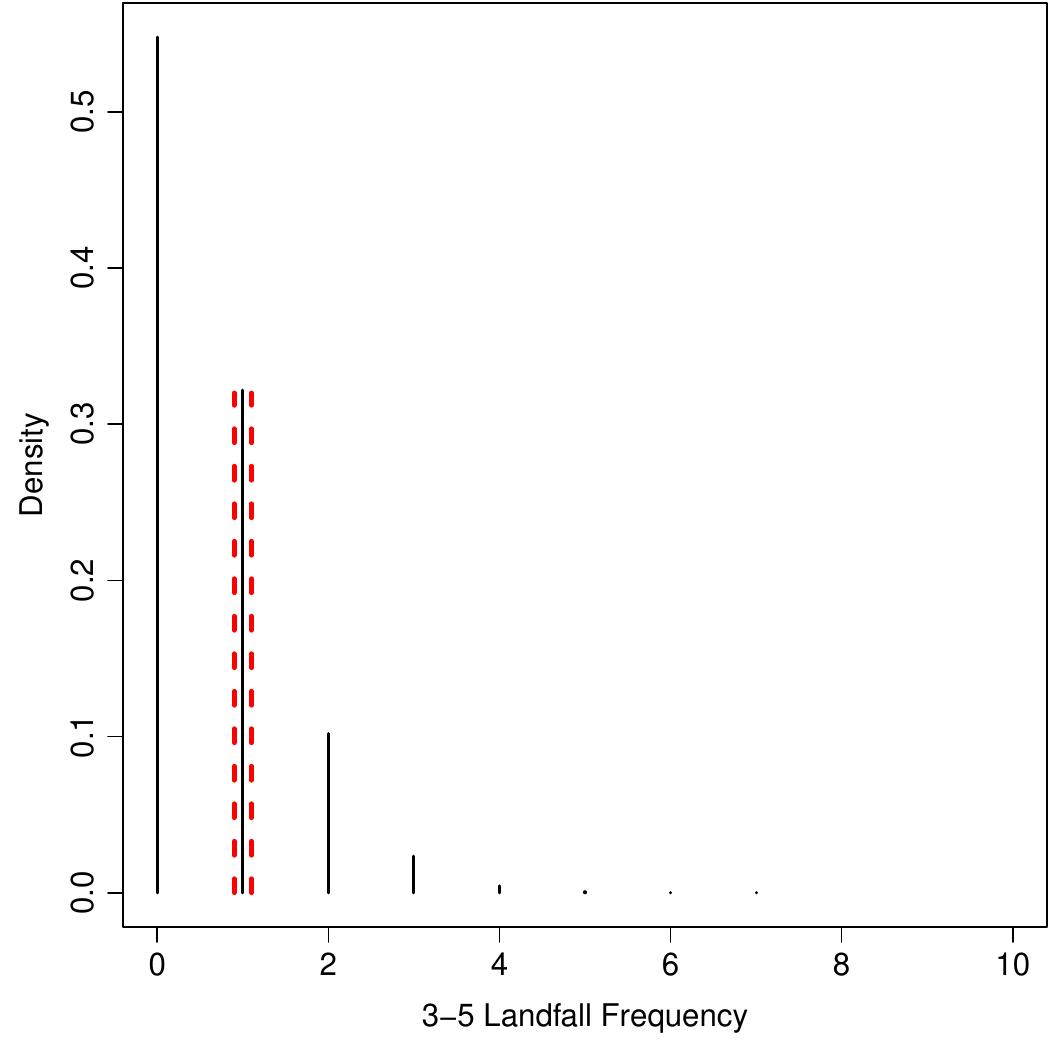}}
&{\includegraphics[width=0.32\textwidth]{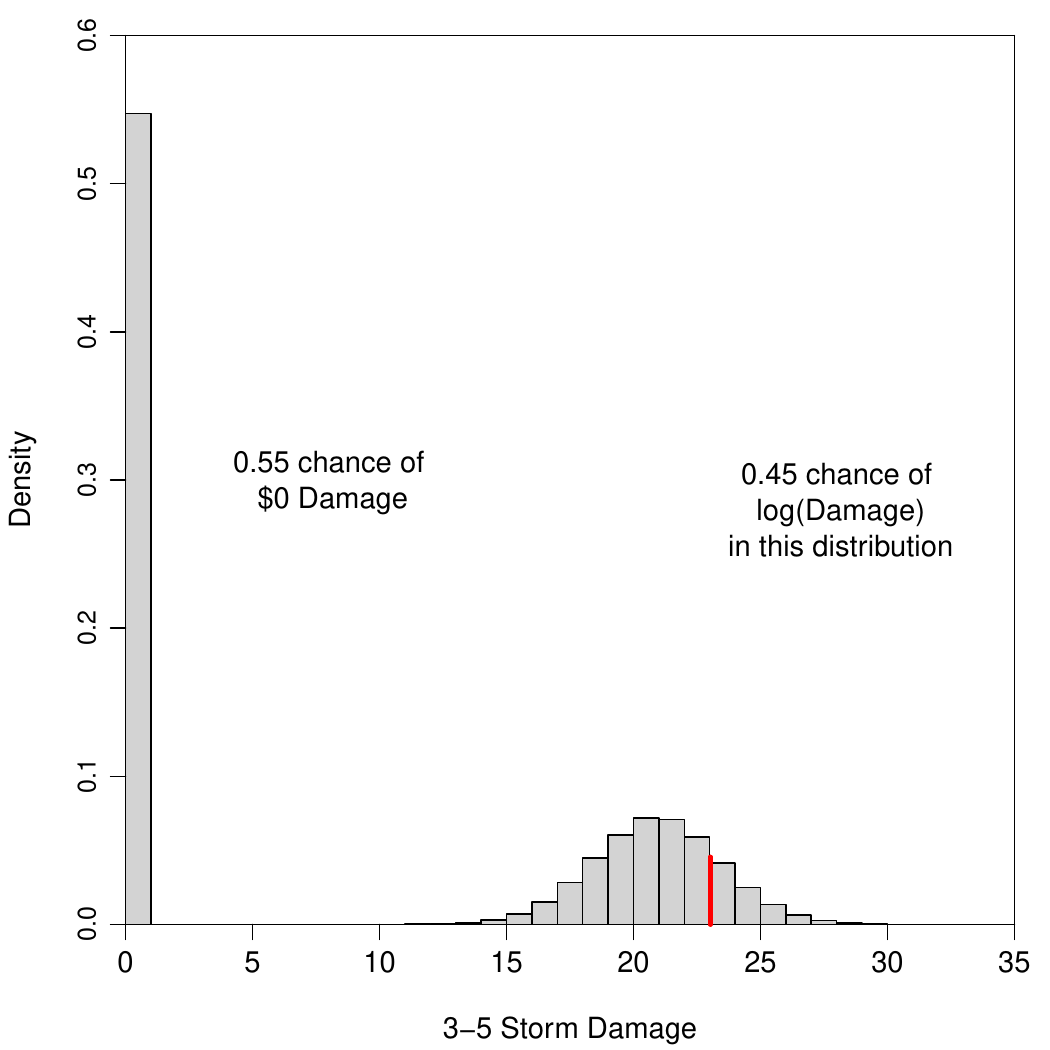}}
\end{tabular}
 \end{figure}

 The results are presented in Figure~\ref{pp12016} for low intensity storms and 
 in Figure~\ref{pp22016} for high intensity storms.
The results for 2016 indicate the model matches well with the data.  The three posterior predictive methods provide similar output except that the second method has slightly more variability and provides a smoother density curve because of the averaging.  As seen in Figure~\ref{pp12016} (three figures in first (left) panel/column), the actual number of storms was well within the predicted distribution for the low intensity category and had at least 30\% of the probability higher than the observed value.  
Similarly, in 
Figure~\ref{pp22016}(first column), high intensity storm occurrence, although falling at the start of the right tail end, was within the limits of the predictive distribution.  

For  low intensity  tropical storms, the predictive distribution in Figure~\ref{pp12016}(three figures in the second (middle) column) indicated that a given storm would inflict damages in 2016, was given approximately a 31-32\% chance in each of the three versions of posterior predictive distributions.  The corresponding figures for  high intensity damage-inflicting storms  in Figure~\ref{pp22016}(second column) are around 31-32\%. There were \$550 million in low intensity damages, thus, 
Figure~\ref{pp12016}(three figures in the third (right) column) reflect the actual amount of damage for 2016 falls well within the predicted distribution
 for the low intensity category with non-zero damages. There were a \$ 1 billion in high intensity damages in 2016. The high intensity damages in 
 Figure~\ref{pp22016}(third column) indicate that the true value of damage falls well within the posterior predictive range for the non-zero values. Although, there is a greater chance of no (\$0) damages (about 55\%) and the log of damage incurred in 2016 had about an 8\% chance of occurence.

  On the other hand, 2019 was a much milder year with no damages. As seen in Figure~\ref{fig:2019_pred}, the actual number of cyclones, landfall frequency and damages are well within the predicted distribution for the low intensity category as well as the high intensity category. 
For 2019, there were no recorded damages in low intensity as well as high intensity cyclones. It can be seen from Figure~\ref{fig:2019_pred} (lower middle plot), that observing zero landfalling high intensity cyclones had the highest chance in at around 36-37\%. Also, the predictive distribution plots in 
Figure~\ref{fig:2019_pred} reflect the considerable chance of no damages, with about a 22\% chance for the low intensity category and 38\% change for the high intensity category, respectively.

\clearpage

 \begin{figure}[h!]
\centering
    \caption[Posterior Predictive Distributions for 2019 Tropical Cyclones]{Posterior Predictive Distributions for 2019 tropical cyclones. 
    The upper row is for the low intensity case, the bottom row is for the 
    high intensity case. The left column displays the probability mass function of the Bayesian predictive distribution for frequency of cyclones, the middle column is the predicted probability mass function of a cyclone to cause damage, and the right column is the predictive density for damages.
    The actual observed values  are displayed with red dashed lines.}
        \label{fig:2019_pred}
\begin{tabular}{ccc}
{\includegraphics[width=0.30\textwidth]{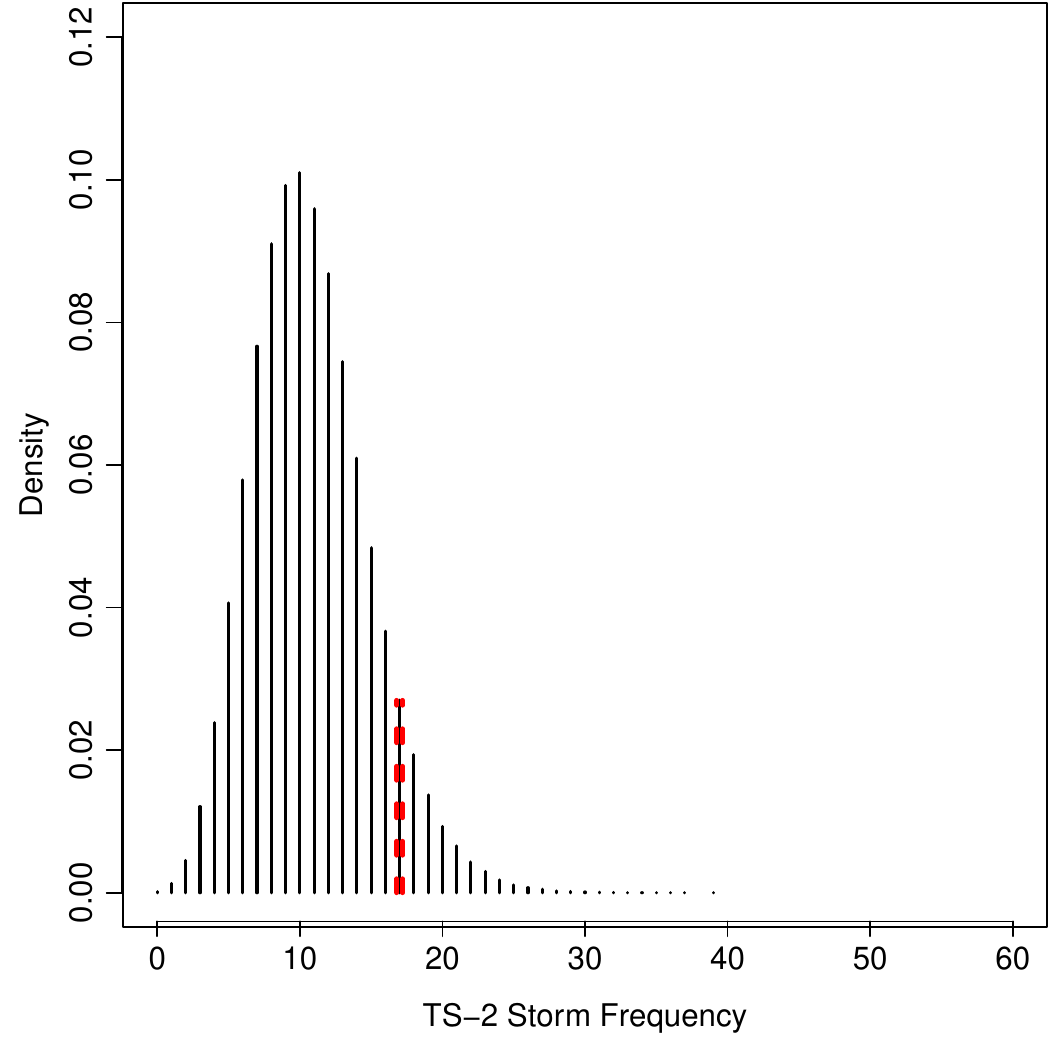}}
&
{\includegraphics[width=0.30\textwidth]{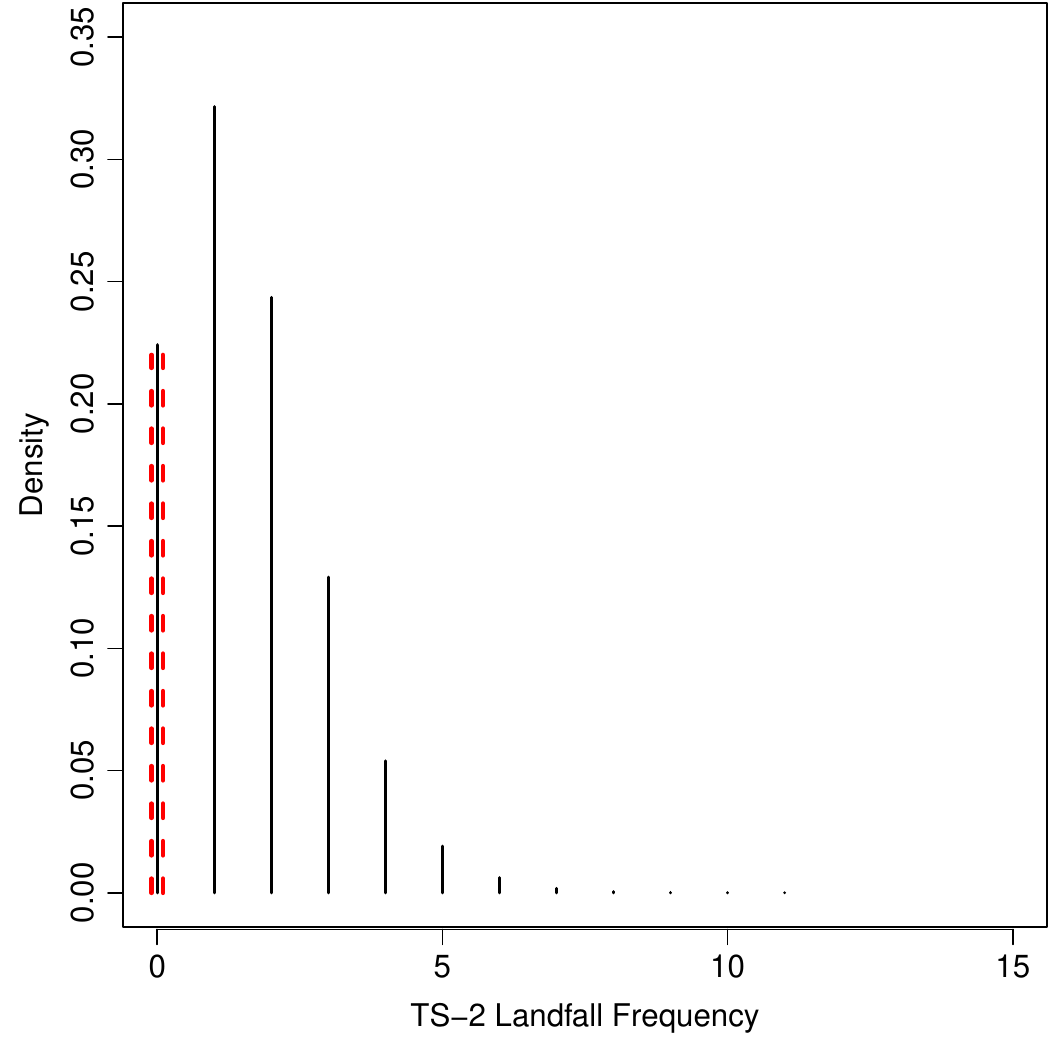}}
&
{\includegraphics[width=0.32\textwidth]{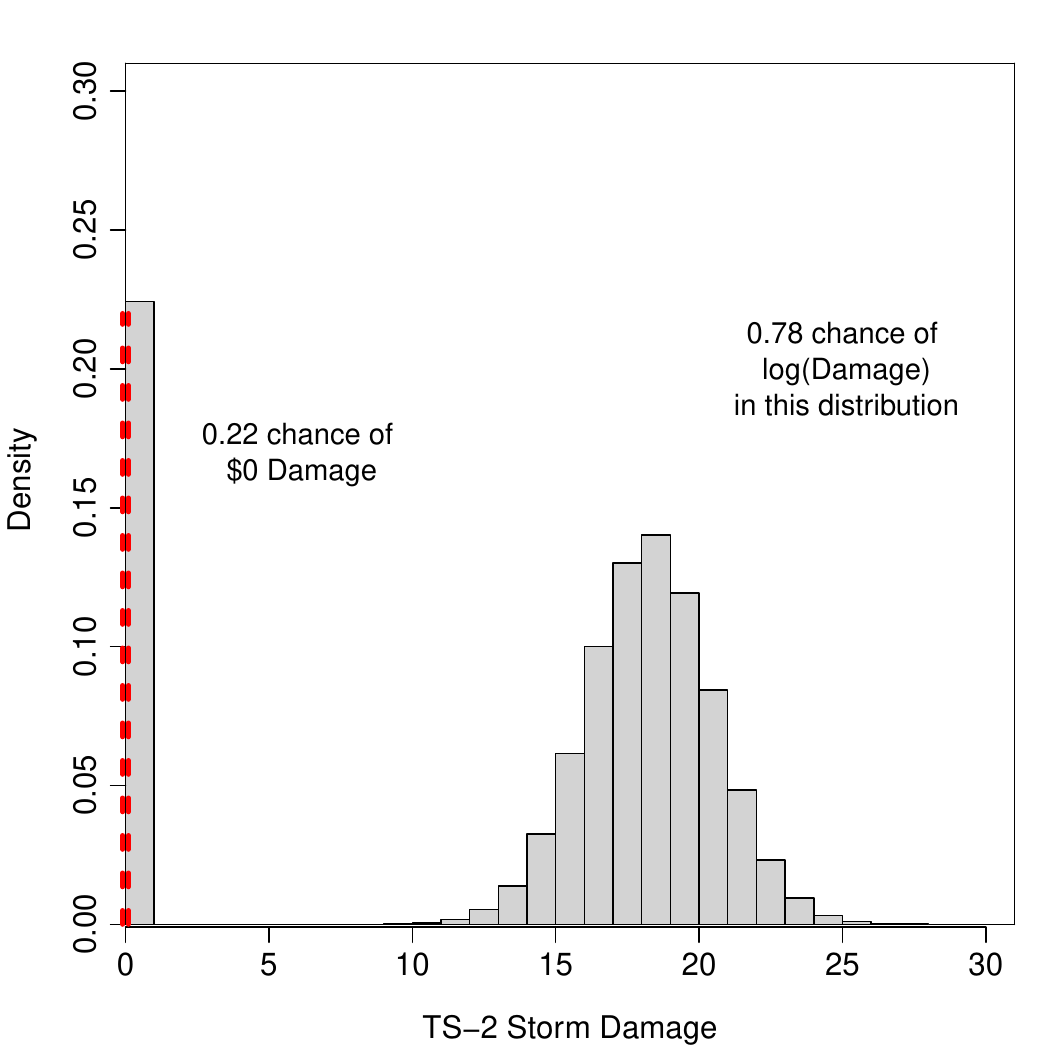}}
\\ 
{\includegraphics[width=0.30\textwidth]{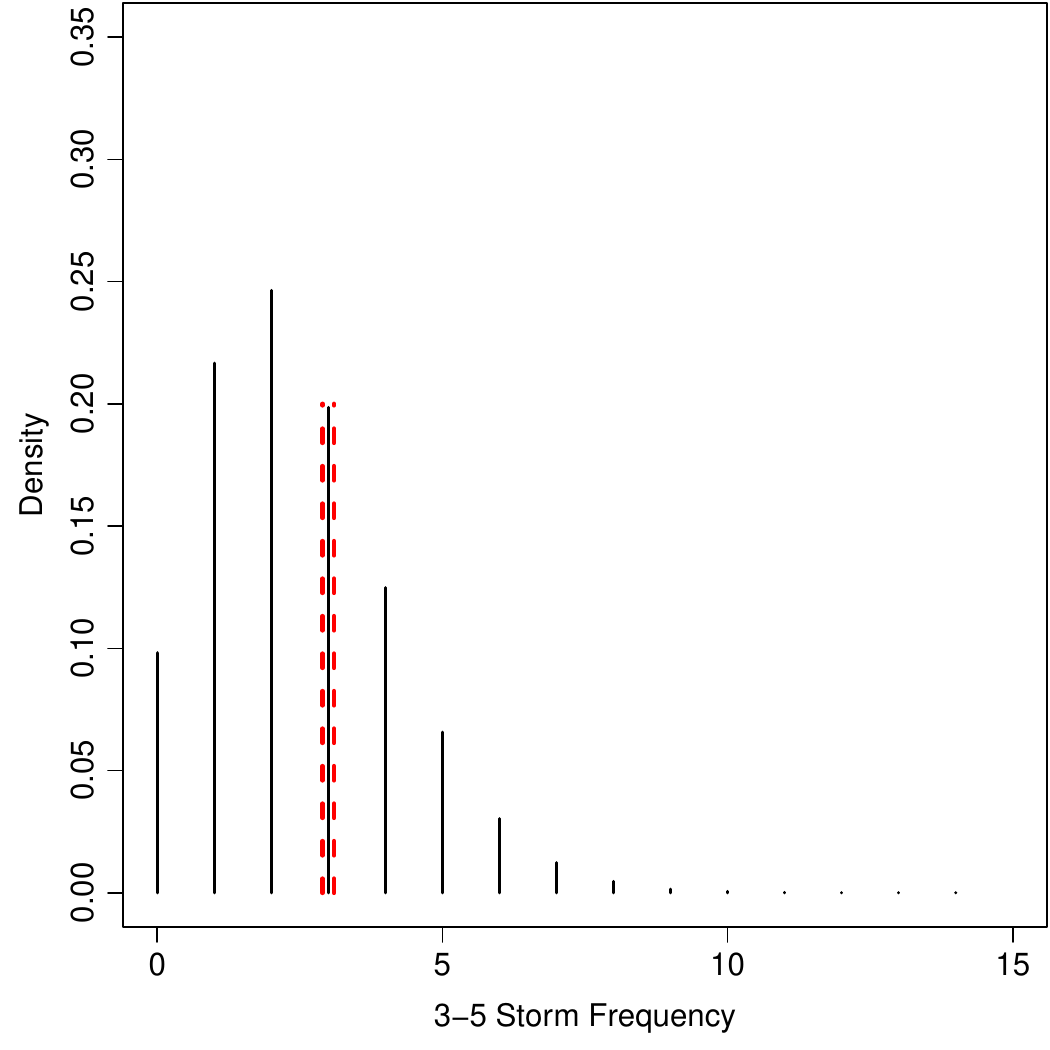}}
&
{\includegraphics[width=0.30\textwidth]{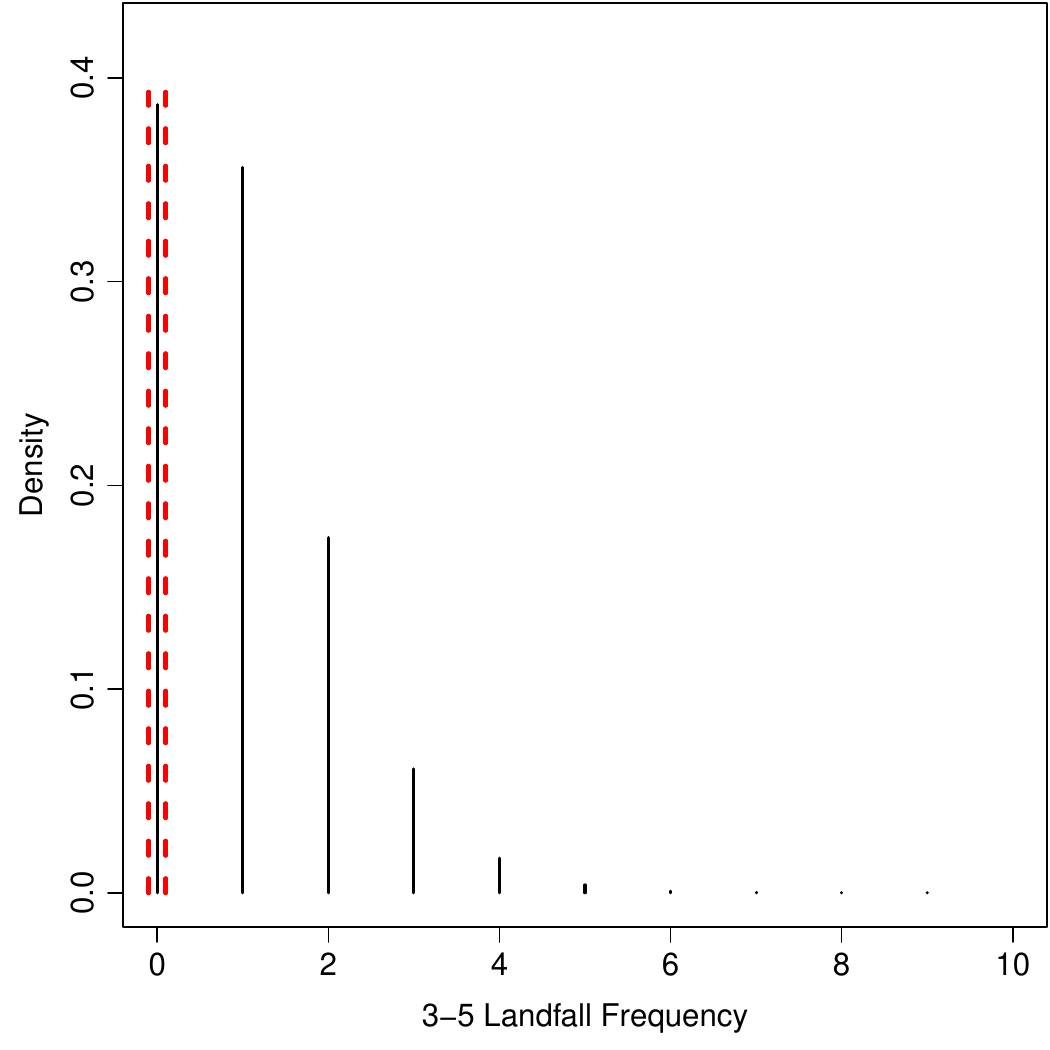}}
&
{\includegraphics[width=0.32\textwidth]{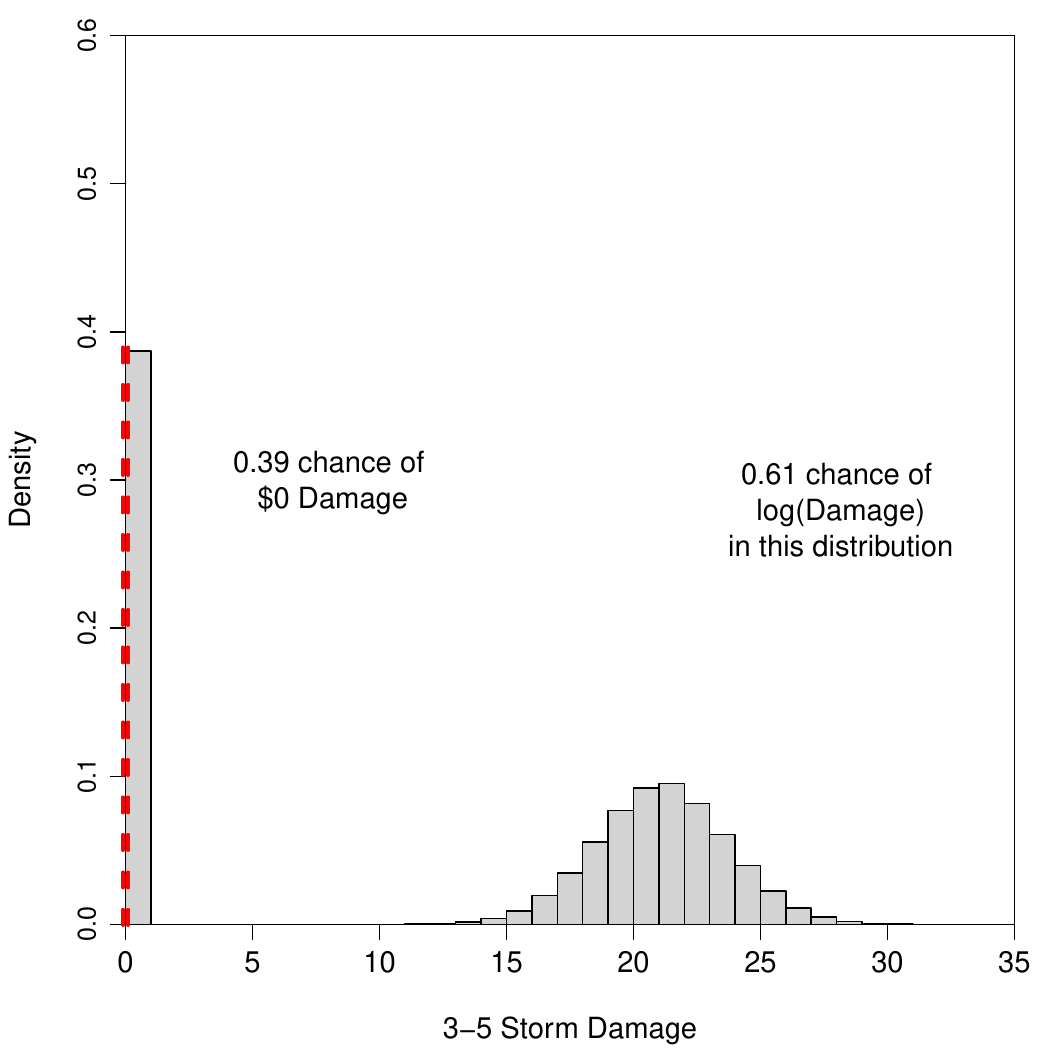}}
\end{tabular}

 \end{figure} 

\subsection{2016 Posterior Prediction Analysis for Individual Storms}
\begin{figure}[h!] 
\centering
    \caption[Posterior Predictive Distributions for Storms in 2016]{Posterior Predictive Distributions for minCP, maxWS and damages of tropical storms Hermine and Matthew in 2016, based on  the hierarchical Bayesian GEV model.  The actual values for the storms  are displayed with red dashed lines.}
        \label{fig:hermine_matthew_hierEVD}
 
\begin{tabular}{ccc}
{\includegraphics[width=0.30\textwidth]{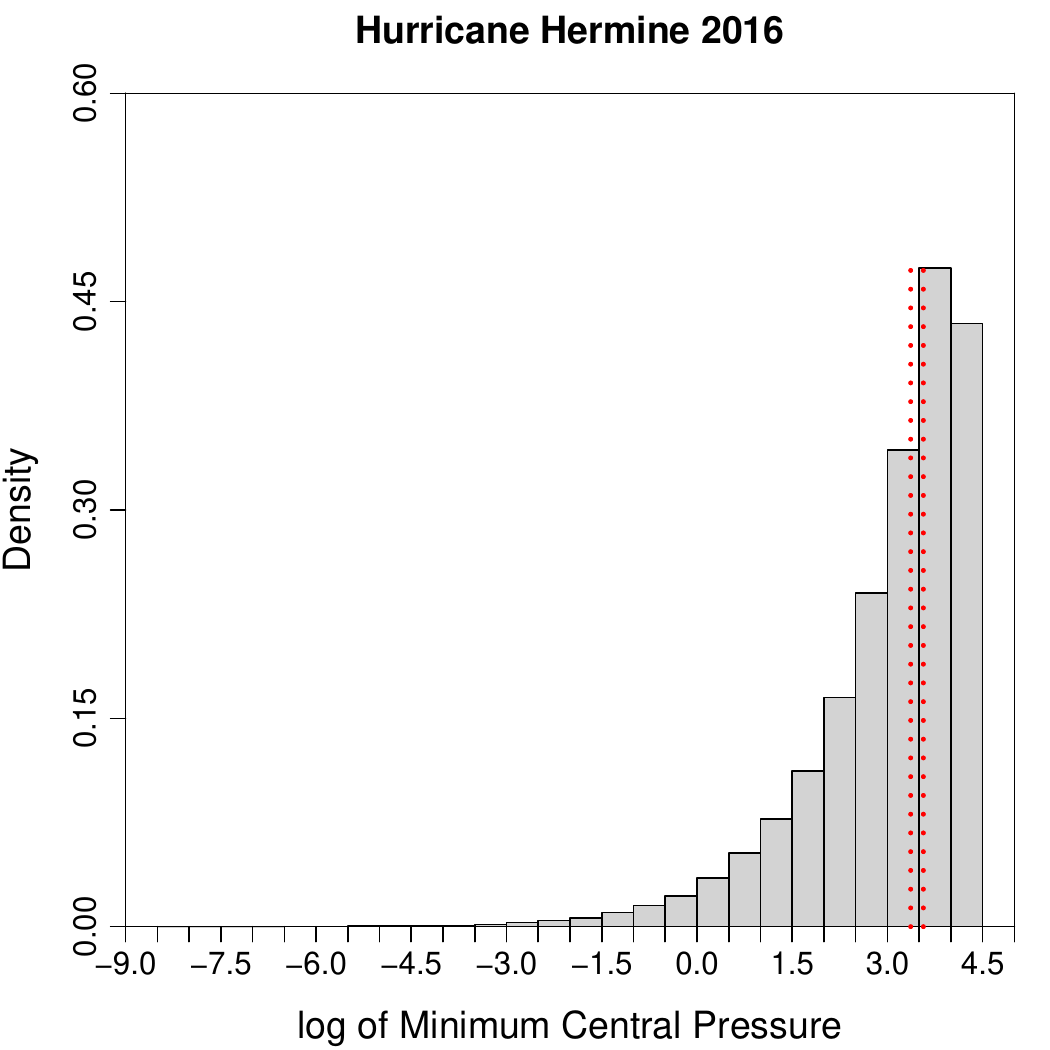}}
&
{\includegraphics[width=0.30\textwidth]{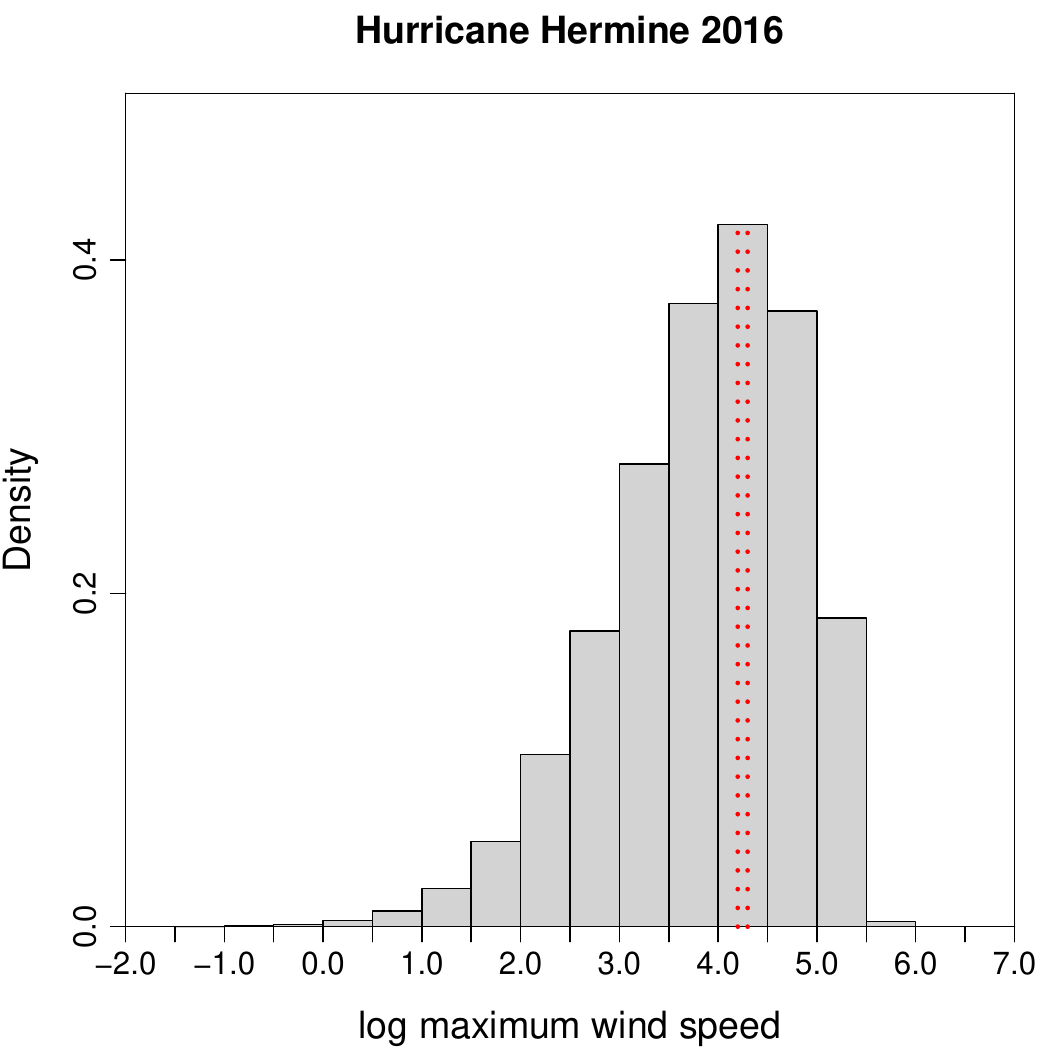}}
&
{\includegraphics[width=0.30\textwidth]{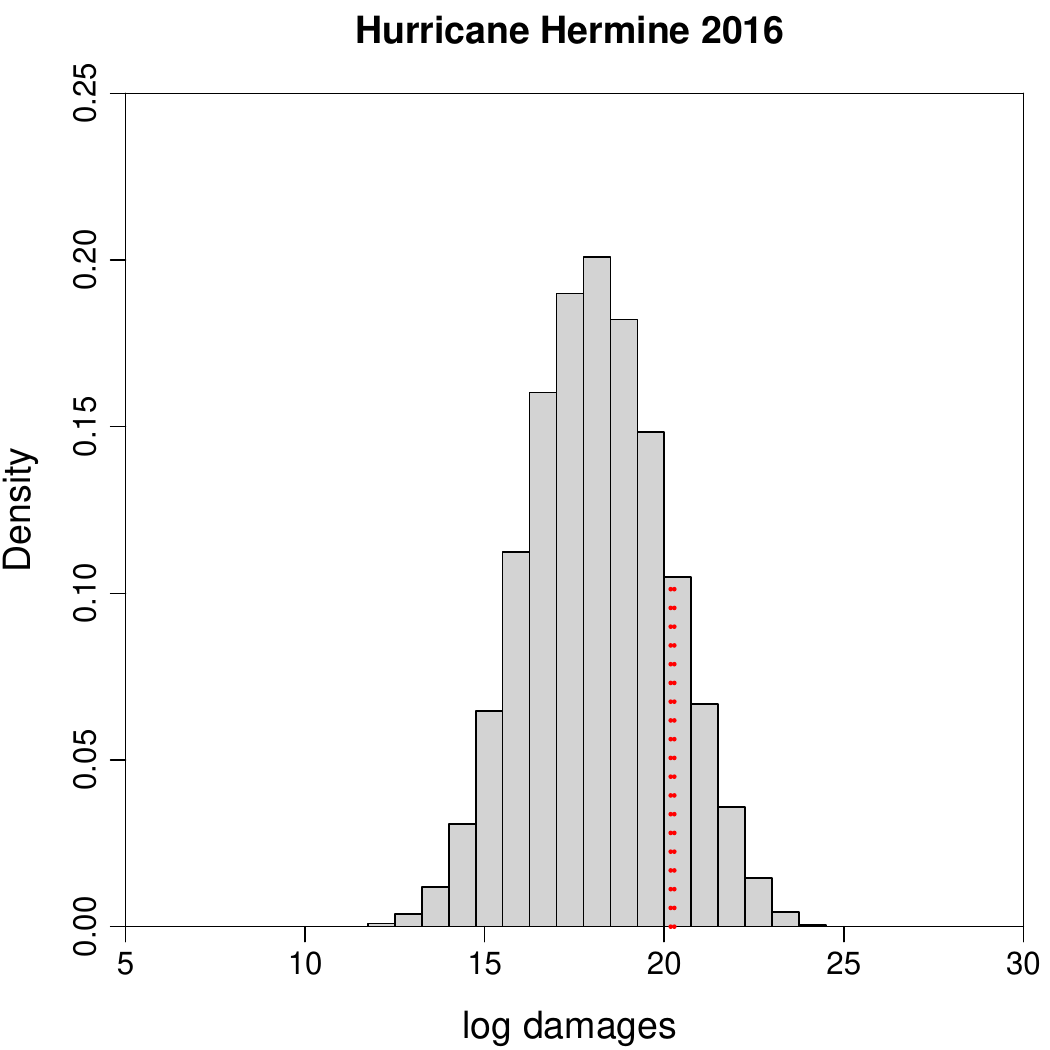}}
\\
{\includegraphics[width=0.30\textwidth]{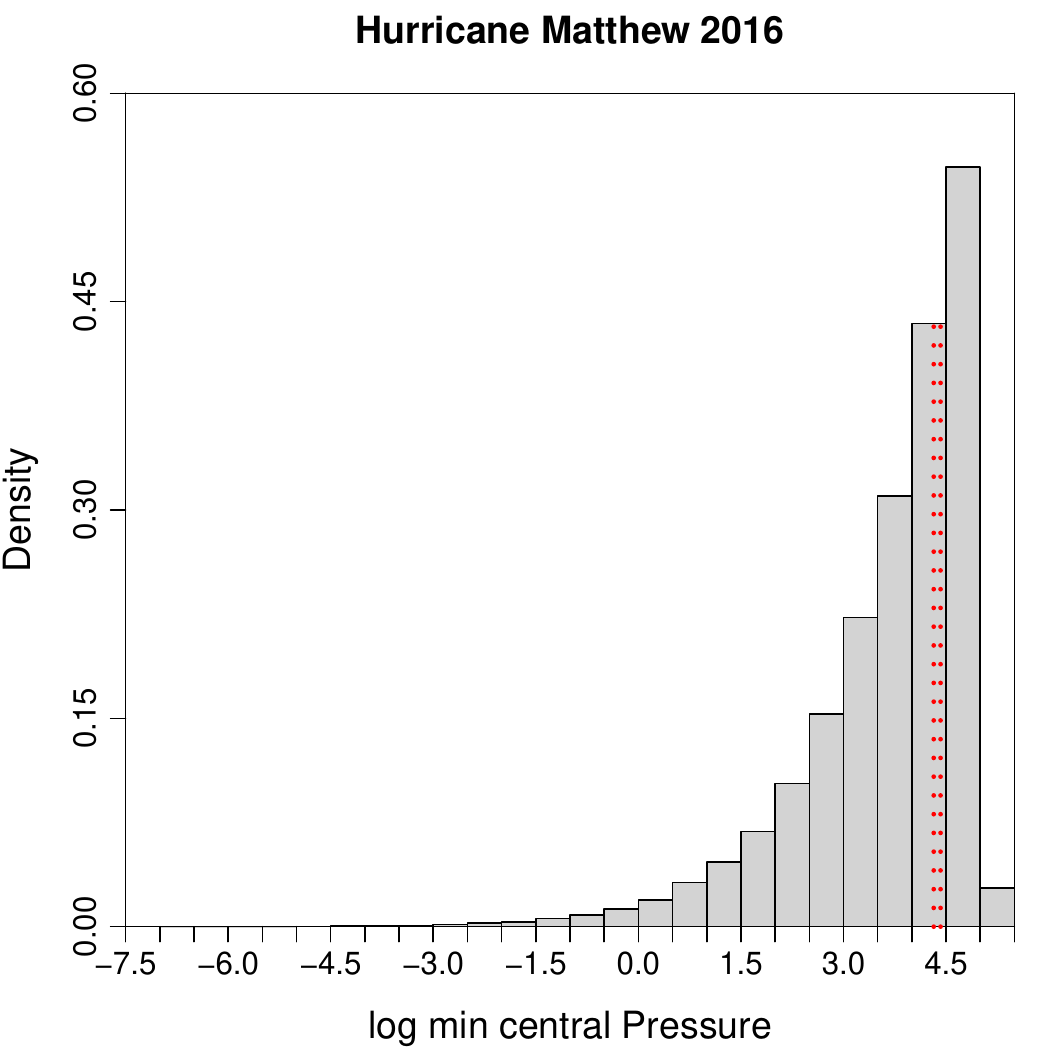}}
&
{\includegraphics[width=0.30\textwidth]{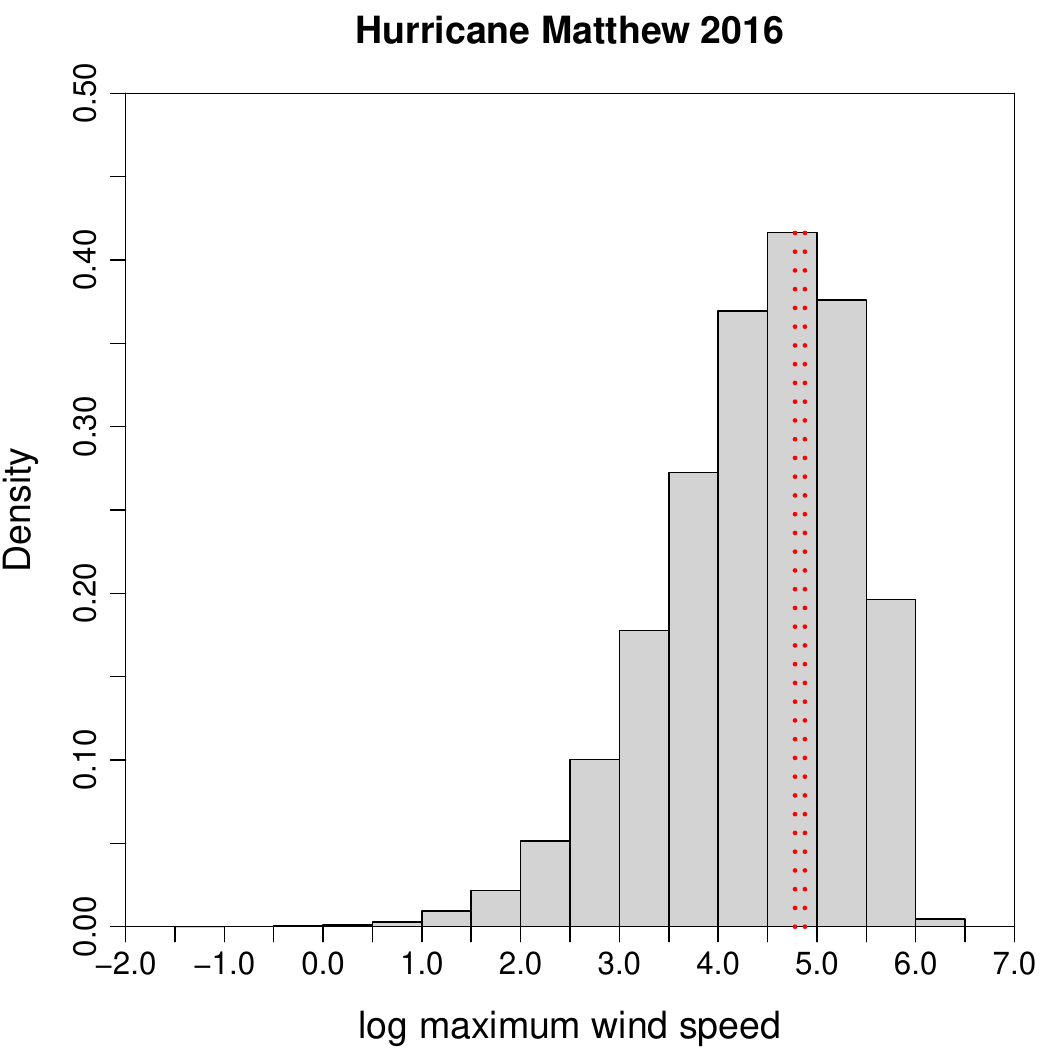}}
&
{\includegraphics[width=0.30\textwidth]{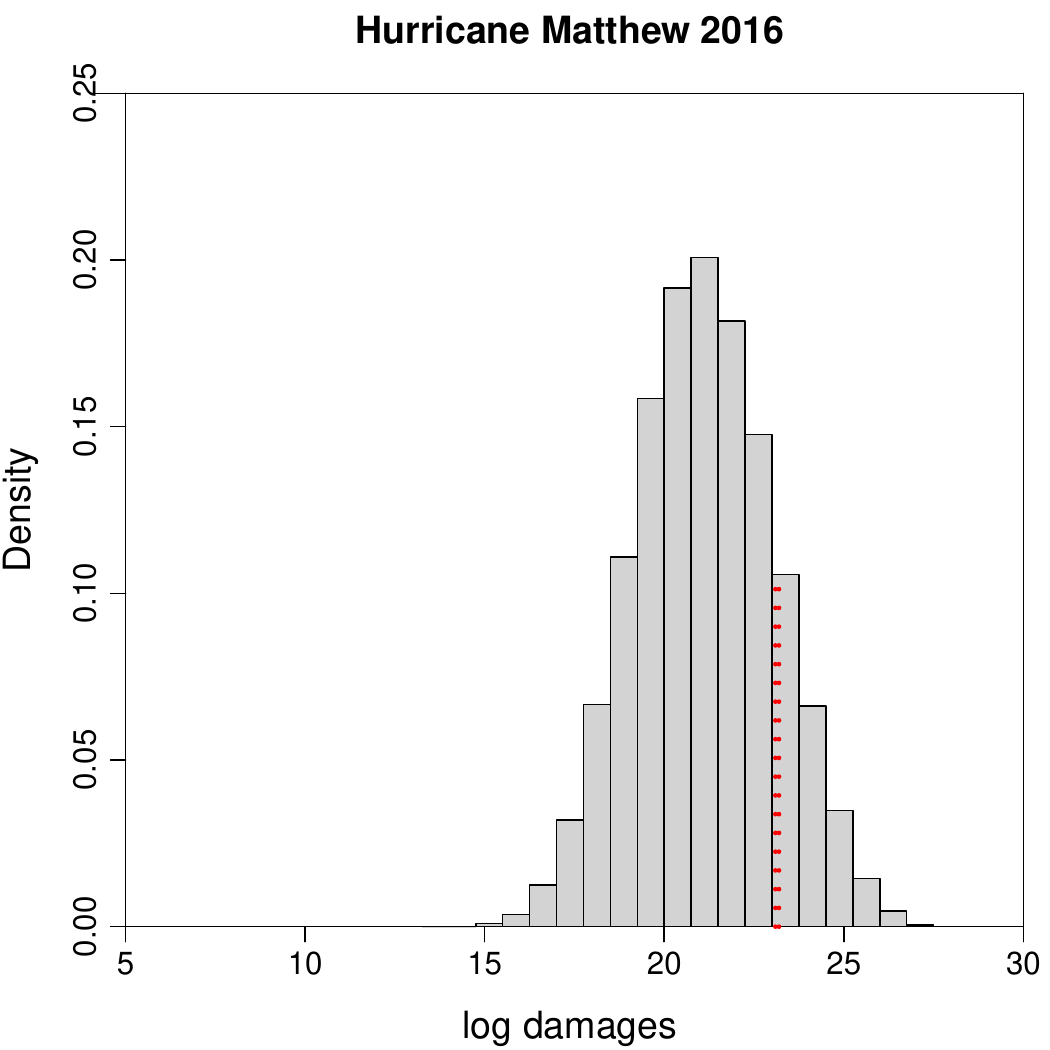}}
\end{tabular}
\end{figure}

There were 2 storms that hit continental U.S. in 2016 with non-zero damages: Hermine and Matthew with normalized damage values 610 million and 11 billion, respectively. We fit a Hierarchical Bayesian GEV model (Section \ref{Sec:Storm_Model}) assess the performance of our model by comparing where the true values for the three variables (maxWS, minCP and damages) fall on the posterior predictive distribution. For each of the two storms, we find the 95\% credible intervals and check whether the actual storm maxWS, minCP and damages  are included in  these intervals or not. It is observed that the true minimum central pressure values, the true maximum wind speeds and true damages were within the 95\% credible interval for both the storms as can be seen from Figure \ref{fig:hermine_matthew_hierEVD}. 
\end{document}